\documentclass[numberedappendix]{emulateapj}
\usepackage{apjfonts}

\newcommand{\bxbm}{BX/BM}
\newcommand{\bmbx}{BM/BX}

\newcommand{\etal}{et al.\,}
\newcommand{\nobs}{80}

\newcommand{\lhaobs}{$L^{\rm obs}({\rm H\alpha})$}
\newcommand{\lhao}{$L^{0}({\rm H\alpha})$}
\newcommand{\lhaoo}{$L^{00}({\rm H\alpha})$}
\newcommand{\lhapred}{$L^{0}_{\rm pred.}({\rm H\alpha})$}
\newcommand{\whasinf}{$W^{\rm rest}_{\rm SINF}({\rm H\alpha})$}
\newcommand{\whabb}{$W^{\rm rest}_{\rm BB}({\rm H\alpha})$}
\newcommand{\whaoo}{$W^{\rm rest,00}_{\rm BB}({\rm H\alpha})$}
\newcommand{\whapred}{$W^{\rm rest}_{\rm pred.}({\rm H\alpha})$}
\newcommand{\luvobs}{$L^{\rm obs}({\rm UV})$}

\newcommand{\sfrha}{$\rm SFR^{0}(H\alpha)$}
\newcommand{\sfrhaoo}{$\rm SFR^{00}(H\alpha)$}
\newcommand{\sfruv}{$\rm SFR^{0}(UV)$}
\newcommand{\sfrsed}{$\rm SFR(SED)$}
\newcommand{\ndet}{63}
\newcommand{\nsins}{62}
\newcommand{\nsinsdet}{52}

\slugcomment{Accepted for publication in the Astrophysical Journal}

\shorttitle{SINS survey of high redshift galaxies}
\shortauthors{F\"orster Schreiber \etal}

\begin{document}

\title{
The SINS survey:
SINFONI Integral Field Spectroscopy of $z \sim 2$ Star-forming Galaxies
\,\altaffilmark{1}}

\author{N. M. F\"orster Schreiber\altaffilmark{2},
        R. Genzel\altaffilmark{2,3},
        N. Bouch\'e\altaffilmark{2},
        G. Cresci\altaffilmark{2},
        R. Davies\altaffilmark{2},
        P. Buschkamp\altaffilmark{2},
        K. Shapiro\altaffilmark{4},        
        L. J. Tacconi\altaffilmark{2},        
        E. K. S. Hicks\altaffilmark{2},
        S. Genel\altaffilmark{2},
        A. E. Shapley\altaffilmark{5},
        D. K. Erb\altaffilmark{6},
        C. C. Steidel\altaffilmark{7},
        D. Lutz\altaffilmark{2},
        F. Eisenhauer\altaffilmark{2},
        S. Gillessen\altaffilmark{2},
        A. Sternberg\altaffilmark{8},
        A. Renzini\altaffilmark{9},
        A. Cimatti\altaffilmark{10},
        E. Daddi\altaffilmark{11},
        J. Kurk\altaffilmark{12},
        S. Lilly\altaffilmark{13},
        X. Kong\altaffilmark{14},
        M. D. Lehnert\altaffilmark{15},
        N. Nesvadba\altaffilmark{16},
        A. Verma\altaffilmark{17},
        H. McCracken\altaffilmark{18},
        N. Arimoto\altaffilmark{19},
        M. Mignoli\altaffilmark{10},
        M. Onodera\altaffilmark{11,20}}
\email{forster@mpe.mpg.de}


\altaffiltext{1}{Based on observations obtained at the Very Large Telescope
                 (VLT) of the European Southern Observatory, Paranal, Chile
                 (ESO Programme IDs
                  070.A-0229, 070.B-0545, 073.B-9018, 074.A-9011,
                  075.A-0466, 076.A-0527, 077.A-0576, 078.A-0055,
                  078.A-0600, 079.A-0341, 080.A-0330, 080.A-0635,
                  080.A-0339).}
\altaffiltext{2}{Max-Planck-Institut f\"ur extraterrestrische Physik,
                 Giessenbachstrasse, D-85748 Garching, Germany}
\altaffiltext{3}{Department of Physics, Le Conte Hall,
                 University of California, Berkeley, CA 94720}
\altaffiltext{4}{Department of Astronomy, Campbell Hall,
                 University of California, Berkeley, CA 94720}
\altaffiltext{5}{Department of Physics and Astronomy, 430 Portola Plaza,
                 University of California, Los Angeles, CA 90095-1547}
\altaffiltext{6}{Department of Physics, University of California at
                 Santa Barbara, Santa Barbara, CA 93106-9530}
\altaffiltext{7}{California Institute of Technology, MS 105-24,
                 Pasadena, CA 91125}
\altaffiltext{8}{School of Physics and Astronomy,
                 Tel Aviv University, Tel Aviv 69978, Israel}
\altaffiltext{9}{Osservatorio Astronomico di Padova,
                 Vicolo dell'Osservatorio 5, Padova, I-35122, Italy}
\altaffiltext{10}{Istituto Nazionale di Astrofisica -- Osservatorio Astronomico
                 di Bologna, Via Gobetti 101, I-40129 Bologna, Italy}
\altaffiltext{11}{Service d'Astrophysique, CEA/Saclay,
                  Orme des Merisiers, 91191 Gif-sur-Yvette, France}
\altaffiltext{12}{Max-Planck-Institut f\"ur Astronomie,
                 K\"onigstuhl, D-69117 Heidelberg, Germany}
\altaffiltext{13}{Institute of Astronomy, Department of Physics,
                  Eidgen\"ossische Technische Hochschule,
                  ETH Z\"urich, CH-8093, Switzerland}
\altaffiltext{14}{Center for Astrophysics, University of Science and
                  Technology of China, 230026 Hefei, China}
\altaffiltext{15}{GEPI,Observatoire de Paris, CNRS, Universit\'e Denis Diderot,
                 5 Place Jules Janssen, 92190 Meudon, France}
\altaffiltext{16}{Institut d'Astrophysique Spatiale, UMR 8617,
                 Universit\'e Paris Sud 11, 91400 Orsay, France}
\altaffiltext{17}{Denys Wilkinson Building, University of Oxford,
                  Keble Road, Oxford, OX1 3RH, UK}
\altaffiltext{18}{IAP,
                  Paris, France}
\altaffiltext{19}{National Astronomical Observatory of Japan,
                  Mitaka, Tokyo 181-8588, Japan}
\altaffiltext{20}{Institute of Earth, Atmosphere and Astronomy, BK21,
                  Yonsei University, Seoul, 120-749 South Korea}

\begin{abstract}

We present the Spectroscopic Imaging survey in the Near-infrared
with SINFONI (SINS) of high redshift galaxies.
With 80 objects observed and \ndet\ detected in at least one rest-frame
optical nebular emission line, mainly H$\alpha$, SINS represents the
largest survey of spatially-resolved gas kinematics, morphologies,
and physical properties of star-forming galaxies at $z \sim 1 - 3$.
We describe the selection of the targets, the observations,
and the data reduction.
We then focus on the ``SINS H$\alpha$ sample,'' consisting of 62
rest-UV/optically-selected sources at $1.3 < z < 2.6$ for which we
targeted primarily the H$\alpha$ and [\ion{N}{2}] emission lines.
Only $\approx 30\%$ of this sample had previous near-IR spectroscopic
observations.
The galaxies were drawn from various imaging surveys with different
photometric criteria; as a whole, the SINS H$\alpha$ sample covers a
reasonable representation of massive
$M_{\star} \ga 10^{10}~{\rm M_{\odot}}$
star-forming galaxies at $z \approx 1.5 - 2.5$, with some bias towards
bluer systems compared to pure $K$-selected samples due to the requirement
of secure optical redshift.  The sample spans two orders of magnitude
in stellar mass and in absolute and specific star formation rates, with
median values $\approx 3 \times 10^{10}~{\rm M_{\odot}}$,
$\rm \approx 70~M_{\odot}\,yr^{-1}$, and $\rm \approx 3~Gyr^{-1}$.
The ionized gas distribution and kinematics are spatially resolved on
scales ranging from $\rm \approx 1.5~kpc$ for adaptive optics assisted
observations to typically $\rm \approx 4 - 5~kpc$ for seeing-limited data.
The H$\alpha$ morphologies tend to be irregular and/or clumpy.
About one-third of the SINS H$\alpha$ sample galaxies
are rotation-dominated yet turbulent disks, another
third comprises compact and velocity dispersion-dominated objects,
and the remaining galaxies are clear interacting/merging systems;
the fraction of rotation-dominated systems increases among the
more massive part of the sample.
The H$\alpha$ luminosities and equivalent widths suggest on average
roughly twice higher dust attenuation towards the \ion{H}{2} regions
relative to the bulk of the stars, and comparable current and
past-averaged star formation rates.
\end{abstract}


\section{INTRODUCTION}
         \label{Intro}

In the now standard model of concordance cosmology, large-scale structure
grows through simple gravitational aggregation and collapse from the initial
fluctuations in the mass density of the early universe.  In this framework,
galaxies form as baryonic gas cools at the center of dark matter halos and
subsequently grow through accretion and mergers, leading to the hierarchical
build-up of galaxy mass.  Increasingly deep and wide-area multiwavelength
surveys in the past decade have established a fairly robust outline of the
global evolution of galaxies over nearly 90\% of the age of the universe.
Rapid evolution is observed at redshifts $z \sim 1 - 4$, with the peak of
(dust-enshrouded) star formation, luminous QSOs, and major merger activity
occurring around $z \sim 2 - 3$ \citep[e.g.,][]{Fan01, Cha05, HB06}.
By $z \sim 1$, roughly half of the stellar mass in galaxies ---
and $> 90\%$ in massive, $\rm \ga 10^{11}~M_{\odot}$ galaxies ---
was assembled \citep[e.g.,][]{Dic03, Fon03, Rud03, Rud06, Gra07, Con07}.
The epochs around $z \sim 1 - 2$ also seem to correspond to a crucial
transition with the emergence of the bimodality and the Hubble sequence
as observed in the present-day galaxy population
\citep[e.g.,][]{Bel04,vdB96,vdB01,Lil98,Sta04,Rav04,Pap05,Kri08b,Wil09}.

The details of {\em how\/} galaxies were assembled and evolved remain,
however, poorly known.  Much of our current knowledge at $z \ga 1$ still
relies heavily on galaxy-integrated spectral energy distributions and
colours, and on global properties such as stellar mass and age, star
formation rate, interstellar extinction, and sizes.
Studies based on integrated spectroscopy (mostly in the optical, much
fewer in the infrared and submillimeter) are still comparatively scarce
but have provided secure redshifts for various photometrically-selected
samples, and first results notably on galactic-scale outflows, dynamical
masses, gas mass fractions, and nebular abundances.  More direct and
detailed constraints are however needed to understand the formation
and evolution of galaxies, involving angular momentum exchange and loss,
cooling, dissipation, dynamical processes, and feedback from star formation
and active galactic nuclei (AGN).  Such constraints are crucial as input
and benchmarks for theories and simulations of galaxy formation and
evolution.

Of particular relevance in this context is the issue of the dominant
mechanisms by which massive galaxies at high redshift assemble their
baryonic mass, and what processes drive their star formation activity
and early evolution.  While major merging is undoubtedly taking place
at high redshift \citep[e.g.,][]{Tac06, Tac08}, new observational results
suggest that rapid but more continuous gas accretion via ``cold flows''
and/or minor mergers likely played an important role in driving star
formation and mass growth of the massive star-forming galaxy population
at $z \ga 1$ \citep[e.g.,][]{Noe07, Elb07, Dad07}.
This is in line with recent theoretical work based on both
semi-analytical approaches and hydrodynamical simulations
\citep[e.g.,][]{Ker05,Dek06,Kit07,Naa07,Guo08,Dave08,Genel08,Dek09a}.
The results from our own SINFONI survey of kinematics of $z \sim 2$ galaxies
(the subject of the present paper), as well as similar studies carried out
by other teams \citep[e.g.,][]{Erb03, Erb06b, Law07b, Law09, Wri07, Wri09}
have provided key evidence in support of this alternative scenario, at
least in a significant number of the galaxies observed.

This emphasizes the crucial role of spatially- and spectrally-resolved
investigations of individual galaxies at early stages of their evolution.
Such studies enable the mapping of kinematics and morphologies, and of
the distribution of star formation, gas and stars, and physical properties
such as chemical abundances and excitation state of the gas.  The constraints
and results can then be fed into studies of larger samples (connecting through
global galaxy parameters such as mass and star formation rate), and theoretical
models and numerical simulations (as observationally motivated ingredients
and assumptions).  Obtaining spatially-/spectrally-resolved data is
however notoriously challenging because of the faintness of high redshift
galaxies, and also because many important spectral diagnostic features are 
redshifted out of the optical bands.  The advent of sensitive near-infrared
(near-IR) integral field spectrometers mounted on $\rm 8-10\,m$ class
ground-based telescopes have recently opened up this avenue
\citep[e.g.][]{FS06a,Gen06,Nes06a,Nes06b,Nes07,Nes08,Swi06,Swi07,
         Law07b,Law09,Wri07,Wri09,Bour08,Sta08,Stark08,Mai08,Epi09}.
These new instruments provide simultaneously the two-dimensional spatial
mapping and the spectrum over the entire field of view.  Operating at near-IR
wavelengths, they enable one to access, for $z \sim 1 - 4$, well-calibrated
spectral diagnostics of the physical properties from rest-frame optical
emission lines such as H$\alpha$, H$\beta$,
[\ion{N}{2}]\,$\lambda\lambda\,6548,6584$,
[\ion{O}{3}]\,$\lambda\lambda\,4959,5007$,
[\ion{O}{2}]\,$\lambda\,3727$,
and [\ion{S}{2}]\,$\lambda\lambda\,6716,6731$.

Using the near-IR integral field spectrometer SINFONI \citep{Eis03a, Bon04} 
at the Very Large Telescope (VLT) of the European Southern Observatory (ESO),
we have carried out a major program of spatially-resolved studies of high
redshift galaxy populations: the Spectroscopic Imaging survey in the
Near-IR with SINFONI, or ``SINS.''
With the rich information provided by SINFONI on individual galaxies,
the key science goals of the SINS survey are to investigate in detail:
(1) the nature and timescales of the processes driving baryonic
 mass accretion, star formation, and early dynamical evolution, 
(2) the connection between bulge and disk formation,
(3) the amount and redistribution of mass and angular momentum within
 galaxies, and
(4) the relative role and energetics of feedback from star formation
 and AGN.

Our initial results, based on about 30 optically- and near-IR-selected
objects at $z \sim 1.5 - 2.5$, revealed a diversity in kinematics and
morphologies of the H$\alpha$ line emission \citep{FS06a, Gen06, Bou07}.
Perhaps the most surprising outcome was the large fraction of systems
with compelling signatures of rotation in disk-like systems.
Quantitative analysis through kinemetry established that about 2/3
of the best-resolved objects with highest signal-to-noise (S/N) data
are disks while 1/3 are clear mergers \citep{Sha08}.
The dynamical mass surface densities, angular momenta, and velocity-size
relation of the disk-like systems favour an ``inside-out'' scenario for
the formation of early disks and little net loss of angular momentum of
the baryons upon collapse from the parent dark matter halo.  
These early star-forming disks have clumpy H$\alpha$ morphologies,
large intrinsic velocity dispersions, and high inferred gas fractions of
$\sim 20\% - 40\%$.  This implies the disks must be globally unstable,
possibly fragmenting into massive star-forming clumps that migrate by
dynamical friction towards the gravitational center where they coalesce
to form a young bulge within $\rm \sim 1 - 2~Gyr$ \citep{Gen08}, as seen
in numerical simulations of unstable gas-rich disks
\citep*{Nog99, Imm04a, Imm04b, Bour07, Dek09b}.
These results suggest that secular processes in non-major merging systems
are an important mechanism for growing galaxies at $z \sim 2$, a conclusion
that we found to also be in agreement with the growth of structure from
merger trees in the Millenium Simulation \citep{Genel08}.

We have collected observations of \nobs\ $z \sim 1 - 3.5$
star-forming galaxies.  In this paper, we present the full sample, the
observing strategy, and the data reduction and maps extraction procedures.
We then focus on the largest sub-sample consisting of \nsins\ optically-
and near/mid-IR selected star-forming galaxies at $z \sim 1.5 - 2.5$,
for which H$\alpha$ was the primary line of interest and which we refer
to as the ``SINS H$\alpha$ sample.''  We describe and analyze their
ensemble H$\alpha$ properties and kinematics.
The development and application of kinematic analysis tools and
dynamical modeling are presented by \citet{Sha08} and \citet{Cre09}.
Further aspects of the kinematics and physical properties are presented in
other papers, including the Tully-Fisher relation at $z \sim 2$ \citep{Cre09},
the detection of faint broad-line H$\alpha$ emission and implications on
feedback processes \citep[e.g.,][]{Sha09}, the line excitation and
gas-phase abundances, the relation between galaxy scaling properties,
and rest-frame optical continuum morphologies
(\citeauthor[][in preparation]{Bus09, Bou09, FS09a}).

The paper is organized as follows.
The selection of all SINS targets is described in \S~\ref{Sect-samples}.
We then focus on the SINS H$\alpha$ sample.
In \S~\ref{Sect-representativeness}, we discuss how well it represents
the $z \sim 2$ star-forming galaxy population.
The SINFONI observations and data reduction are described in
\S~\ref{Sect-obsred} and the extraction of flux and kinematics
from the data in \S~\ref{Sect-extrac}.
The integrated H$\alpha$ properties are presented in \S~\ref{Sect-Ha_prop}
and compared to those of other near-IR spectroscopic samples at similar
redshifts in \S~\ref{Sect-other_samples}.
Taking advantage of the high quality data and large size of our
SINS H$\alpha$ sample, we set constraints on the dust distribution
and star formation histories of the galaxies in \S~\ref{Sect-dust_sf}
and discuss the kinematic properties in \S~\ref{Sect-kinematics}.
The paper is summarized in \S~\ref{Sect-conclu}.
Throughout, we assume a $\Lambda$-dominated cosmology
with $H_{0} = 70\,h_{70}~{\rm km\,s^{-1}\,Mpc^{-1}}$,
$\Omega_{\rm m} = 0.3$, and $\Omega_{\Lambda} = 0.7$.
For this cosmology,
1\arcsec\ corresponds to $\rm \approx 8.4~kpc$ at $z = 2$.
Magnitudes are given in the Vega-based photometric system, unless
explicitly stated otherwise.  All stellar masses and star formation
rates are quoted for a \citet{Cha03} initial mass function (IMF).

\vspace{1ex}
\section{SINS SAMPLE SELECTION}
         \label{Sect-samples}

The galaxies observed as part of our SINS survey were culled
from the spectroscopically-confirmed subsets of various imaging
surveys in the optical, near-IR, mid-IR, and submillimeter regime.
We focussed on the redshift interval $z \sim 1 - 4$.
The photometric selection of the parent samples encompassed a range of
star-forming populations at high redshift, including optically-selected
``BX/BM'' and Lyman-break galaxies at $z \sim 2$ and $z \sim 3$, near-
and mid-IR selected galaxies at $z \sim 1.5 - 2.5$ (with a majority of
``$sBzK$'' objects), submillimeter-bright $z \sim 1 - 3$ sources, and
H$\alpha$ emitters at $z \sim 1 - 2$.  A total of \nobs\ galaxies were
observed, \ndet\ of which were detected in at least one emission line.
This includes two companion sources at the same redshift as the targeted
objects, identified through their line emission in our SINFONI data.
Table~\ref{tab-samples} lists all of the galaxies observed, along with
their redshifts from optical spectroscopy, their $K$-band magnitudes,
their class, and the surveys from which they were drawn.
Figure~\ref{fig-samples} shows the distribution of the full SINS
sample among the different classes and as a function of redshift.

The selection criteria common to all SINS targets were a combination
of target visibility during the observing runs, night sky line avoidance
for H$\alpha$ or [\ion{O}{3}]\,$\lambda\,5007$ depending on the redshift,
and an estimated observed integrated emission line flux of
$\rm \ga 5 \times 10^{-17}~erg\,s^{-1}\,cm^{-2}$.
For about one-third of the galaxies, these line flux estimates
could be directly taken from existing near-IR long-slit spectroscopy.
For the majority of the sample, however, this prior information was not
available.  These were mostly galaxies with $1 < z_{\rm sp} < 2.7$, for
which H$\alpha$ was the main line of interest.  We computed expected
integrated H$\alpha$ fluxes from estimates of the star formation rates
derived from broad-band SED modeling, rest-frame UV luminosities, and/or
Spitzer/MIPS $\rm 24~\mu m$ or SCUBA $\rm 850~\mu m$ fluxes.
The star formation rates were converted to H$\alpha$ fluxes following
the prescription of \citet{Ken98}, corrected to a \citet{Cha03} initial
mass function (IMF) and accounting for interstellar extinction whenever
possible.  Accurate redshifts for the targets was mandatory to ensure that
the emission lines of interest fall within the near-IR atmospheric windows
and between the strong night sky lines.  The density (per wavelength unit),
intensities, and rapid time variability of the sky lines make emission
line redshift determinations in the near-IR fairly inefficient, even
at the spectral resolution of $R \sim 3000 - 4000$ of SINFONI.

Since we were primarily concerned with the ionized gas kinematics and
morphologies as tracers of the dynamical and evolutionary state of the
systems, and with their star formation properties, we generally tried
to avoid known AGN galaxies, although a small number was included.
In total, six SINS galaxies (representing 10\% of the detected sources)
were previously known or suspected AGN from existing optical and/or near-IR
spectroscopy, and X-ray emission or MIPS $\rm 24\mu m$ observations when
available.
The line properties in the individual SINFONI spectrum of these sources
(primarily broad line widths and high [\ion{N}{2}]/H$\alpha$ flux ratios)
reflect the presence of the AGN.  In some of these clear AGN cases, the
line emission associated with the AGN and star-forming components can be
spatially and/or spectrally separated \citep[see][for an example]{Gen06},
allowing us to investigate the dynamics and physical properties of the
host galaxies.

Summarizing, the criteria applied to all of the SINS targets were
a secure optical spectroscopic redshift, night sky line avoidance and a
minimum estimated integrated flux for the primary line of interest, and 
source visibility during the observing runs.  The following subsections 
describe in more detail the selection of galaxies of each class and survey,
and the additional considerations that were in some cases explicitly applied.
In brief, these include: emission line width and indications of velocity
structure or lack thereof (for part of the $\sim 1/3$ optically-selected
targets with prior near-IR long-slit spectroscopy), $B-z$ and $z-K$ colours
(satisfying the ``$sBzK$'' criterion of \citealt{Dad04b}, for eleven targets
or $14\%$ of the full sample), and rest-frame UV and/or optical morphologies
(encompassing irregular, multi-component, disky, and compact morphologies,
for 23 targets or $29\%$ of the full sample).  Any other characteristic
(such as optical or near-IR magnitude cutoff) was inherited from the
different selection specific to each of the parent photometric survey
or source catalogue, as described below.  The consequences of these
combined criteria on the resulting ensemble properties of the SINS
sample are discussed in \S~\ref{Sect-representativeness}.

\subsection{Optically-Selected BX/BM Objects}
            \label{Sub-bmbx}

The \bxbm\ criteria \citep{Ade04, Ste04} are based on observed
optical $U_{n}G\mathcal{R}$ colours and represent an extension to
$z \sim 1.5 - 2.5$ of the classical Lyman-break technique targeting
$z \sim 3$ galaxies \citep{Ste93, Gia98, Ste99}.  The efficient
\bxbm\ and Lyman-break techniques have yielded the first substantial
($> 1000$) samples of spectroscopically-confirmed $z \sim 1 - 3$
galaxies, at $\mathcal{R}_{\rm AB} < 25.5~{\rm mag}$.  By construction,
the \bxbm\ criteria identify primarily actively star-forming galaxies
with moderate amounts of extinction in the ranges $z \sim 2 - 2.5$
(BX objects) and $z \sim 1.5 - 2$ (BM objects).  The properties of
the \bxbm\ population have been extensively discussed in many papers
\citep[e.g.,][]
{Erb03,Ste04,Sha04,Sha05a,Ade05a,Ade05b,Red05,Red06,
Erb06a,Erb06b,Erb06c,Law07a}.
In brief, they have typical stellar ages of $\rm \sim 500~Myr$,
stellar masses $M_{\star} \sim 2 \times 10^{10}~{\rm M_{\odot}}$,
star formation rates $\rm SFR \sim 50~M_{\odot}\,yr^{-1}$, and
extinction $A_{V} \sim 0.8~{\rm mag}$, with a tail extending to
more massive, evolved, and/or dustier galaxies.

We drew our \bmbx\ targets from the near-IR spectroscopic sample of
\citet[][see also \citealt{Erb03,Sha04,Ste04}]{Erb06a,Erb06b,Erb06c}.
This spectroscopic survey was carried out with NIRSPEC at the Keck~II
telescope.
In the initial phases of the SINS survey --- for observational reasons ---
we emphasized brighter sources with spatially resolved velocity gradients,
large velocity dispersions, or spatially extended emission based on the
existing spectroscopy.  At later phases, we also observed compact sources
without indications for velocity gradients and with average or unresolved
H$\alpha$ line widths to expand the range of kinematic properties probed.
We observed a total of 17 galaxies, including 16 BX objects with median
$z = 2.2$ and one BM object at $z = 1.41$.  Emission lines were detected
in all of the objects (with the main line of interest being H$\alpha$).
Two galaxies form a pair at nearly the same redshift
($\rm Q2346-BX404/405$), with relative velocity of $\rm 140~km\,s^{-1}$
and projected separation of $3\farcs 63$ (30.3~kpc).
The results on the first 14 galaxies were presented by \citet{FS06a}.
Since then, we have collected data of three new targets, re-observed
a number of sources leading to longer integration times and higher S/N,
and complemented the $K$-band data targeting H$\alpha$$+$[\ion{N}{2}]
emission with $H$-band data for H$\beta$$+$[\ion{O}{3}] for several
of the $z > 2$ BX objects.

\subsection{Near-/Mid-IR-Selected Galaxies}
            \label{Sub-IRsel}

Near-IR surveys yield important complementary, and to some extent
overlapping samples of $z \ga 1$ galaxies.  Efficient colour criteria
have been devised to isolate high redshift photometric candidates from
$K$-band limited source catalogues, intended to include more specifically
evolved and/or dust-obscured populations that may be underrepresented in
optically-selected samples.  One of the most efficient and widely used
is the ``$BzK$'' selection, introduced by \citet{Dad04b}.  It combines
near-IR and optical colours, defining regions in the $B - z$ versus $z - K$
colour diagram to identify star-forming (``$sBzK$'') or passively evolving
(``$pBzK$'') galaxies at $1.4 < z < 2.5$.  For our SINS survey, $pBzK$
objects are not relevant because, by selection, they are expected to lack
the nebular line emission we are interested in.
The $sBzK$ criterion has the feature of being insensitive to reddening
by dust, and so it selects star-forming galaxies with a wide range of
extinction as well as ages.
There is a significant overlap between near-IR selected $sBzK$
and optically-selected \bxbm\ populations to a given $K$-band limit
(and increasing towards fainter limits), although $sBzK$ objects tend
to include a larger proportion of more evolved and massive systems,
and with higher star formation rates and amounts of extinction
\citep[e.g.][]{Red05, Kon06, Gra07, Dad07}.

More recently, sensitive $\rm 3 - 8~\mu m$ imaging with the IRAC
camera onboard the Spitzer Space Telescope has extended the coverage
of optical/near-IR surveys to longer wavelengths.  This allows in
principle the construction of more genuinely mass-selected samples at
high redshift based on rest-frame near-IR emission, better tracing the
light from stars dominating the stellar mass and less affected by dust
extinction and star formation than the emission at shorter wavelengths.
In the context of this paper, ``near-/mid-IR selection'' refers to
galaxies drawn from $\rm 2.2~\mu m$ ($K$ band) or $\rm 4.5~\mu m$
magnitude-limited surveys
\footnote{
thus excluding, e.g., Spitzer/MIPS $\rm 24~\mu m$ flux-limited samples;
although MIPS observations were carried out for many of the survey fields
from which we drew our SINS targets, none of them were MIPS-selected.
}.

In total, the SINS near-/mid-IR-selected sub-samples count 45
sources; 43 were drawn from various surveys and two were serendipitously
discovered in line emission in our SINFONI data.  The sources span the
redshift range $1.3 < z < 2.6$, with median $z = 2.1$.
Depending on the field/survey, different indicators of star formation
activity were available to estimate the expected observed integrated
line fluxes, and, for some subsets, we also considered colours and/or
morphologies in addition to the criteria applied for all SINS targets
described above.  Eleven sources (from the Deep3a and zCOSMOS surveys)
were specifically chosen to satisfy the $sBzK$ criterion.  However, the
common key features of estimated SFR of $\rm \ga 10~M_{\odot}\,yr^{-1}$
(to ensure H$\alpha$ detectability), brightness in observed $K$ band
(from the magnitude limits of the parent surveys), and redshift range
$\sim 1 - 3$ naturally result in a majority of our near-/mid-IR-selected
targets with $B-z$ and $z-K$ measurements having the colours of $sBzK$
objects (90\%), even if most were not explicitly selected so
(see \S~\ref{Sect-representativeness}).
Morphologies (from high resolution Hubble Space Telescope imaging)
were considered for 23 sources (from the GMASS and zCOSMOS surveys), to
probe a range of types.  This was a secondary factor in that we first
selected based on redshift, expected line flux, and source visibility,
and after looked at the morphologies.

The fraction of the SINS near-/mid-IR-selected galaxies detected in at
least one emission line is 77\% (33 out of 43, excluding our serendipitous
detections described below).  This is driven in part by the fact that the
large majority of these sources had no previous near-IR spectroscopy to
verify a priori the exact line fluxes and wavelengths.  In addition,
some of those sources were observed in poorer conditions for comparatively
short integration times, leading effectively to brighter limiting fluxes
(see \S~\ref{Sect-Ha_prop}).
The properties of these undetected targets are further discussed in
\S~\ref{Sect-representativeness}.

\subsubsection{K20 Targets}   \label{Sub-k20}

We observed the five sources at $z > 2$ presented by \citet{Dad04a},
drawn from the K20 survey \citep[e.g.][]{Cim02a, Cim02b, Cim02c, Mig05}.
The K20 survey was a spectroscopic campaign of 545 $K$-selected objects
at $K_{\rm s} < 20~{\rm mag}$ and with no morphological or color biases,
over two widely separated fields totalling $\rm 52~arcmin^{2}$.
One of them is a $\rm 32~arcmin^{2}$ region in the Chandra Deep Field
South \citep[CDFS,][]{Gia04}, also included in the GOODS South Field
\citep{Dic09}, where all nine galaxies studied by \citet{Dad04a} are located.
These were initially selected on the basis of their photometric redshift
$z_{\rm ph} > 1.7$, and all were spectroscopically confirmed to lie at
$z_{\rm sp} > 1.7$.  For one of them, $\rm K20-ID9$, the optical redshift
of 2.25 was reported as less secure; in our SINFONI data, H$\alpha$ and
[\ion{N}{2}]\,$\lambda\,6584$ are clearly detected at a redshift of
$z = 2.0343$.

All five K20 sources observed for SINS were detected in H$\alpha$ and
[\ion{N}{2}] emission.  They all satisfy the $sBzK$ criterion.
Only one, $\rm K20-ID5$, had been previously observed spectroscopically
at near-IR wavelengths, with the GNIRS spectrograph at the Gemini South
observatory \citep{Dok05}.  The relative intensities of the emission lines
in the GNIRS $\rm 1 - 2.5~\mu m$ single-slit spectrum are characteristic
of either photoionization by an AGN or shock ionization due to a strong
galactic wind.  The evidence from X-ray to radio data available for this
galaxy led \citeauthor{Dok05} to favour the latter interpretation.
Our SINFONI data map fully the two-dimensional emission in H$\alpha$,
[\ion{N}{2}], [\ion{O}{3}], and H$\beta$ at twice the spectral resolution.
The spatially-resolved line ratios and kinematics, as well as AO-assisted
$K$-band imaging with the NACO instrument at the VLT, reveal more clearly
AGN signatures at the nucleus although shock excitation is also inferred
in the outer regions \citep{Bus09}.

\subsubsection{Deep3a Targets}   \label{Sub-deep3a}

We observed seven targets from the $K$-selected
catalogue presented by \citet{Kon06} extracted over the central
$18^{\prime} \times 18^{\prime}$ of the so-called ``Deep-3a'' field.
This region corresponds to the area with deepest near-IR imaging of
a three times wider field imaged as part of the DEEP Public Survey
\citep[DPS;][]{Ols06, Mig07} of the ESO Imaging Survey program
\citep[EIS;][]{Ren97}.  Optical $UBVRI$ imaging from the WFI camera at the
ESO/MPG 2.2~m telescope was complemented with near-IR $JK_{\rm s}$ data
from the SOFI instrument at the ESO NTT 3.5~m telescope.  Additional deep
$BRIz^{\prime}$ optical imaging with Suprime-Cam on the Subaru telescope was
obtained by \citet{Kon06}.  The $5\,\sigma$ $K_{\rm s}$ limiting magnitude
reaches $K_{\rm s} \approx 20.85~{\rm mag}$ (2\arcsec-diameter aperture).
Optical spectroscopic redshifts for a subset of the sources with
$B_{\rm AB} \la 25~{\rm mag}$ were obtained with VIMOS at the ESO VLT
\citep{Dad09}.

All our Deep3a targets were $K_{\rm s} < 20~{\rm mag}$ $sBzK$-selected
objects spectroscopically confirmed at $1.4 < z_{\rm sp} < 2.5$.  All
are fairly bright at 24\micron\ with fluxes $\rm \ga 100~\mu Jy$ from
MIPS data, ensuring H$\alpha$ detectability.
Taking advantage of the Deep3a field size allowed us to pick some of the
sources close to stars suitable for AO-assisted follow-up.  At the time
of our first observations of Deep3a targets, no near-IR spectroscopic data
were available for the $sBzK$ objects.  Three of the sources we targeted
at later stages had been in the meantime observed with SINFONI using the
lower resolution $R \sim 2000$ $H+K$ grating as part of an independent
program (ID 075.A-0439, P.I.: E. Daddi).
The choice of those three sources was driven by H$\alpha$ brightness,
and excluding two bright sources because their redshifts put H$\alpha$
in a region of lower atmospheric transmission at the red edge of the
$H$ band and their H$\alpha +$[NII] characteristics show the emission
originates from unresolved AGN.

\subsubsection{GMASS Targets}   \label{Sub-gmass}

Nineteen of the SINS targets were drawn from the
``Galaxy Mass Assembly ultra-deep Spectroscopic Survey''
(GMASS; \citealt[][see also \citealt{Cim08,Cas08,Hal08}]{Kur09}).
The GMASS sample was selected at $\rm 4.5~\mu m$ with
$m_{\rm 4.5,\,AB} < 23.0~{\rm mag}$ in a $6\farcm 8 \times 6\farcm 8$ area
in the GOODS South field, with $\approx 80\%$ overlap with the Hubble Ultra
Deep Field \citep[HUDF;][]{Bec06}.  A sub-sample at $z_{\rm ph} > 1.4$ and
$B_{\rm AB} < 26.5~{\rm mag}$ or $I_{\rm AB} < 26.5~{\rm mag}$ was then
observed spectroscopically, the optical magnitude cutoffs ensuring feasible
spectroscopy with the FORS2 blue or red grisms.  The key feature of GMASS
is the mid-IR selection based on the very deep GOODS IRAC imaging, which
corresponds to rest-frame near-IR for $z = 1.5 - 2.5$ and should be even
closer to stellar mass selection than rest-frame optical selection.
Together with literature redshifts, about 50\% of the
$\rm 4.5~\mu m$-selected GMASS sample has a spectroscopic redshift.

For our SINFONI observations, we selected galaxies from the GMASS
spectroscopic catalogue at $1 < z_{\rm sp} < 4$ with predicted integrated
H$\alpha$ flux of $\rm \ga 5 \times 10^{-17}~erg\,s^{-1}\,cm^{-2}$ based on
SFR estimates from MIPS $\rm 24\,\mu m$ flux and rest-frame UV luminosity.
We then considered the rest-frame UV morphology based on the GOODS deep
ACS $z_{\rm 850}$ mosaic and, whenever possible, the rest-frame optical
morphology from HUDF deep NICMOS/NIC3 imaging through the F110W and F160W
filters (approximately $J$ and $H$ bands).  We emphasized galaxies with
irregular, multi-component, or disky morphologies (in similar
proportions: 7/5/5 galaxies) in order to sample both merging and
disk-like systems, but we also included two unresolved sources.
$K$-band brightness was not a criterion per se; the SINS GMASS targets
span the range $K_{\rm s} = 19.3 - 21.4~{\rm mag}$.

None of the nineteen sources observed had prior near-IR spectroscopy.
We detected thirteen of them in at least one line (H$\alpha$)
\footnote{
The undetected GMASS sources include four of the irregular and
multi-component systems and the two unresolved sources; the observations
were taken under poorer than average seeing conditions, reducing the
sensitivity for compact sources and/or sub-components.}.
One of the targets, $\rm GMASS-2113$, turned out to have a close companion
$1\farcs 9$ to the east (or 16.0~kpc at the $z = 1.613$ of the GMASS source)
with a 1.6 times brighter emission line at nearly the same wavelength,
$\rm 60~km\,s^{-1}$ bluewards.
No other emission line is detected but given the very slim chances of having
two different emission lines within several 10's of $\rm km\,s^{-1}$ from
two sources close in projection but at different redshifts, the emission line
can be confidently identified with H$\alpha$.  Hereafter, the GMASS source
and this eastern companion will be designated as $\rm GMASS-2113W$ and
$\rm 2113E$, respectively.  $\rm GMASS-2113E$ is not included in the GMASS
catalogue but we cross-identified it in the $K_{\rm s}$-limited FIREWORKS
catalogue of the CDFS by \citet{Wuy08}.  It is 1.3~mag fainter than
$\rm GMASS-2113W$ in $K_{\rm s}$; it is brighter than the magnitude cutoffs
of the GMASS survey ($m_{\rm 4.5,\,AB} = 22.61$, $B_{\rm AB} = 24.58$, and
$I_{\rm AB} = 24.23~{\rm mag}$) but is partly blended with $\rm GMASS-2113E$
in the IRAC $\rm 4.5~\mu m$ map given the $\rm FWHM = 1\farcs 7$ of the
point-spread function (PSF).  The photometric redshift derived by
\citet{Wuy08} from the 16-band FIREWORKS optical to mid-IR photometry
is $z_{\rm ph} = 1.638^{+0.186}_{-0.157}$, fully consistent with our
H$\alpha$ redshift of $z_{\rm sp} = 1.6115$.
Of the total of 20 targets in the GMASS field
(counting the $\rm GMASS-2113W/E$ pair as two), 18 are $sBzK$'s
(including $\rm GMASS-2113E$).

\subsubsection{zCOSMOS Targets}   \label{Sub-cosmos}

We observed four sources as a pilot sample of an on-going
collaboration between the SINS and zCOSMOS teams.
The Cosmic Evolution Survey, or ``COSMOS,'' is currently the widest
multiwavelength survey, with coverage at all accessible wavelengths
from the X-ray to the radio regime, over an area of $\rm 2~deg^{2}$
\citep[][and references therein]{Sco07}.
The zCOSMOS program is the major optical spectroscopic campaign carried
out with VIMOS at the VLT \citep{Lil07}.  It consists of two components:
the ``zCOSMOS-bright'' of a purely magnitude-limited sample at
$I_{\rm AB} < 22.5~{\rm mag}$ over $\rm 1.7~deg^{2}$, and the
``zCOSMOS-deep'' focussing specifically on $1.5 < z < 2.5$ $BzK$-
and \bxbm-selected sources with $B_{\rm AB} < 25~{\rm mag}$ over
the central $\rm 1~deg^{2}$.

The galaxies selected for SINFONI observations were culled among the
$sBzK$ sample confirmed at $1.4 < z_{\rm sp} < 2.5$ from zCOSMOS-deep.
For the pilot observations described here, the targets were originally
drawn from the $K$-band catalogue of \citet{Cap07} based on near-IR
data to $K_{\rm s} \la 20~{\rm mag}$ (deeper near-IR imaging is
available in the meantime; \citealt{McC09}).
The morphology of the targets was a criterion, so as to include
extended and complex, irregular, clumpy, and more compact
and regular systems.  None of the targets had been previously
spectroscopically observed in the near-IR.
Three targets were detected; the non-detection is the most compact
and regular one from the ACS morphology.

\subsubsection{GDDS Targets}   \label{Sub-gdds}

Eight SINS targets were drawn from the Gemini Deep Deep Survey
\citep[GDDS][]{Abr04}.
The GDDS is a redshift survey of $K < 20.6~{\rm mag}$ and
$I < 24.5~{\rm mag}$ objects at $1 < z < 2$ in four widely-separated
$\rm 30~arcmin^{2}$ fields using the GMOS multi-object spectrograph at
the Gemini North telescope.  The survey
targeted passively evolving galaxies at $0.8 < z < 1.8$ (among the reddest
and most luminous photometric candidates, based on the $I - K$ versus $K$
colour-magnitude distribution) as well as galaxies from the remaining high
redshift population, including a wide range of star formation activity.

We selected our targets among the non-AGN GDDS sources in two of the
fields, SA12 and SA15.  Our requirements were a secure redshift at
$1.3 \la z_{\rm sp} \la 2.7$ for H$\alpha$ in the $H$ or $K$ band, and
clear signs of on-going star formation in the rest-frame UV spectrum
(spectral class ``100'', as described by \citealt{Abr04}), although we
did attempt one source with signatures indicative of intermediate-age
to old stellar populations only ($\rm SA12-5836$ of class ``011'') and
another with features characteristic of (nearly) pure evolved stars
($\rm SA12-7672$ of class ``001'').

All six class ``100'' targets have colours or $2\sigma$ limits that
meet the $sBzK$ criterion.  We detected five of them in H$\alpha$; the
non-detection, $\rm SA15-7353$, has a $2\sigma$ limit in $B - z$ colour
that places it just at the boundary between $sBzK$ and non-$sBzK$ objects
and a redshift that implies an observed wavelength for H$\alpha$ in a region
of poorer atmospheric transmission --- this may have prevented line detection
or the H$\alpha$ flux is below the surface brightness limit of our 2\,hr
on-source integration.
Perhaps surprisingly, the class ``001'' source $\rm SA12-7672$ falls in
the $sBzK$ area of the $B - z$ vs $z - K$ colour diagram; it is however
very red in $z - K$.  In our SINFONI data, the continuum is well detected
for this bright $K_{\rm s} = 19.17~{\rm mag}$ source but no emission line
is seen, consistent with the optical spectral classification.
For the class ``011'' non-$sBzK$ source $\rm SA12-5836$, residuals from
particularly strong night sky lines affect importantly the region around
the expected wavelength for H$\alpha$, possibly explaining why we did not
identify line emission.
In our SINFONI data of $\rm SA12-8768$, we detected a faint source from
its line emission at the same wavelength as H$\alpha$ for $\rm SA12-8768$
and $2\farcs 4$ to the northwest; we attribute this detection to H$\alpha$
from a companion galaxy at a projected distance of 19.8~kpc and relative
velocity of $\rm -30~km\,s^{-1}$.
Hereafter, this ``serendipitous'' detection will be referred to as
$\rm SA12-8768NW$.

\subsection{Lyman-Break Galaxies}
            \label{Sub-lbg}

We observed a small collection of Lyman-break galaxies (LBGs) at $z \sim 3$.
Seven of them were taken from the large survey of photometrically-selected
(by the classical Lyman-break technique based on observed $U_{n}G\mathcal{R}$
colours; see \citealt{Ste93, Ste99}) and spectroscopically-confirmed LBGs
carried out by Steidel and coworkers, and described in detail by \citet{Ste03}.
The objects were detected in the optical $\mathcal{R}$ band, and candidate
LBGs at $\mathcal{R}_{\rm AB} \leq 25.5~{\rm mag}$ were followed up with
optical spectroscopy for accurate redshift determination.  Three of them
($\rm Q0201+113~C6$, $\rm Q0347-383~C5$, and $\rm Q1422+231~D81$) had
previous near-IR long-slit spectroscopy with Keck/NIRSPEC and VLT/ISAAC,
presented by \citet{Pet01}.
$\rm Q0347-383~C5$ was well detected and spatially resolved in the
SINFONI data, and is a clear merger \citep{Nes08}.
$\rm Q0201+113~C6$ and $\rm Q1422+231~D81$ were also detected but are
marginally resolved spatially and the data were taken under unfavourable
conditions, so that reliable analysis could only be carried out for the
source-integrated properties \citep{Nes05}.
The other four LBGs from the \citet{Ste03} survey were undetected in
our SINFONI data, which is likely due to the poor observing conditions;
in addition, three of them were observed only once with integration times
of 1 or 2\,hr, which may have been insufficient to detect line emission.

We targeted two other LBGs from different surveys and fields.
One is the so-called ``Arc $+$ core,'' a $z = 3.24$ galaxy behind
the $z = 0.3$ X-ray cluster $\rm 1E\,06576-56$ \citep[e.g.,][]{Meh01}.
The strong lensing (by a factor of $> 20$) together with the spatial
resolution of the SINFONI data resolved the kinematics in the inner
few kpc on physical scales of $\rm \approx 200~pc$ \citep{Nes06a}.
The other one was drawn from the ESO EIS survey of the CDFS field
with spectroscopic $z = 3.083$ obtained from VLT/FORS1 optical
follow-up \citep{Cri00}.  We did not detect line emission in our
9600\,s integration in the $K$-band.

\subsection{Submillimeter-Bright Galaxies}
            \label{Sub-smg}

Six bright submillimeter-selected galaxies (SMGs) were observed,
at redshifts between 1 and 3.  All of these were part of the target
list for a long-term program of CO molecular line mapping carried
out with the IRAM Plateau de Bure mm interferometer
\citep[e.g.][]{Gen03, Ner03, Gre05, Tac06, Tac08, Sma09}.
The SMGs chosen for our SINFONI observations all had accurate radio
positions and optical spectroscopic redshifts
\footnote{These and additional imaging data to help prepare the
SINFONI observations were kindly provided by I. Smail, S. Chapman,
and R. Ivison.}.
Three of the SMGs were originally drawn from the SCUBA Lens Survey
\citep{Sma02}, and are magnified by foreground lensing clusters and,
for $\rm SMM\,J04431+0210$, also by a foreground spiral galaxy.

We detected two of the lensed SMGs in our SINFONI data sets:
$\rm SMM\,J14011+0252$ and $\rm SMM\,J04431+0210$, both well-studied
in the literature \citep[e.g.][]{Frayer99, Frayer03, Ner03, Swi04}.
We used SINFONI in $J$, $H$, and $K$ bands for a detailed study of
the kinematics and physical conditions from the rest-frame optical line
ratios of the merging system $\rm SMM\,J14011+0252$ \citep{Tec04, Nes07}.
Our high quality, high S/N ratio $K$-band data of $\rm SMM\,J04431+0210$
revealed a compact source with kinematics and [\ion{N}{2}]/H$\alpha$ ratio
indicative of a dominant AGN component \citep{Nes05}.
None of the three SMGs in the SSA22 field were detected in our SINFONI
data; the observations of these sources suffered from poor observing
conditions and, moreover, the integration times of $\rm \sim 1 - 2.5\,hr$
may have been too short to detect the emission lines targeted in the $K$
band.

\subsection{Line Emitters}
            \label{Sub-emline}

We targeted the field around the $z = 2.16$ radio galaxy
$\rm MRC\,1138-262$, which was found to have an overdensity of
H$\alpha$ emitters \citep{Kur04}.  Detailed analysis of the SINFONI
data, and in particular focusing on the feedback energetics from the
AGN powering the radio galaxy, are presented by \citeauthor{Nes06b}
(2006b; see also \citealt{Nes05}).

We also observed a pair of line emitters from the NICMOS/HST parallel
GRISM survey of \citet{McC99}.  The slitless G141 grism used spans the
wavelength range $\rm \lambda \approx 1.1 - 1.9~\mu m$ with spectral
resolution $R \equiv \lambda/\Delta\lambda \sim 200$.  In this survey
of random fields covering $\rm 64~arcmin^{2}$, 33 sources were discovered
serendipitously on the basis of detection of an emission line in the grism
data, without biases from colour selection schemes.  \citet{McC99} argued
that the most likely identification is H$\alpha$, which was subsequently
confirmed with detection of [\ion{O}{2}]\,$\lambda\,3727$ emission in
nine of the 14 galaxies by \citet{Hic02}.
The pair we targeted, $\rm NIC\,J1143-8036a/b$ (with projected angular
separation of $0\farcs 8$, not observed by \citealt{Hic02}) would lie at
$z = 1.35$ and 1.36 if the lines seen around $\rm 1.54~\mu m$ are H$\alpha$.
Detection of H$\alpha$ and [\ion{N}{2}]$\,\lambda 6584$~\AA, with a ratio
of [\ion{N}{2}]/H$\alpha$~$=~0.17$, in our $> 10$ times higher spectral
resolution SINFONI data confirms the line identification of
$\rm NIC\,J1143-8036a$ and implies $z = 1.334$.
An emission line is detected at the position of $\rm NIC\,J1143-8036b$
and with velocity offset of $\rm \approx 130~km\,s^{-1}$ (about ten times
smaller than inferred from the NICMOS G141 observations, perhaps due to
lower spectral resolution and more uncertain wavelength calibration of
these data).  While [\ion{N}{2}] emission is not seen in our data for
$\rm NIC\,J1143-8036b$, implying [\ion{N}{2}]/H$\alpha$ $< 0.09$, the
proximity in wavelength and in angular separation of the components
makes it very likely that the two sources are indeed a merging pair
\citep[see][]{Nes05}.

\vspace{1ex}
\section{GALAXY POPULATIONS PROBED BY THE SINS H$\alpha$ SAMPLE}
         \label{Sect-representativeness}

The largest fraction of the SINS galaxies comprises the \nsins\
optically-selected \bxbm\ and near-/mid-IR-selected objects, which span
the range $z = 1.3 - 2.6$ and for which H$\alpha$ was the primary line 
of interest.  This ``SINS H$\alpha$ sample'' makes up 78\% of the total
sample observed, and it is the focus of the analysis in the present
paper.

Having been assembled using disparate selection criteria, it is worth
assessing what part of the high redshift population is represented by
our SINS H$\alpha$ sample with respect to an ``unbiased'' population
of $z \sim 2$ galaxies.
We preliminarily note that the very variety of criteria employed makes 
the resulting sample less biased than any of its constituent sub-samples.  
Perhaps the main bias of our sample comes from the mandatory optical
spectroscopic redshift ($z_{\rm sp}$), which, as explicit in the
previous Section, means in practice an optical magnitude cutoff
(in addition to the primary colour or magnitude selection).
The typical optical cutoff $\rm \sim 25 - 26~mag$ (AB) implies on
average bluer optical to near-IR colours and will miss $\sim 50\%$
of $z \sim 2$ galaxies in the mass range explored in this paper, a
result of them being very faint at rest-UV wavelengths due to either
substantial dust obscuration or dominant evolved stellar populations
with little if any recent star formation
\citep[e.g.,][]{Dok06, Rud06, Gra07}.
Moreover, our requirement of minimum H$\alpha$ flux and our sensitivity
limits are likely to translate into an overall bias towards younger and
more actively star-forming systems.

To place our SINS galaxies in context, we compared their distributions
of redshift, photometric, and stellar properties with those of a purely
$K$-selected sample in the same $1.3 < z < 2.6$ interval.  We chose this
reference sample from the CDFS, one of the best-studied deep survey fields
with extensive multiwavelength coverage, and used the broad-band SEDs
and redshifts from the publicly available $K$-band limited FIREWORKS
catalogue of \citet{Wuy08}.
We selected CDFS sources to $K_{\rm s, Vega} = 22~{\rm mag}$,
which corresponds to the faintest $K$ magnitude among our SINS sample.
Because most of the CDFS sources at $z > 1$ have no $z_{\rm sp}$,
some may scatter in and out of the range $z = 1.3 - 2.6$ due to
photometric redshift ($z_{\rm ph}$) uncertainties, but this does not
have any significant impact on our conclusions (e.g., varying the redshift
limits a little does not change significantly the distribution of properties).
We did not apply any other criterion, so as to have a reference sample
that is as representative as possible of the bulk of $z \sim 2$ galaxy
populations.  In particular, we did not prune based on star-forming
activity, and this is reflected by the presence of massive objects with
low absolute and specific star formation rates.

The stellar properties (ages and masses, star formation rates, and
visual extinctions) were obtained from modeling of the broad-band SEDs.
SED fitting results were not available for all of our SINS targets,
and for those that were, the model ingredients and assumptions vary
between the different studies.  For consistently derived properties,
we thus modeled the SEDs of our SINS galaxies following the procedure
described in Appendix~\ref{App-sedmod}.  In brief, we used the stellar
evolutionary code from \citet{BC03}, and the free parameters were the
age, extinction, and luminosity scaling of the model synthetic SED.
We adopted the \citet{Cha03} IMF and the reddening curve of \citet{Cal00},
and assumed a solar metallicity.  We considered three combinations of
star formation history and dust content: constant star formation rate
(``CSF'') and dust, an exponentially declining star formation rate with
$e$-folding timescale of $\rm \tau = 300~Myr$ and dust (``$\tau 300$''),
and a dust-free single stellar population formed instantaneously (``SSP'').
These are simplistic choices, but many of the galaxies have $4 - 5$
photometric data points, preventing us from constraining reliably
their star formation histories in addition to the other properties.
We adopted the best of those three cases based on the reduced chi-squared
value of the fits (for all SINS galaxies, this is either CSF or $\tau 300$).
SED modeling for the CDFS reference sample was carried out as for
the SINS sample, using the same assumptions and model ingredients
(extensive SED modeling for FIREWORKS, with varying assumptions,
is presented by \citealt{Mar09} and \citealt{FS09b}).

Formal fitting uncertainties of the derived properties are based
on Monte-Carlo simulations, as described in Appendix~\ref{App-sedmod}.
We chose the \citet{BC03} models and the \citet{Cha03} IMF, \citet{Cal00}
reddening curve, and solar metallicity for continuity with previous work
and for more consistent comparisons with other published studies in
\S~\ref{Sect-other_samples} and \S~\ref{Sect-kinematics}.
To explore the effects of variations in SED modeling assumptions
and assess systematic uncertainties, we also used the \citet{Mar05}
models (with a \citealt{Kro01} IMF), and further verified the impact
of changes in stellar metallicity and extinction law on our results
(see Appendix~\ref{App-sedmod}).
While the different assumptions lead to systematic shifts in
the ensemble properties, none of the trends and comparisons in our
analysis is significantly affected.
Results with the \citet{BC03} models and the default set of IMF,
reddening law, and metallicity are reported in all Tables and used
in all Figures; the impact of using the \citet{Mar05} models or of
changes in other parameters are given whenever appropriate.

We did not correct the broad-band SEDs for emission line contribution
for two reasons.  Emission line fluxes are not available for the 
majority of the $K$-selected CDFS sample.  For the SINS galaxies,
existing optical and near-IR spectroscopy provides more information
but not for all relevant emission lines and it is not possible to
correct all bands included in the SEDs for line contamination.
However, we note that H$\alpha$, one of the strongest lines expected 
for star-forming galaxies, contributes on average $\approx 10\%$ of
the measured flux density in the relevant bandpass based on our 
SINFONI data (see \S~\ref{Sect-Ha_prop}).

Figures~\ref{fig-Kz}, \ref{fig-magcol}, and \ref{fig-sedprop} compare
the SINS and CDFS samples.  The relevant magnitudes and colours for
the SINS galaxies, taken from the available photometry, are listed
in Table~\ref{tab-photprop} and the best-fit stellar properties are
given in Table~\ref{tab-sedprop}.  In the plots (and subsequent figures
throughout the paper), the systems classified as disks and mergers from 
kinemetry analysis \citep[][see also \S~\ref{Sect-kinematics}]{Sha08}
are indicated in red and green, respectively.  We also mark the sources
identified as AGN based on their optical (rest-UV) spectra (circles with
6-pointed skeletal star).  Here, we include $\rm K20-ID5$ among the AGN,
although its rest-frame optical emission line spectrum may include a
large (perhaps dominant) contribution from shocks in extra-nuclear
regions (see \S~\ref{Sub-k20}).  Histograms show the projected
distributions of  the samples, and hatched bars, the median values.
Thick lines are used to represent the running median of the property
along the vertical axis as a function of that along the horizontal axis,
in the same bins as employed for the histograms.

Compared to the $K_{\rm s, Vega} < 22~{\rm mag}$ CDFS reference sample,
the SINS sample is brighter by about 1~mag in terms of apparent $K$-band
and absolute rest-frame $V$-band magnitude.  The SINS galaxies have mean
and median $K_{\rm Vega} \approx 20~{\rm mag}$, ranging from 18 to 22~mag,
and mean and median $M_{V, {\rm AB}} \approx -22.5~{\rm mag}$ and between
$-23.9$ and $\rm -21.0~mag$.  The CDFS sample contains a large proportion
of fainter sources towards the lower redshifts of the range considered,
as expected for a magnitude-limited sample.  The SINS sample does not
show this effect because it was constructed very differently, and the
sources were taken from surveys where the $K$-band imaging had widely
varying depths ($5\,\sigma$ $K_{\rm Vega}$ limits from $\sim 20$ to
$\rm \sim 23~mag$).  The bimodal redshift distribution of the SINS
galaxies reflects the gap between the $H$ and $K$ atmospheric windows.

The bias introduced by the necessity of having a well-determined
optical spectroscopic redshift for all our SINS targets (hence a
sufficiently bright optical magnitude) is best illustrated in the
$B_{\rm AB} - K_{\rm Vega}$ versus $K_{\rm Vega}$ diagram of
Figure~\ref{fig-magcol}{\em a\/}.  Here, we have used the $G$-band
magnitude for the \bxbm\ galaxies as proxy for the $B$-band magnitude
\footnote{
The $G$ bandpass has an effective wavelength $\approx 4800$~\AA\,
close to that of the various $B$ bands at $\approx 4400$~\AA,
and the colour term is expected to be small compared to the 
measured colours of the galaxies.}.
While the median $B_{\rm AB} - K_{\rm Vega}$ of the SINS sample as
a whole is nearly the same as the reference CDFS sample, it is clear
that at any $K$ magnitude the SINS galaxies have bluer colours.
Figure~\ref{fig-magcol}{\em b\/} shows the $BzK$ diagram for the
near-/mid-IR selected subset of the SINS H$\alpha$ sample (the 17
\bxbm\ galaxies have no photometry in the $z$ band or another filter
close enough in wavelength).  All but four of the 43 sources with
available $BzK$ photometry satisfy the $sBzK$ criterion for star-forming
systems at $1.5 \la z \la 2.5$ even if only 11 were explicitly selected so.
This results primarily from the requirement of minimum H$\alpha$ flux as
predicted from other available star formation rate indicators (recall
that only four of the near-/mid-IR selected targets had a previous
H$\alpha$ measurement).

Figure~\ref{fig-sedprop} illustrates how the main biases translate
in terms of stellar and extinction properties.  Overall, compared
to the $K_{\rm s} < 22.0~{\rm mag}$, $1.3 < z < 2.6$ CDFS sample,
the SINS galaxies are about three times more massive, 30\% younger, 
0.2~mag more obscured at $V$ band, and five times more actively
star-forming.  At any given mass the SINS galaxies probe the younger
part of the galaxy population, with higher absolute and specific star
formation rate.  Nevertheless, the range of properties encompassed 
by our SINS sample is substantial: two orders of magnitude in stellar
mass, absolute and specific star formation rates, and the entire age
and $A_{V}$ ranges as derived for the CDFS reference sample.
Quantitatively, the median values and ranges of properties for
the SINS galaxies are as follows:
stellar mass $M_{\star} = 2.6 \times 10^{10}~{\rm M_{\odot}}$
($\approx 2 \times 10^{9} - 3 \times 10^{11}~{\rm M_{\odot}}$),
stellar age of 300~Myr (50~Myr $-$ 2.75~Gyr, actually the lower
limit imposed in our SED modeling and the maximum being set by the
restriction of having no galaxy older then the universe at its redshift),
visual extinction $A_{V} = 1.0~{\rm mag}$ ($\rm 0 - 3~mag$),
star formation rate $\rm SFR = 72~M_{\odot}\,yr^{-1}$
($\rm 0.7 - 810~M_{\odot}\,yr^{-1}$), and
specific star formation rate $\rm sSFR = 2.9~Gyr^{-1}$
($\rm 0.1 - 24~Gyr^{-1}$).
\footnote{
For the different model assumptions considered in Appendix~\ref{App-sedmod},
the changes in ensemble properties for the SINS H$\alpha$ and CDFS samples
are comparable, so that the relative differences and ranges between the two
samples remain approximately the same.
The median values for the SINS H$\alpha$ sample using the \citet{Mar05}
models vary as follows: the stellar mass decreases by $\approx 25\%$ to 
$M_{\star} = 2.0 \times 10^{10}~{\rm M_{\odot}}$, the stellar age becomes
roughly twice younger or 130~Myr, the $A_{V}$ is higher by 0.2~mag, and
the median absolute and specific SFRs increase by factors of 1.8 and 2.7,
respectively, to
$\rm SFR = 127~M_{\odot}\,yr^{-1}$ and $\rm sSFR = 7.8~Gyr^{-1}$.}
These ranges are significantly larger than the differences between
the median of the SINS and CDFS distributions.

The sources that we have classified quantitatively as disks and mergers
\citep{Sha08} tend to be among the brighter, more massive, and somewhat
more actively star-forming, a result of the S/N requirements for kinemetry.
However, the disks and mergers do not appear different in global photometric
and stellar properties, except perhaps in optical to near-IR colours.
\footnote{
This is unchanged when using the SED modeling results for the other
assumptions considered in Appendix~\ref{App-sedmod}, and is verified
with the Mann-Whitney $U$ test.}
The disks are $\approx 0.6$ and 0.5~mag redder in $B-K$ and $z-K$,
respectively, but the Mann-Whitney $U$ test indicates the differences
are only marginally significant.
By selection, the surveys from which
we drew our SINS targets are unlikely to have included violent major
mergers in their most extreme star-forming and dust-obscured phases,
such as present among the bright submillimeter-selected population
\citep[e.g.][]{Sma04, Cha05, Swi04, Swi06, Tac06, Tac08}.
Many of the objects that we could not classify by kinemetry are compact
with observed kinematics dominated by large local random motions rather
than rotational/orbital motions \citep{Gen08, Cre09},
a class that appears to be more ubiquitous in the optically-selected
samples studied by \citet{Law07b, Law09} and \citet{Wri09}.
We return to this point in \S~\ref{Sect-other_samples} and
\S~\ref{Sect-kinematics}.

The part of the $z \sim 2$ population that is most clearly absent among
the SINS sample is the massive quiescent tail at low absolute/specific
star formation rates; such objects would be difficult to detect as no
or very faint H$\alpha$ is expected, at least from star formation.
Figure~\ref{fig-sedprop} also indicates that lower mass objects are
underrepresented compared to a pure $K$-selected sample in the same redshift
range.  This results from the different magnitude and redshift distributions
of the SINS sample compared to the CDFS $K_{\rm s, Vega} \leq 22~{\rm mag}$
reference sample, as noted above in discussing Figure~\ref{fig-Kz}.
A large fraction of the low-mass objects in the CDFS sample lie at the
faint end of the magnitude range (in both observed $K$ and rest-frame
$V$ band) and are at the lowest redshifts in the interval considered here 
(i.e. around $z \sim 1.5$); restricting the comparison to $2 < z < 2.6$
reduces (but does not eliminate) the differences at low masses.

Detected and undetected sources in our SINS H$\alpha$
sample are distinguished in Figures~\ref{fig-samples}, \ref{fig-Kz},
\ref{fig-magcol}, and \ref{fig-sedprop}.  There is no obvious trend
with redshift for the undetected galaxies, as may be expected from our
night sky line avoidance and minimum expected H$\alpha$ flux criteria
when choosing our targets.  Non-detections also do not differentiate
in any of the photometric and stellar properties considered here.
For the $B-K$ colours, we restricted the comparison to
near-/mid-IR-selected targets only (to which all non-detections belong)
because these have consistent photometry in similar $B$ bandpasses while
for the optically-selected \bxbm\ targets (30\% of the detections), we 
approximated the $B$ band magnitudes with the $G$ band photometry and
this may bias the comparison.  In all properties as well as in redshift,
the mean and median between detected and undetected sources differ by
less than one standard deviation of the detected sources, and the
Mann-Whitney $U$ test confirms that the two sub-samples do not have
significantly different distributions, irrespectively of the SED
modeling assumptions.
Non-detections therefore do not seem to be related to the global
photometric and stellar properties of the targets; observing conditions
and strategy together with the H$\alpha$ surface brightness distribution
are likely the dominant factors
(see \S~\ref{Sect-Ha_prop}).

We conclude from this section that in spite of the diversity in
selection criteria (from the parent surveys/catalogues and the additional
specific criteria considered in choosing our targets), the SINS H$\alpha$
sample provides a reasonable representation of massive actively star-forming
galaxies with $M_{\star} \ga 10^{10}~{\rm M_{\odot}}$ at $z \sim 2$
in the following sense.  While it is by construction not complete in
a magnitude- or volume-limited sense, and it emphasizes bluer objects
in optical to near-IR colours as expected for samples with optical
spectroscopic redshifts, it does span a wide range in the photometric
and stellar properties examined above.  The small fraction (16\%) of
undetected targets do not stand out in any of these properties.

\vspace{1ex}
\section{SINFONI OBSERVATIONS AND DATA REDUCTION}
         \label{Sect-obsred}

\subsection{SINFONI Observations}
            \label{Sub-obs}

The observations of the SINS H$\alpha$ sample
were carried out with SINFONI \citep{Eis03a, Bon04}
mounted at the Cassegrain focus of the VLT UT4 telescope.
SINFONI consists of the near-IR cryogenic integral field spectrometer
SPIFFI \citep{Eis03b} and of a curvature-sensor adaptive optics (AO)
module called MACAO \citep{Bon03}.  A set of mirror slicers in SPIFFI
splits the focal plane in 32 parallel slitlets and rearranges them in
a pseudo long-slit fed into the spectrometer part of the instrument.
The light is then dispersed onto the $\rm 2K^{2}$ HAWAII detector.
The width of each slitlet corresponds to the projected angular size
of two pixels, resulting in effective spatial pixels (``spaxels'')
with rectangular shape.  Spatial dithering of on-source exposures by
an odd or fractional number of pixels during the observations allows
full sampling of the spatial axis perpendicular to the slitlets.
Pre-optics enable selection between pixel scales of 125, 50, and
$\rm 12.5~mas\,pixel^{-1}$.
Three gratings cover the full $J$, $H$, and $K$ atmospheric windows
and a lower resolution grating covers the $H+K$ bands simultaneously.
The nominal FWHM spectral resolution for the pixel scales relevant to
our SINS observations are as follows:
$R \approx 1900$, 2900, and 4500 for $J$, $H$, and $K$ at
$\rm 125~mas\,pixel^{-1}$, and
$R \approx 2700$ and 5000 at $\rm 50~mas\,pixel^{-1}$.
SINFONI can be operated in pure seeing-limited mode, in which case the AO
module acts as relay optics.  For AO-assisted observations, the correction
can be performed using a natural guide star (NGS-AO mode) or an artificial
star created by the Laser Guide Star Facility (LGS-AO mode), including the
sodium laser system PARSEC \citep{Rab04, Bon06}.

The data were collected during 24 observing campaigns between 2003
March and 2008 July, as part of Guest Instrument and MPE guaranteed
time observations.  In addition, data of several GMASS targets were obtained
under normal program allocations as part of a collaboration between the
SINS and GMASS teams.  The observing conditions were generally good to
excellent, with clear to photometric sky transparency and typical seeing 
at near-IR wavelengths with  $\rm FWHM = 0\farcs 5 - 0\farcs 6$.
Table~\ref{tab-runs} lists all the observing runs.  Table~\ref{tab-obs}
summarizes the observations for each target, with the band/grating, pixel
scale, and observing mode used, the total on-source integration time, the
spatial resolution of the data (see below), and the runs during which the
data were taken.  For completeness, we list observations for the entire
SINS survey, although specific details given hereafter refer to the
H$\alpha$ sample only.

To map the H$\alpha$ and [\ion{N}{2}]\,$\lambda\lambda\,6548,6584$
line emission of the SINS H$\alpha$ sample galaxies, we used the higher
resolution $H$ or $K$ gratings, depending on the redshift of the sources.
For a subset of twelve, we also obtained observations of 
[\ion{O}{3}]\,$\lambda\lambda\,4959,5007$ and H$\beta$, and of
[\ion{O}{2}]\,$\lambda\,3727$ for one them, accessible through
different bands.
The majority of the observations were carried out in seeing-limited mode
with the largest pixel scale of $\rm 125~mas\,pixel^{-1}$ giving a FOV
of $8^{\prime\prime} \times 8^{\prime\prime}$.
We observed a total of eight targets with AO (twelve when including
the $z \sim 3$ LBGs), which have suitable reference stars for NGS-AO
and, at later times, also for LGS-AO.  For five of them (seven when
counting the LBGs), we selected the intermediate $\rm 50~mas\,pixel^{-1}$
scale with FOV of $3\farcs 2 \times 3\farcs 2$ to take full advantage of
the gain in angular resolution provided by the AO, achieving FWHM
resolution of $0\farcs 15 - 0\farcs 25$ (see Figure~\ref{fig-psfdet})
\footnote{
We note that the diffraction-limited scale of SINFONI, with pixel size
of 12.5~mas and FOV of $0\farcs 8 \times 0\farcs 8$, is not suitable for
our faint high redshift targets with spatially-extended emission because
the smaller pixel scale results in too large read-noise penalty and the
FOV is generally insufficient to cover the entire source.}.
Except for one ($\rm Q1623-BX502$), all those targets were first observed
at the $\rm 125~mas\,pixel^{-1}$ scale to verify the accuracy of the blind
offsets and the appropriate observing strategy for the AO-follow up with
the smaller pixel scale and FOV.
For the other targets with AO data, we used the larger pixel scale
as trade-off between enhanced angular resolution and sensitivity.

Depending on the source, we adopted one of two observing strategies:
an efficient ``on-source dithering'' where the object was kept within the
FOV in all exposures but at different positions, and an ``offsets-to-sky''
strategy where the exposures for background subtraction were taken at
positions away from the target.  The ``on-source dithering'' was used for
the majority of the sources.  In this scheme, the data were taken in series
of ``AB'' cycles, with typical nod throws of about half the SINFONI FOV so as
to image the source in all frames, and jitter box widths of about one-tenth
the FOV to minimize the number of redundant positions on the detector array.
A typical ``observing block'' (OB) consisted of six such dithered on-source
exposures.  For the ``offsets-to-sky'' scheme, the telescope pointing was
alternated between the object (``O'') and adjacent sky regions (``S'')
empty of sources usually in an ``O-S-O-O-S-O'' pattern for each OB.
The pointing on the object and sky positions was varied by about one-tenth
of the FOV, thus ensuring adequate independent sampling of the sky signal
subtracted from each of two object frames sharing the same sky frame.

The individual exposure times varied between 300\,s, 600\,s, and 900\,s
depending mainly on the variability and intensity of the background and
night sky line emission, in order to optimize the background subtraction
and remain in the background-limited regime in the wavelength regions
around the lines of interest.  The total on-source integration times range
from 20\,m to 10\,hr, with an average of 3.4\,hr spent per band and pixel
scale for each target.  The total integration times were driven by the
surface brightness of the sources and by our aim of mapping the line
emission and kinematics out to large radii.  In general, if a new target 
was not detected after $\rm 1 - 2\,hr$, we did not observe it further.
Therefore, the non-detections among our SINS sample may have line emission
but fainter than the relatively shallow sensitivity limits of these data
sets.  Also, for a few targets with the shortest integration times, more
observations were not obtained because of various factors including
weather conditions, observing run duration, and target priorities.
The consequences on the distributions of H$\alpha$ properties of these
observational strategies and constraints are investigated in
\S~\ref{Sect-Ha_prop}.

Exposures of the acquisition stars used for blind offsetting to the
galaxies were taken to monitor the seeing and the positional accuracy
(generally one ``O-S-O'' set per science OB).  For flux calibration and
atmospheric transmission correction, we observed late-B, early-A, and G1V
to G3V stars with near-IR magnitudes in the range $\rm \sim 7 - 10~mag$.
These telluric standards data were taken every night, as close in time 
and airmass as possible to each target observed during the night.
Acquisition stars and telluric standards were always observed with the
same instrument setup as for science objects (band and pixel scale).

\subsection{SINFONI Data Reduction}
            \label{Sub-datared}

We reduced the data using the software package {\em SPRED} developed
specifically for SPIFFI \citep{Sch04, Abu06}, complemented with additional
custom routines to optimize the reduction for faint high redshift targets.
The data reduction is analogous to standard procedures applied for near-IR
long-slit spectroscopy but with additional processing to reconstruct the
three-dimensional (3D) data cube.  The main reduction steps applied to
each science target for a given instrument band and pixel scale setup
were as follows.

The night sky line and background emission as well as the dark current
were first removed from the science data.  This was done by subtracting
(without shifting) the raw science frames pairwise for data sets taken
with the ``on-source dithering'' pattern, or subtracting the sky frame from
its adjacent object frames for those obtained with the ``offsets-to-sky''
sequence.  The data were then flatfielded with exposures of a halogen
calibration lamp.  Bad pixels identified from the dark and flat-field
frames were corrected for in the science data by interpolation, completing
the pre-processing stage.
Arc lamp frames were used to generate the ``wavemap,'' and to trace
the edges and curvature of the slitlets.  The arc lamp frames were
reconstructed to data cubes to verify the 3D reconstruction parameters.

The pre-processed science data frames were then reconstructed into cubes,
corrected for distortion, and flux-calibrated and transmission-corrected
as described below.
All science cubes within a given OB were spatially aligned according to
the dither offsets sequence used for the observations.  Our high redshift
targets were always too faint to allow the determination of spatial shifts
from centroiding or cross-correlating of single exposures.  However, the
small offsets applied within an OB for dithering or offsets-to-sky are
very accurate at the VLT, as they are performed relative to the telescope
active optics guide star.
We verified this whenever possible by comparing the morphology of the
targets between individual combined OBs (i.e., averaging the aligned 
exposures within an OB).  The relative offsets between different OBs
(often taken on different nights) were determined based on the measured
position of the acquisition star observed for each OB and the known offsets
applied to go on target, from the centroid position of the sources in the
individual combined OBs when sufficiently bright, or from the relative
offsets between OBs for those that were taken successively without
re-acquisition in between.  
All aligned, flux- and transmission-calibrated science cubes of a given
target were finally combined together by averaging with sigma-clipping
(i.e., iteratively removing data points deviating from the mean, typically
clipping at the $2.5\,\sigma$ level).  This step also generated a
``sigma-cube,'' recording the standard deviation of the values for
a given pixel in the 3D data cube across all cubes combined.

The reduced data often showed very large residuals from the strong night
sky lines because our individual exposure times of $\rm 5 - 15~min$ are 
comparable or longer than the variability timescales of the sky lines.
Moreover, due to effects of instrumental flexure while tracking the targets,
the exact wavelength calibration for individual science exposures can deviate
from the master wavemap based on the arc lamp data by up to $\sim 1/2$ of
a pixel or more along the dispersion axis, leading to asymmetric residual
profiles.
To reduce these residuals, the wavelength calibration and sky subtraction
steps were repeated with optimization following the method described by
\citet{Dav07a}.  In brief, the wavelength calibration of the individual
exposures was refined using the positions of the night sky lines in the raw
frames (before sky subtraction).  This ensured that all science frames were
interpolated to a common wavelength grid, with an accuracy better than $1/30$
of a pixel.  A more sophisticated sky subtraction procedure was applied next,
which involves scaling each transition group of telluric OH lines separately
in order to further reduce the residuals around the emission lines of
interest for a given source.

The data of the telluric standard stars and the acquisition stars
were reduced in a similar way as the science data.  Flux calibration
was performed on a night-by-night basis using the broad-band magnitudes
of the telluric standards.  Correction for atmospheric transmission 
was done by dividing the science cubes by the integrated spectrum of
the telluric standard.

Broad-band images of the acquisition stars were created by averaging
together all wavelength channels of the reduced cubes, with $\sigma$-clipping
to exclude the strongest residuals from night sky lines.  The resulting
angular resolution for a given target in a given instrument setup was
determined on the combined image obtained from the acquisition star's
data associated with all of the target's OBs, providing the effective
spatial PSF of the data sets.  For both no-AO and AO-assisted data,
and for the purpose of characterizing the angular resolution of the
data, the effective PSF shape is well approximated by a Gaussian.
In Appendix~\ref{App-psfs}, we investigate in more detail the PSF
characteristics and quantify the (small) impact of uncertainties
on the extraction and modeling of kinematics maps.
Table~\ref{tab-obs} lists the FWHMs of the best-fit two-dimensional
Gaussian profiles to the effective PSFs.  Figure~\ref{fig-psfdet} shows
the distribution of the H$\alpha$ data sets as a function of their PSF
FWHM for all objects observed and for the detected ones.  Since the PSF
calibrations were not obtained simultaneously with the science data
(and, for AO-assisted observations, were taken on-axis), these values 
represent approximately the effective angular resolution of the data sets.
Inspection of the individual images of the stars interleaved between the
science OBs indicate typical variations in PSF FWHM of $\approx 20\%$
for OBs of a given galaxy, but these variations do not significantly
affect the results (as we show in Appendix~\ref{App-psfs}).

The data reduction affects the resulting spectral resolution.
To determine the effective resolution at the reconstructed data cube level,
we applied a similar reduction procedure but without sky subtraction and
spatial registration, thus creating ``sky'' data cubes.  We extracted the
night sky spectrum by integrating over a square aperture of $\approx 30$
pixels on a side (the results are little sensitive to the size of the region).
The line shape of unblended sky lines is well approximated by a Gaussian
profile, and the FWHM in wavelength units is constant across each band.
For the instrument setups relevant to the H$\alpha$ line measurements
discussed in this paper, the effective FWHM spectral resolution
corresponds to $\rm \approx 85$ and $\rm 120~km\,s^{-1}$ in $K$ and $H$
at the $\rm 125~mas\,pixel^{-1}$ scale, and $\rm \approx 80~km\,s^{-1}$
in $K$ at the $\rm 50~mas\,pixel^{-1}$ scale.

\vspace{2ex}
\section{EXTRACTION OF EMISSION LINE AND KINEMATIC PROPERTIES}
         \label{Sect-extrac}

This Section describes the method applied to extract the emission
line properties, including fluxes, kinematics, and sizes.
The procedure makes explicit use of the noise properties of the
data, which are characterized in Appendix~\ref{App-noise}.
Figures~\ref{fig-maps_bmbx1} to \ref{fig-maps_gdds2} in
Appendix~\ref{App-allmaps} show for each galaxy a subset of
all extracted results following the procedures detailed below: 
the velocity-integrated H$\alpha$ emission line map,
the position-velocity diagram along the major axis,
and the integrated spectrum.

\subsection{Line Emission and Kinematics Maps}
            \label{Sub-extrmaps}

We extracted maps of the velocity-integrated line fluxes, relative
velocities, and velocity dispersion from the reduced data cubes
using line profile fitting.  We employed the code {\em LINEFIT}
developed by our group specifically for SINFONI applications
\citep{Dav09}.  It is an evolved version of the procedure applied
in our previously published work \citep[e.g.,][]{FS06a}.
The key (and new) features of {\em LINEFIT} include the following:
{\em (i)\/} the spectral resolution is implicitely taken into account by
convolving the assumed intrinsic emission line profile and a template line
shape for the effective instrumental resolution before performing the fits,
{\em (ii)\/} weighted fits are performed according to three possible schemes
based on an input noise cube: uniform (effectively no weighting), Gaussian,
or Poisson weighting, and
{\em (iii)\/} formal fitting uncertainties are computed from 100 Monte Carlo
simulations, where the points of the input spectrum at each spatial pixel
are perturbed assuming Gaussian noise properties characterized by the rms
from the input noise cube.

The weighted fits lead to more robust measurements of the line fluxes,
central wavelengths, and widths for our near-IR spectroscopic data where
the noise level varies strongly as a function of wavelength (due to the
increased noise level at wavelengths of strong night sky lines).
The complex noise properties complicate analytic error propagation to
compute measurements uncertainties.  After experimentation, we found
than an empirical approach based on Monte Carlo simulations leads to
realistic estimates of the formal uncertainties on all fitted parameters.  
The underlying assumption is that while the noise behaviour is not Gaussian 
across wavelengths and as a function of aperture size, it is for a given
wavelength channel and a given aperture size.  We verified this in our
data sets from an analysis of the the pixel-to-pixel rms and of the 
distribution of counts measured in non-overlapping apertures randomly
placed in regions empty of source emission in the reduced data cubes.  
These distributions are indeed well described by Gaussian functions
for a given aperture size and spectral channel.
This analysis provides the average pixel-to-pixel rms noise at each 
wavelength over the effective field of view (the region of overlap of
all science exposures for a target) and the appropriate scaling as a
function of aperture size (most relevant for the integrated spectra and
axis profiles extracted in \S~\ref{Sub-extrspec} and \ref{Sub-profiles}),
which we used as input noise spectrum in the line fitting.  The details 
of the noise analysis are given in Appendix \S~\ref{App-noise}.

Before fitting, we median-filtered the data cubes to slightly increase
the S/N ratio with a 2~pixel-wide filter along each of the three axes or,
for some of the most diffuse and extended sources, with a 3~pixel-wide
filter spatially (e.g., $\rm K20-ID9$).
In the fitting procedure, we always assumed a single Gaussian line profile,
which we found to be appropriate on a pixel-to-pixel basis for our galaxies.
With this assumption, our fits are sensitive to the dominant emission line
component and, in particular, are negligibly influenced by a possible faint
broad underlying component (either intrinsic to the galaxies or due, for
instance, to beam-smearing at larger radii of a central high velocity
dispersion source).  To account for instrumental spectral resolution, we used
the average of unblended night sky line profiles for the corresponding band
and pixel scale that was determined empirically from the ``sky'' data cubes
(see \S~\ref{Sub-datared}).  After initial sigma-clipping rejection of the
strongest outliers, we applied Gaussian weighting $\rm \propto 1/{\rm rms}^2$
in the line fitting procedure using the associated noise cube.  Pixels with
formal $\rm S/N < 5$ on their velocity-integrated line flux or with obvious
bad fits were masked out in the output velocity and velocity dispersion maps.

In performing the line fits, a continuum component was subtracted.
This component was determined as the best-fit first-order polynomial
through adjacent spectral intervals free from possible line emission
from the galaxies and from residuals ($> 2-3\,\sigma$) from night sky
lines.  This generally leads to rather noisy continuum maps for the
fainter continuum sources, because of the limited wavelength range
used for the linear fit to the continuum, although the line fluxes
are not significantly affected in those cases.
In order to obtain more robust continuum maps, or at least detect the
region(s) of peak continuum emission, we took advantage of the full
band coverage of SINFONI and computed the continuum from an iterative
procedure, where we averaged the data spectrally excluding channels
corresponding to strong night sky lines and those including line
emission from the galaxies.  The iterations consisted of varying
the threshold applied to exclude spectral channels based on the noise
cube, so as to optimize for S/N ratio of the resulting continuum map.

For the detected galaxies of our SINS H$\alpha$ sample,
the effective angular resolution of the H$\alpha$ maps obtained from
seeing-limited observations is typically $\approx 0\farcs 6$ (median
value) and ranges from $0\farcs 40$ to $1\farcs 15$ (from the near-IR
seeing FWHM and accounting for the spatial median filtering applied when
extracting the maps).  This corresponds to typically $\rm \approx 5~kpc$
and a range of $\rm 3.5 - 10~kpc$ at the respective redshift of the sources.
For the objects observed with AO at the $\rm 125~mas\,pixel^{-1}$ scale,
the resulting resolution is $\approx 0\farcs 33$ and, after median filtering,
$\approx 0\farcs 41$ or $\rm \approx 3.4~kpc$.  For the AO-assisted data sets
at the $\rm 50~mas\,pixel^{-1}$ scale, the resolution is about three times
better than for the seeing-limited data: $0\farcs 17$ and, after smoothing,
$0\farcs 20$ or $\rm \approx 1.6~kpc$.

\subsection{Integrated Spectra and Properties}
            \label{Sub-extrspec}

We measured for each galaxy the global emission line properties from
spatially-integrated spectra from the unsmoothed reduced data cubes.
The spectra were integrated in circular apertures centered on the
centroid of the line emission as determined from the line maps.
Significant noise levels, especially towards the noisier edges of
the effective field of view and for the fainter sources, complicated
the definition of more optimum integration regions such as isophotal
apertures.  The radius of the circular apertures was taken to enclose
$\geq 90\%$ of the total H$\alpha$ flux based on curve-of-growth analysis,
carried out from the spectra integrated in apertures of increasing radius
for each galaxy.
This radius is typically $1\farcs 0 - 1\farcs 25$ for the (mostly
seeing-limited) data sets at the $\rm 125~mas\,pixel^{-1}$ scale,
and $0\farcs 5 - 0\farcs 75$ for most of the AO-assisted data sets
at the $\rm 50\,mas\,pixel^{-1}$ scale.

The fits to the primary line of interest, H$\alpha$, were performed
in the exact same manner as described in \S~\ref{Sub-extrmaps}
\footnote{
More generally, for other lines that are also detected in our data,
we fixed the wavelength and width to those implied by the redshift
and width of H$\alpha$.  Fixing these parameters helped in extracting
fluxes for lines in spectral regions affected by higher noise levels 
and/or that are intrinsically weaker but still detected.  In most cases,
we found that these assumptions were generally appropriate.
}.
In some of the sources, the integrated line profiles exhibit significant
asymmetries, double-peaked profiles or faint blue-/redshifted tails
(e.g., $\rm Q2343-BX389$ in Figure~\ref{fig-maps_bmbx3},
$\rm K20-ID9$ in Figure~\ref{fig-maps_k20},
$\rm D3a-15504$ in Figure~\ref{fig-maps_deep3a2},
$\rm ZC-1101592$ in Figure~\ref{fig-maps_zcosmos})
or a superposition of narrow and broad components
(e.g., $\rm Q1623-BX663$ in Figure~\ref{fig-maps_bmbx2},
$\rm K20-ID5$ in Figure~\ref{fig-maps_k20}).
In those cases, multi-component fits or moments calculation would be more
appropriate \citep[see, e.g.,][]{Sha09}.  For the purpose of this paper,
we kept with the simple approach of fitting a single Gaussian profiles
and verified that the results did not differ substantially from those
obtained with more sophisticated methods.

Weighting and derivation of the formal uncertainties of the best-fit fluxes,
relative velocities and redshift, and velocity widths were based on an input
noise spectrum calculated for the aperture size over which the spectra were
spatially integrated.  This was done according to the empirically derived
noise model described in Appendix~\ref{App-noise}, which accounts for the
fact that the effective noise properties in our reduced SINFONI data cubes
are not purely Gaussian.
For undetected lines, we determined $3\,\sigma$ upper limits on the line
fluxes based on the noise spectrum calculated for a circular aperture of
$1\farcs 0$ radius, assuming an intrinsic line width equal to the mean
of the detected sources (i.e., for $\rm \sigma = 130~km\,s^{-1}$) and
a central wavelength as expected from the optical redshift of the galaxy.

The resulting uncertainties of the line flux, velocity, and velocity
width from the Monte Carlo simulations using the noise from the empirical
model represent formal measurements errors.
The uncertainties from the absolute flux calibration are estimated to be
$\sim 10\%$ and those from the wavelength calibration, $\la 5\%$.  Other
sources of uncertainties include continuum placement and the wavelength
intervals used for line and continuum fits.
To gauge the effects of such possible systematics, we compared
measurements in a subset of the sources obtained by varying slightly
the continuum and line intervals, and also computed total line fluxes
by summing over all pixels in the line maps within the aperture adopted
to integrate the spectrum.  Together with results from curve-of-growth
analysis described above, this suggests that systematics amount to
$20\% - 30\%$ typically, and in some data sets with lowest S/N up to
$\sim 50\%$.

Table~\ref{tab-Hameas} lists the H$\alpha$ flux, vacuum redshift, and
velocity dispersion derived from the integrated spectrum of each source,
along with the formal fitting uncertainties and the radius of the circular
aperture used for the measurements.  The table also lists the fractional
contribution $f_{\rm BB}({\rm H\alpha})$ from the integrated H$\alpha$
line flux to the total broad-band flux density (from the $H$ or $K$
band magnitudes for sources at $z < 2$ and $z > 2$, respectively;
Table~\ref{tab-photprop}).
The integrated velocity dispersion (and all measurements of velocity
dispersions throughout the paper) is corrected for instrumental
spectral resolution (implicitely done within {\em LINEFIT}).
We emphasize that the measurements of integrated velocity dispersion
used in this paper, which we denote as $\sigma_{\rm int}({\rm H\alpha})$,
refer to the width of the emission line in the spatially-integrated spectrum
{\em without\/} any shifting of the spectra of individual spatial pixels to
a common or systemic velocity.  It thus includes possible contributions
from velocity gradients or non-circular gas motions that may be present
in the galaxies.
Determination of the ``intrinsic'' velocity dispersion free from
large-scale velocity gradients due, e.g., to rotation in a disk,
requires detailed kinematic modeling, a velocity correction for
individual co-added pixel spectra, or, at least for disks, to map
out the velocity dispersion profile well outside of the central
regions affected by beam-smearing of the steep inner rotation
curve \citep[e.g.,][]{FS06a,Gen06,Gen08,Wri09,Cre09,Law09}

\subsection{Axis Profiles and Position-Velocity Diagrams}
            \label{Sub-profiles}

We determined the position angle (P.A.) of the morphological major
axis by fitting a two-dimensional elliptical Gaussian to the H$\alpha$
line maps while we took the kinematic major axis as the direction of
steepest gradient in the H$\alpha$ velocity fields.
On average, the respective P.A.'s agree within $\approx 20^{\circ}$.
With one exception, the largest differences are for marginally resolved 
sources (i.e., with morphological major axis FWHM $\approx$ PSF FWHM,
where the major axis FWHM is measured as explained in \S~\ref{Sub-extrsizes}),
including $\rm Q1623-BX455$, $\rm Q1623-BX599$, $\rm Q2346-BX416$, and
$\rm SA12-8768NW$ with P.A. differences larger than $\approx 45^{\circ}$.
$\rm Deep3a-6004$ is well resolved (morphological major axis FWHM 
$\approx 3\,\times$ PSF FWHM; Figure~\ref{fig-maps_deep3a1}) but
its morphological P.A. is nearly orthogonal to the kinematic P.A.
($\approx 80^{\circ}$).  This may be due to a variety of reasons
--- one of them being an asymmetric distribution of the most intense
star-forming regions traced by H$\alpha$ and/or of the obscuring dust
within the galaxy.  In general, we adopted as major axis of the galaxies
the kinematic P.A. except for the cases with too poor quality velocity 
fields from low S/N, or simply no clearly apparent velocity gradient,
where we took the morphological P.A. instead.

To extract profiles of the flux, velocity, and velocity dispersion along
the major and minor axis of each galaxy, we applied the same procedure as
described in \S~\ref{Sub-extrspec}.  We computed spectra integrated from
the unsmoothed reduced data cubes in circular apertures spaced equally
along the major and minor axes (with typical diameters of 6 pixels and
separations of 3 pixels, roughly $\approx 1.5$ and 0.75 times the PSF FWHM
of the data sets).  Weighting and formal uncertainties in the line fitting
procedure were based on the noise spectrum for the corresponding aperture
size inferred from the empirical noise model (Appendix~\ref{App-noise}).
We extracted position-velocity diagrams in 6~pixel-wide synthetic slits
along the major axis of the galaxies, integrating the light along the
spatial direction perpendicular to the slit orientation.

\subsection{Size Estimates}
            \label{Sub-extrsizes}

We determined the intrinsic half-light radii $r_{1/2}({\rm H\alpha})$ of
the galaxies from the H$\alpha$ curves-of-growth extracted as described
in \S~\ref{Sub-extrspec}, and corrected for the respective PSF FWHMs.
We also measured the intrinsic FWHMs of one-dimensional Gaussians fitted
to the major axis H$\alpha$ light profiles and corrected for spatial
resolution, which is appropriate for getting an estimate of the linear
size of inclined disk systems \citep[see also][]{FS06a, Bou07}.  For the
detected sources that are spatially resolved (i.e., with morphological
major axis $\rm FWHM(H\alpha) >$ PSF FWHM), we find an average
$0.5 \times {\rm FWHM(H\alpha)} / r_{1/2}({\rm H\alpha}) \approx 1.45$,
and a median ratio of $\approx 1.3$.
Clearly, the assumption of Gaussians is simplistic since the spatial
distribution of the line emission is generally irregular and often
asymmetric and clumpy for our SINS galaxies.  Nevertheless, inspection
of the fits and of the radial distributions of pixel values about the
center indicates that Gaussian profiles are reasonable representations
of the overall observed light profiles (modulo clumps or other prominent
features that produce obvious substructure in the radial distributions).

While both methods are affected by the spatial resolution, the FWHMs 
of Gaussian fits to the major axis light profiles are more sensitive to
differences between different data sets due to the fixed synthetic slit 
width.  This is less of a concern for data taken with similar effective
PSFs or for sources that are well resolved (e.g., with AO), as was the
case for our previous studies \citep[e.g.,][]{FS06a, Bou07, Gen08, Cre09}.
For this paper, however, we consider all H$\alpha$ data sets with a
range of spatial resolution (see Figure~\ref{fig-psfdet}).
For consistent analysis of the full SINS H$\alpha$ sample, we thus
adopted the $r_{1/2}(\rm H\alpha)$ estimates throughout most of this
paper.

Table~\ref{tab-Hameas} lists the $r_{1/2}({\rm H\alpha})$ and
$\rm FWHM(H\alpha)$ values.
For some objects, the size measurements (major axis FWHM or twice the
half-light radius) are smaller than the resolution element from the
estimated PSF.  This is most likely due to variations in time of the
actual seeing conditions, and which may not be reflected in the PSF
calibration since the acquisition stars used for that purpose were
not observed simultaneously with the science data.  In those cases,
we consider the observed sizes as upper limits to the intrinsic sizes.

Uncertainties on the size estimates were determined as follows.
Inspection of the PSFs associated with individual OBs of a given object
suggests that typical seeing (or effective resolution for AO data sets)
variations are of order 20\%.
In addition, in correcting for beam smearing, we assumed the PSF
is axisymmetric, which is not always the case (although the average
ellipticity of $\approx 0.1$ and $0.06$ for the seeing-limited and
AO effective PSFs indicates this is a reasonable assumption; see
\S~\ref{App-psfs}).
We computed the uncertainties taking 20\% as a representative uncertainty
on the measurements of observed galaxy sizes themselves from possible PSF
variations during the observations, and using the ellipticity of the
effective PSF for each galaxy as a measure of the error from the PSF
shape.  These were propagated analytically in calculating the resulting
uncertainties on the intrinsic (physical) sizes of galaxies.
The errors derived from the PSF ellipticities are typically 20\%
(with rms scatter 10\%) for the seeing-limited data sets, 30\% for
the AO-assisted data sets at $\rm 125~mas\,pixel^{-1}$, and 13\% for
the AO-assisted data sets at $\rm 50~mas\,pixel^{-1}$.  Overall, the
size uncertainties are $\approx 30\% - 35\%$ (with rms scatter 20\%).
These do not account for other sources of errors that are essentially
impossible to quantify, such as the dependence on the sensitivity of
the data and the actual surface brightness distribution of each
individual source (but see \S~\ref{Sub-lineprop}).

More detailed derivations of morphological parameters possible
for the sources with higher S/N data were presented elsewhere
\citep{Gen08, Sha08, Cre09}.
There, whenever possible, we verified the morphological parameters
from the H$\alpha$ line maps against the continuum maps from our
SINFONI data or available sensitive ground-based imaging of 
comparable resolution (e.g., the ISAAC imaging of CDFS).
While the line and continuum morphologies can differ in detail,
the centroid, major axis, and sizes typically agree reasonably well.
This is further confirmed from the higher resolution NICMOS/NIC2 F160W
($H$ band) data obtained for six of the BX galaxies and AO-assisted
$K$-band imaging for a few more sources obtained with NACO at the
VLT \citep{FS09a}.

\vspace{1ex}
\section{INTEGRATED H$\alpha$ PROPERTIES OF THE SINS H$\alpha$ SAMPLE}
         \label{Sect-Ha_prop}

\subsection{H$\alpha$ Detections and Flux Sensitivity Limits}
            \label{Sub-linedet}

Of the \nsins\ galaxies from the SINS H$\alpha$ sample, 60 were targets
drawn from photometric surveys, 50 of them were detected in H$\alpha$
and two ``serendipitous'' companions were identified from their line
emission.  All of the non-detections are for galaxies without prior near-IR
spectroscopy that would have provided line fluxes and guided us in the
choice of sufficiently bright targets (see Figure~\ref{fig-samples}).
Some of the non-detections can probably be attributed to the moderate to
poor seeing conditions under which the data were taken, with estimated
PSF FWHM $\approx 0\farcs 8 - 1\farcs 45$ at near-IR wavelengths
(Figure~\ref{fig-psfdet}).  Because we did not repeat observations of
targets undetected after the first $\rm 1 - 2\,hr$, these data sets
have relatively bright limiting fluxes compared to the depth achieved
in our typical SINS H$\alpha$ data sets.  In some cases, this short
integration may have been insufficient for even moderately bright
but very extended sources, with lower average surface brightness.

A ``typical detection limit'' for our data is not straightforward to
quantify as the sensitivity varies strongly with wavelength and our
data sets have a wide range of integration times.
The faintest sources have observed integrated H$\alpha$ fluxes of
$\rm \approx 2 \times 10^{-17}~erg\,s^{-1}\,cm^{-2}$
and averaged surface brightness of
$\rm \sim 1 \times 10^{-17}~erg\,s^{-1}\,cm^{-2}\,arcsec^{-2}$;
their integrated line emission is detected at $\rm S/N \approx 5$.
The $3\,\sigma$ upper limits for the undetected sources, calculated
from the noise spectrum within a circular aperture of radius 1\arcsec\
and assuming an intrinsic integrated velocity dispersion of
$\rm 130~km\,s^{-1}$, range from
$8 \times 10^{-18}$ to $\rm 8 \times 10^{-17}~erg\,s^{-1}\,cm^{-2}$,
with mean and median of $\rm \approx 4 \times 10^{-17}~erg\,s^{-1}\,cm^{-2}$.
Some of those limits are higher than the flux of the faintest sources
detected, again because of the important variations of noise levels with
wavelength (the optical redshift of several of the undetected sources
place their H$\alpha$ line close to regions of bright night sky lines,
poorer atmospheric transmission, or higher thermal background emission)
together with the short integration times of $\rm 40\,min - 2\,hr$.

In terms of H$\alpha$ luminosities (uncorrected for extinction),
the faintest SINS galaxies have integrated
$L({\rm H\alpha}) \approx 3 \times 10^{41}~{\rm erg\,s^{-1}}$.
With half the total light within an area of radius $r_{1/2}({\rm H\alpha})$
and accounting for beam smearing in computing the intrinsic physical area,
the sources with faintest H$\alpha$ surface brightness have central averaged
luminosity surface densities of
$\sim 5 \times 10^{39}~{\rm erg\,s^{-1}\,kpc^{-2}}$.
Using the conversion of \citet{Ken98}, corrected to a \citet{Cha03}
IMF (see \S~\ref{Sub-intrinsic_prop}), these imply lowest observed
star formation rates and star formation rate surface densities of
$\rm \approx 1~M_{\odot}\,yr^{-1}$ and
$\rm \sim 0.03~M_{\odot}\,yr^{-1}\,kpc^{-2}$.
The median integrated H$\alpha$ flux and surface brightness are about five
times higher, or $\rm \approx 1 \times 10^{-16}~erg\,s^{-1}\,cm^{-2}$ and
$\rm \sim 5 \times 10^{-17}~erg\,s^{-1}\,cm^{-2}\,arcsec^{-2}$
(median values for luminosities and star formation rates are
about 15 times higher than the faintest/lowest ones).

\subsection{Distributions of H$\alpha$ Fluxes, Velocity Dispersions, and Sizes}
            \label{Sub-lineprop}

Figure~\ref{fig-HaSigmaRHa} shows the distribution of integrated velocity
dispersions and half-light radii as a function of total H$\alpha$ line
fluxes for our SINS H$\alpha$ sample.  The velocity dispersions are
corrected for instrumental spectral resolution, and the radii are
corrected for beam-smearing accounting for the effective PSF.
The H$\alpha$ fluxes are not corrected for extinction.
While one could use, e.g., the best-fit extinction $A_{V}$ derived from
the SED modeling, it is here more appropriate to stick to uncorrected
quantities as the velocity dispersions and sizes were measured from the 
emergent line emission.  Correcting for extinction would rely on the 
assumption that the obscured regions have the same kinematics and spatial 
distribution as those suffering less extinction and dominating the observed 
properties, a hypothesis that we cannot verify with current available data.
For the \nsinsdet\ detected sources, the total observed H$\alpha$ fluxes
$F({\rm H\alpha})$ cover a range by a factor of $\approx 40$ with median
$\rm 1 \times 10^{-16}~erg\,s^{-1}\,cm^{-2}$. 
The integrated intrinsic velocity dispersions span
$\sigma_{\rm int}({\rm H\alpha}) \approx 35 - 280~{\rm km\,s^{-1}}$
with mean and median of $\rm 130~km\,s^{-1}$.
The intrinsic half-light radii inferred from the curve-of-growth
analysis range from $\rm \la 1.5$ to 7.5~kpc with mean and median
of 3.4~kpc and 3.1~kpc, respectively.

Despite differential cosmological dimming given the redshift range
spanned by the sources, with clear bimodality (Figure~\ref{fig-samples}),
Figure~\ref{fig-LHaSigmaRHa} shows qualitatively similar distributions
in terms of total absolute H$\alpha$ luminosities (uncorrected for
extinction) as obtained as a function of total apparent fluxes.
The observed H$\alpha$ luminosities of the detected sources range
from $L({\rm H\alpha}) = 2.5 \times 10^{41}$ to
$\rm 1.3 \times 10^{43}~erg\,s^{-1}$, with mean and median of
4.2 and $\rm 3.5 \times 10^{42}~erg\,s^{-1}$.
Figure~\ref{fig-MstarSigmaRHa} shows again the integrated intrinsic
velocity dispersions and half-light radii now as a function of stellar
masses from the SED modeling (the three galaxies for which broad-band
photometry is unavailable are excluded).  Overall, similar trends of
increasing $\sigma_{\rm int}({\rm H\alpha})$ and $r_{1/2}({\rm H\alpha})$
with increasing $M_{\star}$ are seen as when plotting against
$F({\rm H\alpha})$ and $L({\rm H\alpha})$.
\footnote{
This holds also when using the stellar masses derived with the different
SED modeling assumptions considered in Appendix~\ref{App-sedmod}.}
This is not surprising
given that more massive star-forming galaxies also tend to have higher
H$\alpha$ luminosities, reflecting the empirical $M_{\star} - {\rm SFR}$
relationship among star-forming galaxies (see \S~\ref{Sect-dust_sf},
and also, e.g., \citealt{Dad07}).
The various trends discussed above, however, can be affected by
the sensitivity limits resulting from observational strategies,
which we quantify in \S~\ref{Sub-detlim}.

In Figures~\ref{fig-HaSigmaRHa} to \ref{fig-MstarSigmaRHa},
the disks and mergers classified by kinemetry \citep{Sha08}
tend to lie at brighter H$\alpha$ fluxes/luminosities (and stellar
masses) and at larger sizes, driven by the necessity of sufficient
S/N and number of resolution elements for reliable kinemetry.
However, as was the case for the photometric and stellar properties 
(\S~\ref{Sect-representativeness}, and Figures~\ref{fig-magcol} and
\ref{fig-sedprop}), our data do not show any evidence that these
fairly massive, large, bright disks and mergers at $z \sim 2$ are
different in terms of their integrated H$\alpha$ fluxes/luminosities,
velocity dispersions, or sizes.
The mean and median values for these disks and mergers are the
same within $20\%$ or less, and the Mann-Whitney $U$ test indicates
that they do not differ significantly in global H$\alpha$ properties.
Obviously, it will be important to
investigate this issue further with larger numbers of sources.
Likewise, the four AGN do not appear to be strong outliers in their
H$\alpha$ properties, although the sample is very limited and this
should not be overinterpreted.  Consequently, median values and
ranges in the integrated H$\alpha$ properties of our SINS sample
change very little when excluding the four known AGN.

With the full 2D mapping of the line emission and kinematics of
our SINFONI data, we can examine in more detail the possible AGN
contribution to the global H$\alpha$ properties, as measured from
the dominant narrow component to which our line extraction procedure
is sensitive (\S~\ref{Sub-extrspec}; see also \citealt{Sha09} for a
discussion of the presence of broad underlying H$\alpha$ emission).
For $\rm D3a-15504$, a fairly large system observed with AO at the
$\rm 50~mas\,pixel^{-1}$ scale, the AGN affects only the central
few kpc while star formation in giant kpc-size sites all across
the rotating disk clearly dominate the global H$\alpha$ properties 
(the $\sigma_{\rm int} = 148~{\rm km\,s^{-1}}$ is close to the 
average of the SINS sample, and the integrated [\ion{N}{2}]/H$\alpha$
line ratio is 0.35; \citealt{Gen06, Gen08}).
For $\rm K20-ID5$, the AGN (or shock excitation) appears to dominate
the H$\alpha$ line emission and kinematics, given its very bright
central peak showing one of the broadest line width among our sample
($\sigma_{\rm int} = 281~{\rm km\,s^{-1}}$) and a fairly high global
[\ion{N}{2}]/H$\alpha$ ratio of $\approx 0.6$.
The other two are probably intermediate cases; both show indications
of a small monotonic velocity gradient, with $\rm Q1623-BX663$ having
higher H$\alpha$ line width but lower integrated [\ion{N}{2}]/H$\alpha$
ratio ($\rm 172~km\,s^{-1}$ and 0.3) and vice-versa for $\rm D3a-7144$
($\rm 140~km\,s^{-1}$ and 0.8).

\subsection{Impact of Sensitivity Limits and Observing Strategy
            on Trends with H$\alpha$ Velocity Dispersions and Sizes}
            \label{Sub-detlim}

Our SINS data in Figures~\ref{fig-HaSigmaRHa} to \ref{fig-MstarSigmaRHa}
suggest possible trends of increasing velocity dispersion and size with
H$\alpha$ flux or luminosity and with stellar mass, although with
significant scatter increasing in $\sigma_{\rm int}$ at low fluxes, 
and towards small sizes.  Taken at face value, this would imply that
the most intense (unobscured) star formation activity takes place in
the larger and more massive systems, and that the integrated H$\alpha$
velocity dispersion is related to galaxy stellar mass despite varying
contributions by large-scale velocity gradients, non-circular motions,
and intrinsic gas turbulence \citep[see also, e.g.,][]{Erb06b, FS06a}.
However, it is important to assess carefully the impact of our
detection limits and observational strategy on the apparent trends.

To evaluate our sensitivity limits in terms of size,
we did the following simulations based on the real SINFONI data.
We first determined the average S/N per spatial resolution element
for each galaxy and calculated the necessary increase in half-light
radius for the S/N to drop below $\rm S/N = 3$, keeping the other
properties constant (total H$\alpha$ flux and integrated line width).
The sizes thus derived, $r_{1/2}^{\rm ~det.lim. data}$, correspond
to the actual surface brightness sensitivity limits of the data sets,
with their respective integration times.  The running median through
$r_{1/2}^{\rm ~det.lim. data}$ as a function of $F({\rm H\alpha})$
in logarithmic units follows closely a straight line with slope
$\beta \approx 0.5$, implying approximately
$r_{1/2}^{\rm ~det.lim. data} \propto F({\rm H\alpha})^{0.5}$
(dashed line in Figure~\ref{fig-HaSigmaRHa}{\em b\/}).
The standard deviation of the individual $r_{1/2}^{\rm ~det.lim. data}$
about this line is $\rm \approx 0.25~dex$, comparable to that of the
measured $r_{1/2}({\rm H\alpha})$ themselves about their running median.
The line of sensitivity limit of the data lies about $\rm 0.6~dex$
above the locus of data points, or a factor of 4; this is also the
average ratio of $r_{1/2}^{\rm ~det.lim. data}/r_{1/2}$ of the
individual galaxies.
The immediate implication is that the H$\alpha$ sizes we measured
are little affected by the sensitivity of our data sets.

However, as already mentioned, when observing we followed the strategy
that if a source was not detected within $\rm 1 - 2~hr$ irrespective of
actual morphology, we did not reobserve it.
This is a potential concern for interpreting possible trends, as it
means that we may have missed the more extended and/or diffuse sources
for a given flux.  To mimic this effect in our simulations, we re-did
the same calculations but this time normalizing the S/N per resolution
element for each source to a common integration time of 1\,hr
(this is a rather conservative estimate as in some cases we observed
up to 2.5\,hr before abandoning sources that were not detected).
The running median of the resulting
$r_{1/2}^{\rm ~det.lim. 1hr}$ versus $F({\rm H\alpha})$ follows
a line nearly parallel to that for $r_{1/2}^{\rm ~det.lim. data}$ 
but is offset by $\rm \approx 0.3~dex$ towards smaller sizes and
corresponds roughly to the upper envelope of the data points
(solid line in Figure~\ref{fig-HaSigmaRHa}{\em b\/}).  
This suggests that the apparent trend possibly results partly 
from our observational strategy.
The standard deviation of the individual $r_{1/2}^{\rm ~det.lim. 1hr}$
about the running median line is $\rm \approx 0.25~dex$, similar to that
for $r_{1/2}^{\rm ~det.lim. data}$.  This indicates that the scatter in
limiting size estimates reflects the different sensitivities in different
wavelength regions more than the different integration times among the
sources.

We evaluated our sensitivity limits in terms of velocity dispersion
in a similar manner, increasing the intrinsic velocity dispersion while
keeping the total H$\alpha$ line flux and half-light radius constant.
The criterion for detection limits was here $\rm S/N = 3$ per spectral
resolution element.  The lines through the running median of
$\sigma_{\rm int}^{\rm ~det.lim. data}$ and
$\sigma_{\rm int}^{\rm ~det.lim. 1hr}$ as a function of
$F({\rm H\alpha})$ are plotted as dashed and solid lines in
Figure~\ref{fig-HaSigmaRHa}{\em a\/}.  The logarithmic slopes are
$\beta \approx 0.8$, significantly steeper than that for the measured
$\sigma_{\rm int}({\rm H\alpha})$ (about 0.2), and the lines lie well
above that of the measurements by $\sim 0.9$ and 0.3~dex, respectively
(taking the average over the range of fluxes of our SINS galaxies).
This analysis indicates that our observing strategy may have prevented
the detection of faint sources with broad lines at the lowest flux levels
but that it is most likely not a limiting factor in the ensemble, as the
$\sigma_{\rm int}^{\rm ~det.lim. 1hr}$ typically are up to a factor of
$\sim 5 - 10$ higher than the measured $\sigma_{\rm int}({\rm H\alpha})$
for the brighter half of the $F({\rm H\alpha})$ range.

Similar lines indicating the inferred limits in velocity dispersion
and size of the data sets and resulting from the observing strategy
are plotted in Figures~\ref{fig-LHaSigmaRHa} and \ref{fig-MstarSigmaRHa}.
Inspection of these Figures leads to the same conclusions about the
apparent trends with observed H$\alpha$ luminosity and stellar mass
as with observed H$\alpha$ flux.
The lines of $r_{1/2}^{\rm ~det.lim. data}$ versus $F({\rm H\alpha})$
correspond to limiting H$\alpha$ luminosities and star formation rates
per unit intrinsic area at $z = 2$ (uncorrected for extinction) of
$\rm 6 \times 10^{39}~erg\,s^{-1}\,kpc^{-2}$ and
$\rm 0.03~M_{\odot}\,yr^{-1}\,kpc^{-2}$ for our data sets
with average integration times of $\rm 3.4\,hr$.
Normalized to a total integration time of 1\,hr, the corresponding
limits are $\rm 2 \times 10^{40}~erg\,s^{-1}\,kpc^{-2}$ and
$\rm 0.1~M_{\odot}\,yr^{-1}\,kpc^{-2}$.

In summary, the above analysis (albeit simplistic) indicates that the
H$\alpha$ sizes and integrated velocity dispersions of our SINS H$\alpha$ 
sample galaxies are well determined and not affected by the depth of the
respective data sets.  
However, we cannot exclude that an observational bias shapes the 
upper envelope of our distributions of $r_{1/2}({\rm H\alpha})$ with
H$\alpha$ flux/luminosity and stellar mass.  This limits our ability
to assess relationships between these properties based on our data.
On the other hand, this does not seem to be a limiting factor for
$\sigma_{\rm int}$ towards the brighter and more massive end and
the apparent trend could reflect a real physical relationship.

\vspace{1ex}
\section{COMPARISON WITH OTHER $z \sim 2$ SPECTROSCOPIC SAMPLES}
         \label{Sect-other_samples}

\subsection{Comparison with the NIRSPEC BX/BM Sample}
           \label{Sub-sinf_nirspec}

Near-IR spectroscopic samples at $z \sim 2$ are still rare, especially
for spatially-resolved 2D mapping.  Our SINS H$\alpha$ sample is the
largest to date based on integral field spectroscopy.  With \nsins\
objects, it fares well compared to the NIRSPEC sample of 114
\bmbx-selected sample of \citet{Erb06b}, the largest long-slit
spectroscopic survey.  It is thus interesting to compare the
integrated H$\alpha$ properties between both surveys, also given that
our SINS sample includes $\sim 2/3$ of near-/mid-IR selected galaxies.
Moreover, our 17 optically-selected \bxbm\ were drawn from the 
NIRSPEC sample and are also the only ones with previously existing
near-IR spectroscopy, allowing direct comparisons of the properties 
measured with different types of instrument.

Figure~\ref{fig-sins_nirspec_comp} plots the integrated H$\alpha$
fluxes, intrinsic velocity dispersions, and intrinsic sizes measured
with SINFONI against those measured with NIRSPEC.
Here, we use the fluxes as reported in Table~4 of \citet{Erb06b}, 
without any aperture correction.  There is a tight relation between
the fluxes, with scatter of 0.21~dex but a systematic offset
corresponding to higher fluxes with SINFONI by a factor of 1.6.
Based on narrow-band imaging for H$\alpha$ of sources in the
$\rm Q1700$ field, \citet{Erb06b} had estimated an average factor
of $\sim 2$ ``aperture correction'' accounting for slit losses
and slit misalignment possibly missing part of the sources' emission.
Absolute flux calibration is challenging for both narrow-band imaging
and long-slit spectroscopy.
For SINFONI, the full coverage of the atmospheric bands with each
of the gratings allows us to synthetize the broad-band fluxes of
our telluric standards, which should help reduce uncertainties of
the absolute flux calibration.  In addition, with the full 2D mapping,
we recover the total fluxes of the sources irrespective of their sizes,
P.A., or morphologies.
If the comparison with our SINS \bmbx\ galaxies applies to the
NIRSPEC sample as a whole, the average aperture correction used
by \citet{Erb06b} might have been overestimated by $\approx 25\%$.

In terms of integrated velocity dispersion and half-light radii,
the SINFONI measurements are on average 10\% larger (scatter 0.18~dex)
and 5\% smaller (scatter 0.13~dex).  The agreement is remarkable
considering that the NIRSPEC slit apertures may have missed part of
the galaxies, and the sizes were estimated also from the long-slit data.
When significant, the differences between fluxes, velocity dispersions,
and sizes of individual objects can be attributed to possible slit
misalignment with respect to the major axis of the galaxies, or to
the proximity of bright night sky lines which may have affected more
the lower spectral resolution NIRSPEC data.

Figure~\ref{fig-sins_nirspec} now compares the distributions
of H$\alpha$ properties of the full SINS and NIRSPEC samples,
as a function of stellar masses.
The fluxes for the NIRSPEC sample in these plots have been scaled by the
factor of two aperture correction estimated by \citet{Erb06c}; for the
purpose of this comparison, the $25\%$ difference with the factor of
1.6 we inferred above has little impact.
Overall, the SINS sample covers the same range of H$\alpha$ fluxes,
luminosities, intrinsic integrated velocity dispersions, and half-light
radii as the NIRSPEC sample.  The median values in all these properties
are nearly the same to within $\leq 15\%$ (ignoring the possible $25\%$
adjustment for aperture correction).
In terms of stellar masses, both samples cover a very comparable range,
with the median for the NIRSPEC sample being only $\approx 30\%$ lower
than the median for the SINS sample (reflecting the somewhat more
important tail at $M_{\star} \la 4 \times 10^{9}~{\rm M_{\odot}}$).
In all of these properties, the differences in median values between
the SINS and NIRSPEC samples are much smaller than the ranges covered.

In view of the significant overlap (to a given $K$ magnitude)
between the \bmbx\ and $sBzK$ populations (to which almost all our
SINS near-/mid-IR selected galaxies belong even if not explicitly
selected so), it may not be surprising to find large overlap in
integrated H$\alpha$ properties between the SINS and NIRSPEC
samples.  In addition, for both studies, the same requirement
of an optical spectroscopic redshift was applied in the target
selection, leading to similar biases towards the bluer and brighter
part of the high redshift population compared to a pure $K$-selected
sample (\S~\ref{Sect-representativeness} and \citealt{Erb06b}).
Perhaps the most interesting outcome of the comparisons above,
based on real data sets, is that even if the total fluxes are
subject to significant uncertainties from aperture corrections,
slit spectroscopy seems to be reliable in recovering the integrated
emission line widths and even the sizes of faint high-redshift
galaxies under typical observing conditions.
This seems to be encouraging for studies of spatially-integrated
kinematics using near-IR multi-object spectrographs such as 
MOIRCS on Subaru, or in the near future LUCIFER at the Large
Binocular Telescope and MOSFIRE at Keck.

\subsection{Comparison with Other IFU Samples}
            \label{Sub-sinf_ifus}

Figure~\ref{fig-sins_ifus} makes a similar comparison as above with
the NIRSPEC sample of \citet{Erb06b} but with data from other studies
using near-IR integral field spectrometers and for galaxies in the
redshift interval $1.3 \la z \la 2.6$.
This is not an exhaustive comparison as we restricted ourselves
to samples with published integrated H$\alpha$ fluxes, and 
intrinsic velocity dispersions and/or half-light radii measurements.
We included the Keck/OSIRIS observations of
\citet[][12 objects in the relevant redshift range]{Law07b, Law09} and of
\citet[][nine objects, counting merger components separately]{Wri07, Wri09}.
These were all optically-selected \bxbm\ sources.
We also included the SINFONI data of
\citet[][six sources, consisting of MIPS $\rm 24~\mu m$-selected
  and/or morphologically large disks at $z \sim 2$]{Sta08}
and of
\citeauthor{Epi09} (2009, nine $I$-band-selected objects with strong
[\ion{O}{2}]\,$\lambda\,3727$ line emission from the VIMOS VLT Deep Survey).
Other samples exist (the bright SMGs of \citealt{Swi06},
or the massive $K < 20$ spectroscopic sample of \citealt{Kri07})
but the relevant set of measurements were unfortunately
not available in the literature.

As can be seen from Figure~\ref{fig-sins_ifus}, these samples
together cover the ranges of stellar masses, H$\alpha$ fluxes and
luminosities, and integrated velocity dispersions spanned by our SINS
galaxies.  There are little differences in terms of H$\alpha$ fluxes and
luminosities, with large overlap among all samples.  The main difference
is in the mass ranges, with the $\rm 24~\mu m$-detected large disks of
\citet{Sta08} lying at the high mass end and the \bmbx\ objects of
\citet{Law07b, Law09} and \citet{Wri07, Wri09} towards lower masses, 
which then translates into different ranges of velocity dispersions
given the trend in $\sigma_{\rm int}$ versus $M_{\star}$.  Each sample
covers roughly $1/2 - 2/3$ of the total mass range of our SINS sample,
with the \citet{Epi09} galaxies spanning a more intermediate interval.
No H$\alpha$ sizes are available for the \citet{Sta08} galaxies.
As these were mostly selected to be morphologically large disks based
on deep near-IR imaging, these would probably sit at the top right
of the distribution in $r_{1/2}({\rm H\alpha})$ versus $M_{\star}$.

There is a striking difference between the \citet{Law07b, Law09}
sample with respect to that of \citet{Wri07, Wri09}, \citet{Epi09},
and the SINS galaxies: they all appear to be significantly smaller.
This may be a natural consequence of the higher resolution of the
OSIRIS$+$AO data, at least concerning the difference with the mostly
seeing-limited SINS and \citet{Epi09} samples.  On the other hand,
this could reflect an observational bias or surface brightness sensitivity
effects, as discussed by \citet{Law09} and for our SINS sample in
\S~\ref{Sub-detlim}.
Their eleven non-detections were observed for comparatively short
integration times, and include two of our brightest SINS sources 
($\rm Q1623-BX663$ and $\rm Q2343-BX389$, with the latter particularly
extended) although poor observing conditions may have conspired.  
\citet{Law09} discussed that this may explain in part the discrepancy
by a factor of $\sim 2$ between their H$\alpha$ fluxes and those of
\citet[][from whose sample their targets were drawn]{Erb06b} after
aperture correction, but also note that for the most compact objects
the aperture correction may simply be too large.  Our SINFONI flux
measurements show excellent agreement for $\rm Q1623-BX502$
($23\%$ difference between OSIRIS and SINFONI).
For $\rm Q2343-BX513$, we measured a total flux twice higher than
\citet{Law09}.  In both cases, in fact, the NIRSPEC fluxes without
aperture correction agree very well with our SINFONI fluxes,
suggesting that for those compact sources, the aperture correction
may indeed have been overestimated.
For SINS, while we did emphasize somewhat larger brighter objects
at early stages, we took care subsequently (for about 2/3 of the
final sample) of minimizing such a selection bias.
It seems unclear what factors played what role in the overall size
differences, and it could reflect a genuine difference of
the galaxies properties (we discuss this point further in
\S~\ref{Sect-kinematics}).

\vspace{1ex}
\section{DUST DISTRIBUTION AND STAR FORMATION PROPERTIES}
         \label{Sect-dust_sf}

Our SINS H$\alpha$ sample combines two key aspects: full spatial mapping
of the H$\alpha$ emission for a sizeable sample of 62 $z = 1.3 - 2.6$
systems.
In this section, we explore, by combining the integrated H$\alpha$
measurements and the global properties derived from the SED modeling,
constraints on the dust distribution and star formation histories of
the SINS galaxies --- two of the most important but elusive galaxy
properties.  These issues have been addressed in previous near-IR
studies of $z \sim 2$ galaxies, but the results were limited because
of small sample sizes or potentially affected by uncertain aperture
corrections \citep[e.g.][]{Dok04, Erb06c, Kri07}.

Throughout, we assume that the measured H$\alpha$ line emission
originates from star-forming regions, with no contribution from
AGN or shock-ionized material (other sources should be negligible
for actively star-forming galaxies; e.g., \citealt{Bri04}).
This is supported by the rest-frame optical line ratios from our
SINFONI data (including [\ion{N}{2}]/H$\alpha$ for all galaxies and
[\ion{O}{3}]/H$\beta$ for a small subset; \citealt{Bus09}) as well as
by their rest-UV spectra, except for the four known AGN.  As argued in
\S~\ref{Sub-lineprop}, in at least one of them star formation nonetheless
appears to dominate the integrated H$\alpha$ flux.  Based on stacking
analysis, \citet{Sha09} suggest evidence for a broad underlying H$\alpha$
component ($\rm FWHM \ga 1500~km\,s^{-1}$) in our SINS sample, which,
along with the dominant star formation activity producing the narrower
component, could be due to either lower-level or obscured AGN activity
or powerful starburst-driven galactic outflows.  However, the line fitting
method applied in this paper is little sensitive to such a component.
The presence of low-luminosity or obscured AGN would also affect to some
extent the broad-band SEDs and thus the derived stellar properties but
except for very few sources, including some of the known AGN, the SEDs
do not show evidence from AGN contamination.

\subsection{Intrinsic H$\alpha$ Luminosities, H$\alpha$ Equivalent Widths,
            and Star Formation Rates Estimates}
            \label{Sub-intrinsic_prop}

In this subsection, we detail our derivation of the various intrinsic
quantities used in the following discussion.  The results are reported
in Tables~\ref{tab-LHaWHa} and \ref{tab-SFRs}.

As we do not have direct estimates of the dust attenuation applicable
to the H$\alpha$ line emission (e.g., from measurements of the Balmer
decrement), we used the best-fit extinction values $A_{V,\,{\rm SED}}$
from the SED modeling (Table~\ref{tab-sedprop}).  Quantities corrected
for this amount of extinction are denoted with the superscript ``0.''
From studies of local star-forming and starburst galaxies, there is
evidence that on average the Balmer line emission is more attenuated
by a factor of $\sim 2$ than the starlight that dominates the optical
continuum emission
\citep*[e.g.][]{Fan88, Cal94, Cal96, Cal00, Mas99, May04, Cid05}.
This is usually interpreted as indicating that the young hot ionizing
stars are associated with dustier regions than the bulk of the (cooler)
stellar population across the galaxies.  Plausibly, direct absorption
of Lyman continuum photons by dust grains present inside the \ion{H}{2}
regions may also account (at least in part) for the observed net effect
\citep*[e.g.][]{Pet72, Spi78, McK97}.
We therefore also accounted for this possibility in our SINS $z \sim 2$
galaxies and adopted the relation from \citet{Cal00}, which implies
$A_{V,\,{\rm neb}} = A_{V,\,{\rm SED}} / 0.44$.
H$\alpha$-derived properties corrected for this extra attenuation
relative to the stars are denoted with a ``00'' superscript.
We assumed the \citet{Cal00} reddening law to represent the wavelength
dependence of the extinction, giving $A_{\rm H\alpha} = 0.82\,A_{V}$.
The \citeauthor{Cal00} law acts as an effective foreground screen of
obscuring dust, and so the extinction correction at the wavelength
of H$\alpha$ is $e^{0.76\,A_{V,{\rm neb}}}$ or, equivalently,
$10^{0.33\,A_{V,{\rm neb}}}$.

We neglected Balmer absorption from the stellar populations in
estimating the intrinsic H$\alpha$ luminosities and equivalent
widths.  For a wide range of star formation histories and for
plausible IMFs, the stellar absorption at H$\alpha$ is always
$< 5$~\AA\ \citep[see also, e.g.,][]{Bri04}, small compared to
the equivalent widths of the feature in emission in our
SINS galaxies and to other sources of uncertainties.
We also neglected Galactic extinction, as the correction for H$\alpha$
observed in $H$ or $K$ band is $< 5\%$ for all fields where our sources
are located.

The intrinsic H$\alpha$ luminosities \lhao\ and \lhaoo\ were calculated
from the observed luminosities \lhaobs\ by correcting for dust attenuation
for the two cases described above.  The corresponding star formation rates
\sfrha\ and \sfrhaoo\ were then computed based on the \citet{Ken98}
conversion:
\begin{equation}
 \log ({\rm SFR(H\alpha)}~[{\rm M_{\odot}\,yr^{-1}}]) =
 \log (L({\rm H\alpha})~[{\rm erg\,s^{-1}}]) - 41.33,
\label{Eq-LHaSFR}
\end{equation}
where the constant includes a factor of 1.7 adjustment between our
adopted \citet{Cha03} IMF and the \citet{Sal55} IMF used by \citet{Ken98}.

We estimated the rest-frame H$\alpha$ equivalent widths in two ways.
We computed the ratio of our integrated H$\alpha$ line fluxes to the
broad-band flux densities (from the total $H$ or $K$ magnitudes for
sources at $z < 2$ and $z > 2$, respectively) after subtracting the
contribution by H$\alpha$
($f_{\rm BB}({\rm H\alpha})$; Table~\ref{tab-Hameas}).
For the majority of our SINS H$\alpha$ sample galaxies, the flux ratio
[\ion{N}{2}]\,$\lambda\,6584$/H$\alpha$ $< 0.4$ and since the average 
$f_{\rm BB}({\rm H\alpha}) \approx 10\%$ (median $\approx 7\%$), we
neglected contamination by other lines than H$\alpha$.
These estimates are denoted \whabb.
Given the uncertainties involved (e.g., accurate aperture corrections
and varying emission line contamination not fully accounted for),
we also computed the equivalent widths using measurements of the
line-free continuum within $\rm \pm 5000~km\,s^{-1}$ of H$\alpha$ in our
integrated SINFONI spectra, denoted \whasinf.  For sources undetected
in the continuum, we adopted $3\,\sigma$ upper limits on the continuum
flux density.

Figure~\ref{fig-ews_comp} compares the equivalent widths obtained
with each method.  The agreement is good, with best-fit line
to the data (excluding limits) of slope 0.995, median
$W^{\rm rest}_{\rm SINF}({\rm H\alpha}) /
 W^{\rm rest}_{\rm BB}({\rm H\alpha}) = 1.28$,
and scatter of the relation (in logarithmic units) of
$\rm \approx 0.35~dex$.
For many of the SINS galaxies, our SINFONI data are less sensitive
to the continuum emission than the available broad-band imaging, and
uncertainties in the continuum determination can be significant (e.g.,
from the background subtraction, affecting directly the continuum level).
We therefore adopted the \whabb\ values for the analysis.
The rest-frame H$\alpha$ equivalent widths computed from the
observed line and continuum fluxes are insensitive to extinction
if $A_{V,\,{\rm neb}} = A_{V,\,{\rm SED}}$.
For the case of extra attenuation towards the \ion{H}{2} regions,
we derived \whaoo\ applying the differential extinction implied by
$A_{V,\,{\rm neb}} = A_{V,\,{\rm SED}} / 0.44$.

Part of our analysis relies on the comparison of measured quantities
with predictions based on the best-fit stellar population implied by
the SED modeling.
The predicted intrinsic H$\alpha$ luminosity \lhapred\ was calculated
from the rate of H ionizing photons in the \citet{BC03} models for the 
best-fit parameters derived from the SED modeling.  We then converted
the H ionizing rates to intrinsic H$\alpha$ luminosities applying
the recombination coefficients for case B from \citet{Hum87},
for an electron temperature $T_{\rm e} = 10^{4}~{\rm K}$ and
density of $n_{\rm e} = 10^{4}~{\rm cm^{-3}}$, which gives:
\begin{equation}
 \log (L({\rm H\alpha})~[{\rm erg\,s^{-1}}]) =
 \log (N_{\rm Lyc}~[{\rm s^{-1}}]) - 11.87,
\label{Eq-LHaLyc}
\end{equation}
where $N_{\rm Lyc}$ is the production rate of Lyman continuum photons
from the stars.  Alternatively, one can apply the widely used \citet{Ken98} 
relation between star formation rate and H$\alpha$ luminosity, although it
was derived from different synthesis models, with somewhat different
ingredients and assumptions.  Converting the best-fit star formation
rates from our SED modeling through equation~(\ref{Eq-LHaSFR}) leads
to differences in predicted H$\alpha$ luminosities of $\leq 10\%$ for
our SINS galaxies.

To compute the \whapred, we combined the \lhapred\ together with 
the synthetized Bessel $R$ band ($\lambda \approx 6600$~\AA) magnitude
from the same \citet{BC03} models.  Line emission is not included
in the \citet{BC03} models so the broad-band magnitudes represent
the average stellar continuum flux density in the bandpass.  Again,
stellar absorption is neglected as it is a minor effect for our sample.

We also derived star formation rates from rest-frame UV luminosities.
To this end, we used the observed total $B$ or $G$ band magnitudes 
(all corrected for the foreground Galactic extinction; see
Table~\ref{tab-photprop}).
For $z = 1.3 - 2.6$ spanned by our galaxies, the $B$ and $G$ bandpasses
probe the rest-frame $\lambda = 1200 - 2100$~\AA\ range.  Assuming that
the rest-frame UV spectra of the galaxies are dominated by the light
from young OB stars, the intrinsic continuum emission is relatively flat
in $f_{\nu}$ over this interval, after accounting for dust extinction.
Table~\ref{tab-SFRs} lists the observed \luvobs\ calculated from the
$B$ or $G$ magnitudes and those corrected for the best-fit $A_{V}$
from the SED modeling.  Here, we assumed a rest-UV wavelength of
1500\,\AA\ and, with the \citet{Cal00} reddening law, the extinction
correction is $e^{2.35\,A_{V,{\rm SED}}}$ or $10^{1.02\,A_{V,{\rm SED}}}$.
The intrinsic rest-UV derived star formation rates \sfruv\ were
then computed via the \citet{Ken98} conversion adjusted for
our adopted IMF:
\begin{equation}
 \log ({\rm SFR(UV)}~[{\rm M_{\odot}\,yr^{-1}}]) =
 \log (L({\rm UV})~[{\rm erg\,s^{-1}\,Hz^{-1}}]) - 28.08,
\label{Eq-LuvSFR}
\end{equation}
where modeling differences between this work and \citet{Ken98}
have little impact.

The \sfruv's and \sfrsed's are not fully independent,
since the SED modeling involves the optical photometry.
We obtain a very tight linear correlation between \sfruv\ and \sfrsed\
with logarithmic slope of 1.05 and standard deviation of the residuals
of 0.14~dex.
\footnote{
Other SED modeling assumptions (see Appendix~\ref{App-sedmod}) lead to
similar slopes within $5\%$ of unity, and similar standard deviations
of residuals of $\rm 0.13 - 0.19~dex$.}
The low scatter seems surprising, as our approach to derive
the \sfruv's is admittedly very crude and ignores, e.g., any $K$-correction
or the individual star formation histories, which are explicitly taken
into account in the SED modeling.
The tightness of the relation likely reflects the degree to which
the rest-UV fluxes and colours drive the SED fits for our SINS galaxies,
which tend to have bluer optical to near-IR colours compared to a less
biased $K$-selected sample (\S~\ref{Sect-representativeness} and
Figure~\ref{fig-magcol}).  This limits in practice the usefulness
of \sfruv\ as additional estimate in our analysis but we nevertheless
consider it for comparisons with the literature.

Another widely used star formation rate indicator is the $\rm 24~\mu m$
flux as measured with the Spitzer/MIPS instrument, probing the rest-frame
$\rm \sim 8~\mu m$ PAH emission at $z \sim 2$.  However, MIPS data are
available for too small a fraction of our SINS sample to allow for
meaningful comparisons and so we do not include these estimates in 
our analysis.

\subsection{Constraints on the Dust Distribution}
            \label{Sub-dust}

Constraints on the dust distribution within galaxies, and in particular
towards the \ion{H}{2} regions relative to the bulk of the stars, ideally
require independent estimates of the extinction to the photoionized nebulae
($A_{V,\,{\rm neb}}$) from H recombination line ratios, which can be compared
to that applicable to the stellar light obtained from broad-band colours or
SED modeling ($A_{V,\,{\rm stars}}$).  At high redshift, H$\alpha$ is the
most easily observed H recombination line
\footnote{The resonantly scattered Ly$\alpha$ line is very sensitive
to radiative transfer effects, which complicates its use to constrain
the dust obscuration.}.
H$\beta$ measurements are in practice quite challenging because the
line is fainter, the underlying stellar absorption is more important
(with equivalent width roughly twice that for H$\alpha$), and because
of the requirement of having H$\alpha$ and H$\beta$ simultaneously falling
within atmospheric transmission windows and in spectral regions free from
bright night sky lines.  Any other H line is either fainter, or redshifted at
wavelengths $\rm \lambda > 3~\mu m$ that are, with current instrumentation,
hardly accessible for faint distant galaxies.

We thus follow an indirect approach to explore whether we can set
useful constraints on the dust distribution within our SINS galaxies.
Figure~\ref{fig-LHaEWsVSpred} compares the measured intrinsic H$\alpha$
luminosities and rest-frame equivalent widths with those predicted from
the best-fit model to the SEDs.
Panels {\em a\/} and {\em c\/} show the case of no differential
extinction between the \ion{H}{2} regions and the bulk of the stars.
To first order, there are two obvious effects that can lead
to deviations from a one-to-one relationship in these plots.
Non-negligible contributions from other sources than star formation
would result in measured \lhao\ and \whabb\ exceeding the predicted
\lhapred\ and \whapred.  A few data points lie above the one-to-one
relation (and are explained in \S~\ref{Sub-sfhs}) but the large
majority of the SINS galaxies clearly lies below.
The other effect would naturally explain this, namely that nebular
photons experience on average more extinction than starlight (and
possibly also that part of the ionizing radiation is absorbed by
dust within the \ion{H}{2} regions), as inferred in local 
star-forming and starburst galaxies.

At $z \sim 2$, \citet{Dok04} and \citet{Kri07} observed a similar
effect in their non-AGN massive $K$-bright star-forming objects.
Based on the same analysis as carried out above, \citet{Dok04}
found that an additional extinction of $\Delta A_{V} \sim 1~{\rm mag}$
brought their measured and predicted H$\alpha$ luminosities and
equivalent widths in good agreement.
In contrast, from comparisons of SFR estimates derived from H$\alpha$
and other indicators, \citet[][see also \citealt{Erb03}]{Erb06c}
did not find evidence for such differential extinction from their
$\sim 100$ \bxbm-selected galaxies but noted that if the aperture
correction by a factor of two applied to their NIRSPEC long-slit
H$\alpha$ observations was overestimated, there could be room for
additional extinction towards the \ion{H}{2} regions.
Direct comparison of our SINFONI H$\alpha$ fluxes with theirs for the
17 objects in common suggests a lower correction of a factor of 1.6
(see \S~\ref{Sub-sinf_nirspec}), which, if applicable to the full
NIRSPEC sample, would allow for a small amount of extra attenuation.

In Figure~\ref{fig-LHaEWsVSpred}{\em b\/} and {\em d\/} we
compare again the measured and predicted quantities but now
assuming $A_{V,\,{\rm neb}} = A_{V,\,{\rm SED}} / 0.44$.
This reduces the scatter of the data points by a factor of
$\approx 1.5$ and the resulting distributions are well represented 
by linear relationships with slope close to unity.  Quantitatively,
and in logarithmic space, the Spearman's rank correlation coefficient 
for \lhao\ vs \lhapred\ is $\rho = 0.41$ and the correlation significance
is at the $2.8\,\sigma$ level.  For \lhaoo\ vs \lhapred, $\rho = 0.76$ and
with correlation significance of $5.3\,\sigma$, implying a stronger positive
correlation.  A robust linear bisector fit to the data with the extra
attenuation gives a slope of 1.04 and standard deviation of the residuals
of 0.30~dex (excluding limits).
For \whabb\ vs \whapred, our data give $\rho = -0.13$ and $0.8\,\sigma$
significance, or hardly any correlation.  For \whaoo\ vs \whapred, the
data become positively correlated with $\rho = 0.28$ and $2\,\sigma$
significance, and the best-fit line has a slope of 1.10 with standard
deviation of the residuals of 0.36~dex.

This behaviour is also seen when adopting other SED modeling
assumptions (see Appendix~\ref{App-sedmod}) and using the corresponding
extinction laws when correcting the observed H$\alpha$ fluxes for dust
obscuration.  Specifically, the case of extra attenuation towards the
\ion{H}{2} regions always results in lower scatter of the data points
by factors of $\approx 1.3 - 1.5$ and best-fit lines with slopes within
$\approx 20\%$ of unity, and tends to increase the correlation significance
to similar levels as reported above for the solar metallicity \citet{BC03}
models with the \citet{Cal00} reddening law.  We note however that the
factor of $1/0.44$ for extra attenuation may not be appropriate for other
extinction laws because it was derived for a \citet{Cal00} reddening law.
The impact of metallicity is further addressed below.

With the assumption of $A_{V,\,{\rm neb}} = A_{V,\,{\rm SED}} / 0.44$,
the extinction-corrected H$\alpha$ luminosities and equivalent widths
are overall about 30\% higher than the model predictions.  This offset
is smaller than the scatter, but it is still possible that other sources
unrelated to the young massive ionizing stars make a moderate contribution
to the observed H$\alpha$ line emission (which would also cause some scatter).
However, none of the four galaxies with known AGN has any significant
excess in measured intrinsic properties compared to the predictions.
Because the sources with AGN do not stand as outliers in the distributions,
the quantitative results above are hardly changed when excluding them.

As alternative to an $A_{V}$-dependent scaling of the extra
attenuation towards the \ion{H}{2} regions, one could invoke
a constant amount of additional extinction for all sources.
For our SINS galaxies to move as an ensemble onto the one-to-one
relations in Figure~\ref{fig-LHaEWsVSpred}{\em a\/} and {\em c\/}
would require $\Delta A_{V} \sim 1~{\rm mag}$.  However, an additive
correction for extra attenuation does not alter the scatter of the
distributions, while a multiplicative correction does.
An $A_{V}$-dependent correction means that the global
differential extinction between the \ion{H}{2} regions and the
stellar populations depends on the averaged gas column density.
Given that this behaviour is observed in local star-forming 
galaxies and starbursts \citep[e.g.][]{Cal94, Cal00, Cid05},
it does not seem implausible that this may apply to high
redshift star-forming galaxies as well.

Obviously, the tightening of correlations should not be over-interpreted
and the quantities compared in Figure~\ref{fig-LHaEWsVSpred}{\em a\/},
{\em b\/}, and {\em d\/} are not strictly independent as the correction
applied to the H$\alpha$ measurements relies on the best-fit extinction
derived from the SED modeling.  This introduces some degree of artificial
correlation.  However, \whabb\ vs \whapred\ in panel {\em c\/} does not
have this drawback since no extinction correction is applied to the
measurements.  We note that a similar distribution in this diagram
is obtained if we use \whasinf\ instead, which then involves only
SINFONI measurements for both H$\alpha$ and the continuum.
The offset in observed H$\alpha$ equivalent width versus the model
predictions is therefore a robust result.

What other effects could lead to lower intrinsic H$\alpha$ luminosities
and equivalent widths compared to the predictions (or overestimated
predicted quantities)?  It is well known that metallicity influences
the H ionizing rate relative to the stellar rest-UV/optical photospheric
emission (e.g., \citealt{Pau01, Bri04, Lei99}).  Higher metallicities
would decrease the predicted quantities.  Assuming $Z = 2.5\,{\rm Z_{\odot}}$,
the \citet{BC03} models indicate this is an effect at the $\sim 20\% - 30\%$
level \citep[see also, e.g.,][]{Lei99} and, if anything, our $z \sim 2$
SINS galaxies are expected to have sub-solar abundances on average
\citep{Erb06a, Bus09}.  Lower metallicities would produce the opposite
effect, increasing further the mismatch between measured and predicted
quantities.  

If the \ion{H}{2} regions in our galaxies were density-bounded
and if Lyman continuum photons can escape the galaxies through paths
cleared by star formation-driven outflows (ubiquitous at high redshift;
e.g., \citealt{Pet01, Sha03, Sma03}), not all H ionizing photons from
the massive stars would lead to nebular H$\alpha$ emission, resulting
in lower values inferred from the measurements.
The fraction of ionizing radiation thus escaping is difficult
to constrain observationally.  Estimates for local and $z \sim 3$
star-forming galaxies suggest however $\sim 10\%$ or less
\citep*[e.g.,][and references therein]{Leh99, Ino05, Ber06, Sha06}.

Possibly the most efficient factor is an IMF biased against
high-mass stars, since H$\alpha$ is primarily sensitive to the
mass range $\ga 10~{\rm M_{\odot}}$ while the continuum and SEDs
probe the light from lower-mass, longer lived stars.
For instance, a lower upper-mass cutoff ($\rm \sim 30~M_{\odot}$
compared to our adopted $\rm 100~M_{\odot}$) or a significantly steeper
slope at the high-mass end (with power-law index $\alpha \approx -3$
in ${\rm d}N/{\rm d}m$, e.g., \citealt{Sca86}, instead of
$\alpha \approx -2.3$ for \citealt{Cha03}, \citealt{Kro01},
or \citealt{Sal55} IMFs) can reduce the H ionizing rates and H$\alpha$
equivalent widths by up to $\sim 1$ order of magnitude
\citep[e.g.][]{Ken94, Lei99}.
Detailed studies of massive young stellar clusters in the
Milky Way and neighbouring galaxies and of nearby starburst
systems are generally inconsistent with such forms of the IMF
\citep[and references therein]{FS03, Man07}.
Moreover, there is an increasing amount of theoretically and
observationally motivated suggestions that the IMF may evolve
with cosmic time and such as to be more weighted toward high
mass stars at high redshift
\citep[e.g.][]{Lar98, Lar05, McK07, Bau05, Dok08, Dave08, Che09}.
In this light, an IMF biased against high-mass stars does not
seem a likely explanation.

By this very sensitivity of H$\alpha$ and of the stellar continuum
and SEDs to different stellar mass ranges, the relations between 
measured and predicted intrinsic H$\alpha$ luminosities and
equivalent widths depend on the star formation history.
Our treatment of the star formation history in the SED modeling 
is very simplistic (because of the limited photometric data points 
for the SEDs of a significant fraction of our targets), assuming only
three cases and thus very sparse sampling of this parameter space.  
If our models have systematically overestimated the timescales, 
the predictions would be systematically higher than the measurements.
We examine this possibility in the next subsection.

\subsection{Constraints on the Star Formation Histories}
            \label{Sub-sfhs}

We focus on the H$\alpha$ equivalent widths, which provide a measure
of the current star formation rate as traced by H$\alpha$ relative to
an average over the galaxies' lifetimes as traced by stars dominating
the underlying continuum.
Figure~\ref{fig-EWsSEDprop} plots the \whabb\ and \whaoo\ as a function
of best-fit age and specific star formation rate from the SED modeling
for our SINS H$\alpha$ sample galaxies.  Model curves computed from the 
\citet{BC03} synthesis code as described in \S~\ref{Sub-intrinsic_prop} 
are shown with solid lines for different star formation histories: constant
star formation rate (CSF), and exponentially declining SFRs with $e$-folding 
timescales $\tau = 300$, 30, and 10~Myr (as representative cases).
The curves are plotted for ages of $\rm 10^{7} - 6 \times 10^{9}~yr$;
over this range, our calculations agree well with predictions
from the synthesis codes STARBURST99 \citep{Lei99, Vaz05} or STARS
\citep*{Ster98, Ster03, FS03, Dav07b} for similar IMF and solar metallicity
\footnote{Differences are small compared to the scatter of our data
and qualitatively of no consequence for the discussion presented here.}.

In Figure~\ref{fig-EWsSEDprop}{\em a\/}, the distribution of our SINS
galaxies occupies a large region of the diagram consistent with a
wide range of constant to declining star formation histories, and
that might suggest our SED modeling with three cases was indeed
too simplistic.  There are four galaxies that lie well above the
CSF model curve.  For each of them, the \whabb\ and \whasinf\ both
indicate consistently very large values.  These are in fact the four
sources with largest contribution from H$\alpha$ to the broad-band
magnitude ($\rm Q1623-BX599$, BX543, BX455, and BX502 with
$f_{\rm BB}({\rm H\alpha}) = 23\%$, 28\%, 36\%, and 57\%, respectively;
Table~\ref{tab-Hameas}).  Since we did not correct the SEDs for line
contamination in our modeling, this most likely drove the fits towards
older ages.  Indeed, the SED modeling by \citet{Erb06b}, based on the
same photometry, evolutionary synthesis code, and assumptions but
including correction for H$\alpha$ line emission, leads to much lower
ages for all four sources, as well as typically higher $A_{V}$ and SFRs
and lower $M_{\star}$ (these authors considered a wider range of star
formation histories but found a best-fit CSF, as we do).  Younger ages
would bring these sources in better agreement with the model CSF curve.

In Figure~\ref{fig-EWsSEDprop}{\em b\/}, for the case of extra
attenuation towards the \ion{H}{2} regions, the distribution of data
points tightens about the CSF model curve (albeit with significant
scatter, to which the various uncertainties in measurements and SED
modeling contribute).  To some extent, the shift in data points
between panels {\em a\/} and {\em b\/} reflects the well-known
degeneracies between age, extinction, and star formation history 
in SED modeling
\citep[see, e.g.,][for detailed discussions]{Pap01,Sha01,Sha05a,FS04,Erb06b}.

Figure~\ref{fig-EWsSEDprop}{\em c\/} and {\em d\/} show the same
but in terms of specific SFR.  To first order, we expect a tight
relationship irrespective of star formation history as H$\alpha$
measures the star formation rate through the ionizing rate of hot
stars and the continuum is dominated by the light from lower-mass
stars dominating the stellar mass.  One can use this behaviour to
discriminate between the effects of dust and star formation history
on our measurements. The model curves indeed run closely to each other
in the plots.  Moreover, none of the model curves (not even for an SSP
if we plot it) passes in the lower right part of the diagram occupied
by a significant fraction of our SINS galaxies when using the observed
\whabb, i.e. effectively assuming $A_{V,\,{\rm neb}} = A_{V,\,{\rm SED}}$.
Again, a super-solar metallicity, a high fraction of escaping ionizing
radiation, or an IMF biased against high mass stars do not provide
plausible explanations.  We note that a time-varying IMF (such as
one becoming more ``bottom-light'' at higher redshift; see, e.g.,
\citealt{Dok08, Dave08, Che09} and references therein) would tend
to shift the model curves along paths roughly parallel to the tracks
shown, and so would presumably not help.  The quantitative effects
would require detailed modeling, which is beyond the scope of this paper.

We therefore conclude in favour of differential dust extinction.
In Figure~\ref{fig-EWsSEDprop}{\em d\/}, the data become overall
more consistent with the models curves with
$A_{V,\,{\rm neb}} = A_{V,\,{\rm SED}}/0.44$.
The trends observed in Figure~\ref{fig-EWsSEDprop} are
qualitatively unchanged for the other SED modeling assumptions
we considered.
Assuming additional attenuation towards the \ion{H}{2} regions,
the results discussed in this section do not provide evidence for
an important decline in global SFRs for the ensemble of the SINS
galaxies over their past history (at least as measurable from the
diagnostics available).
A similar conclusion was reached by \citet{Erb06c} based on their
NIRSPEC sample of $z \sim 2$ \bxbm\ galaxies although with the
difference that they did not require extra attenuation towards
the \ion{H}{2} regions (if the aperture correction applied to
their long-slit data is not overestimated).

The analysis above can be recast in terms of the Scalo birthrate
parameter $b$, which measures the ratio of current SFR over the 
past-averaged SFR \citep[e.g.][]{Sca86, Ken94}.
Figure~\ref{fig-bMstar} shows the values of $b$ versus $M_{\star}$
for our SINS sample, where we took the current SFR as computed from
the extinction-corrected H$\alpha$ and the past-averaged SFR as the
ratio of stellar mass and age derived from the SED modeling.
Again, panels {\em a\/} and {\em b\/} compare the cases without
and with extra attenuation towards the \ion{H}{2} regions.
The median and mean $b$ parameter (excluding limits) of the SINS
galaxies is 0.4 and 0.8 for the former case, and 1.2 and 1.8 for
the latter.
\footnote{
Ranges in median values are, for other SED modeling assumptions,
$0.3 - 0.9$ and $0.9 - 1.3$ for the cases without and with extra
attenuation.}
In the local universe, normal spiral galaxies span a range of
$b < 0.1$ for early-type Sa/Sab to $\sim 1$ for late-type Sc/Sd
or irregular galaxies, while values $\sim 1 - 10$ are found in the
central regions of starburst systems \citep[e.g.][]{Gal84, Ken94, May04}.

Stochasticity is expected from the particular history of each object,
but on the whole, our SINS galaxies appear to have either undergone
a decrease by about half, or to have maintained roughly the same star 
formation activity level since the bulk of the stars observed
in them has been formed.
A few galaxies have $b$ parameters $5 - 10$ times higher than the average.
This includes the four galaxies noted above, with significant contribution
from H$\alpha$ to their $K$-band magnitude and hence with possibly
overestimated past-averaged $<{\rm SFR}> = M_{\star} / {\rm Age}$.
One of the known AGN also stands out ($\rm Q1623-BX663$).
Perhaps more surprisingly, two large massive disks with important
evolved stellar population inferred from their old best-fit ages
also have $b$ parameters much higher than the average
($\rm Q2343-BX389$ and $\rm Q2343-BX610$, at
$M_{\star} = 6.7 \times 10^{10}$ and $\rm 11.3 \times 10^{10}~M_{\odot}$,
and with $f_{\rm BB}({\rm H\alpha}) = 0.18$ and 0.11, respectively).
These systems may have experienced a recent episode of ``starburst''
activity triggered by enhanced gas accretion, possibly through
cold flows or minor mergers, and/or the onset of instabilities
in a fragmenting gas-rich disk, as we argued in
\citet[][see also, e.g., \citealt{Bour07, Genel08, Dek09a, Dek09b}]{Gen08}.
Interestingly, optically-selected \bxbm\ galaxies among our sample
appear distinct from the near-/mid-IR-selected galaxies, with median
$b = 0.9$ to $1.6$ depending on the extinction correction applied,
compared to median $0.2$ to $0.8$ for all other sources.

\subsection{Star Formation Rate Estimates}
            \label{Sub-sfrs}

The star formation rates derived from our H$\alpha$
luminosities through equation~(\ref{Eq-LHaSFR}) and corrected
for $A_{V,\,{\rm neb}} = A_{V,\,{\rm SED}}$ have median and mean
\sfrha\ of 32 and $\rm 46~M_{\odot}\,yr^{-1}$ (excluding limits) 
and range from $< 1.6$ to $\rm 213~M_{\odot}\,yr^{-1}$.
When applying extra attenuation towards the \ion{H}{2} regions
with $A_{V,\,{\rm neb}} = A_{V,\,{\rm SED}} / 0.44$, the median and
mean \sfrhaoo\ are 85 and $\rm 182~M_{\odot}\,yr^{-1}$, and range
up to $\rm 1500~M_{\odot}\,yr^{-1}$.
As is directly implied by Figure~\ref{fig-LHaEWsVSpred}{\em a\/} and
{\em b\/}, the estimates without the extra attenuation are overall
a factor of $\sim 2$ lower than those from the SED modeling
(median and mean of 72 and $\rm 141~M_{\odot}\,yr^{-1}$ and range
of $\rm 0.7 - 809~M_{\odot}\,yr^{-1}$) while those with the extra
attenuation are in better agreement, in both the ensemble, being
overall $\approx 30\%$ higher, as well as individually with about
$1.5 \times$ lower scatter about the relationship.
\footnote{
The median and ranges vary for the other SED modeling assumptions
by up to factors of $\approx 3$ for \sfrsed\ as well as \sfrha\ and
\sfrhaoo\ (because the extinction correction is based on the best-fit
$A_{V}$ from the SED modeling) but in all cases, the \sfrha's are
overall significantly lower than the \sfrsed's (by $\approx 30\%$ to
a factor of $\sim 4$) while applying the extra attenuation leads to
\sfrhaoo's in better agreement with \sfrsed's (to $\leq 30\%$) with
$1.3 - 1.5$ times lower scatter in the relationship.}
The highest SFR estimates are for $\rm K20-ID5$, one of the known AGN
for which our SINFONI data (including line ratios) as well as the
broad-band SED indicates clear contributions from non-stellar
emission, driving the intrinsic SFRs to large values.

With our H$\alpha$-derived SFRs, we briefly look at the resulting
$M_{\star} - {\rm SFR}$ relation in Figure~\ref{fig-SFRallMstar}.
Panels {\em a\/} and {\em b\/} show the relations for the two cases
of extinction, and panels {\em c\/} and {\em d\/} show those obtained
from \sfruv\ and \sfrsed.  We find quite good agreement between the 
relations using \sfrhaoo, \sfruv, and \sfrsed.  As expected, the relation
with \sfrha\ is offset by $\rm \approx 0.3~dex$ to lower SFRs, and appears
to be somewhat flatter.  The scatter in our relations ranges from 0.38~dex
with \sfrha\ to 0.47~dex with \sfrhaoo, and 0.6~dex for the others.
\footnote{
Changes in our SED modeling assumptions affect the zero point and scatter
of the $M_{\star} - {\rm SFR}$ relationship by factors of $\sim 2 - 3$;
the resulting best-fit slopes have a power-law index consistent with unity
to $\rm \pm 0.25~dex$.
}
As reference, we also overplot the slope and rms scatter from
\citet{Dad07} but we caution that a direct and detailed comparison
with our results should not be overinterpreted, as both the stellar masses
and the SFRs are derived differently; there might be complex systematics
that affect the slope, for instance, and our SINS H$\alpha$ sample is
not sufficiently large to reliably investigate such effects.
All these relations obviously apply for actively star-forming galaxies;
passive systems or those with rapidly declining star formation rate
would lie below the locus of actively star-forming galaxies.
This seems to be the case for $\rm SA12-5836$, which has the lowest
\sfrsed\ and \sfruv, and was not detected in our H$\alpha$ observations.
The colours of this target in fact do not satisfy the $sBzK$ criterion
and its spectral features from the GDDS optical spectroscopy are
indicative of intermediate-age to old stellar populations \citep{Abr04}.

The $M_{\star} - {\rm SFR}$ relation and its evolution with cosmic time
has been the focus of several recent studies at high and low redshift.
In particular, there appears to be a significant discrepancy between
the empirical relation derived from various indicators and that 
derived from semi-analytical and hydrodynamical cosmological 
simulations of galaxy formation.  At $z \sim 2$, the empirical
relation lies a factor of several above that from simulations
at the high-mass end $M_{\star} \ga 10^{10}~{\rm M_{\odot}}$
\citep[e.g.][]{Dad07, Dave08, Dam09}.  This persists with the 
H$\alpha$-derived SFRs for our SINS sample, even without extra 
attenuation towards the \ion{H}{2} regions.
The generally low scatter of the empirical relations has been
interpreted as indicative of smoothly and slowly varying or
roughly constant SFRs.
The overall consistency between the $M_{\star} - {\rm SFR}$ relations
from various indicators, sensitive to different stellar populations
and thus different epochs in the star formation history of the
galaxies, further supports this interpretation.  Our H$\alpha$
data from SINS simply add to the previous evidence.

SFRs derived from observations suffer from potentially important
uncertainties, as do those derived from theoretical models and
numerical simulations of galaxy formation.  These have been
extensively discussed elsewhere
\citep[e.g.,][among many others]
  {Bri04, Erb06c, Red06, Pap07, Dad07, Dave08, Che09, Cal09}.
Some of them obviously apply to our estimates as well.  However, a
clear strength of our SINS H$\alpha$ sample is that, for the first time,
we have reliable measurements of the {\em total} H$\alpha$ fluxes for a 
large sample of $\sim 60$ star-forming galaxies at $z \sim 2$, providing
a robust basis for comparisons and future investigations.

\vspace{1ex}
\section{SPATIALLY-RESOLVED H$\alpha$ KINEMATICS AND KINEMATIC DIVERSITY}
         \label{Sect-kinematics}

Detailed analysis of the H$\alpha$ velocity-integrated flux maps and of the
kinematics have been presented for various subsets of the SINS H$\alpha$
sample in other papers \citep{FS06a, Gen06, Bou07, Sha08, Cre09}.
In the following, we build on the findings reported in our published
quantitative studies of the kinematics and on the results presented in the
previous Sections of this paper.  We focus on a general overview of the
ensemble properties, based specifically on the H$\alpha$ kinematics and
spatial distributions.

The kinematic diversity among our SINS H$\alpha$ sample is illustrated
in Figure~\ref{fig-vfs}.  The figure shows, all on the same angular scale,
the H$\alpha$ velocity fields for 30 of the \nsinsdet\ detected objects.
The galaxies are approximately sorted from top to bottom according
to whether their kinematics are disk-like or merger-like, and from
left to right according to whether they are ``rotation-dominated'' or
``dispersion-dominated.''  About $\approx 30\%$ of the detected objects
(or half of the galaxies in Figure~\ref{fig-vfs}) can be classified through
quantitative methods.  For the remaining galaxies (with lower S/N and/or 
fewer resolution elements across the systems), we followed a qualitative
approach or used alternative and more approximate diagnostics.  For five
of those systems, the S/N per resolution element is still too low to
extract spatially-resolved kinematic information, and so are excluded
in this Section.
\footnote{
These are $\rm D3a-7429$, $\rm GMASS-2454$, $\rm GMASS-2550$,
$\rm ZC-772759$, and $\rm SA12-8768NW$.}
The criteria we applied are described in \S~\ref{Sub-kinclass};
each kinematic class is further discussed in \S~\ref{Sub-disks},
\S~\ref{Sub-compact}, and \S~\ref{Sub-mergers}.

Our disk-/merger-like classification relies on the {\em gas kinematics}
of the galaxies, specifically on the degree of (a)symmetry in the H$\alpha$
velocity fields and velocity dispersion maps as explained below.
Given the frequently strongly clumpy and asymmetric spatial distribution
of the light (in H$\alpha$ or stellar continuum) and the complications
from $K$ corrections, classical morphological classification schemes
may not be reliable for our $z \sim 2$ galaxies
\citep*[e.g.,][]{Lot04, Law07a, Pet07, Elm07, Con08, Lot08}.
Notwithstanding, inspection of high resolution broad-band optical
and/or near-IR imaging indicates that the kinematically identified
(major) mergers also clearly show evidence for merging in their
morphology \citep[see also][]{Sha08, FS09b}.

\subsection{Kinematic classification}
           \label{Sub-kinclass}

The distinction between disk- and merger-like kinematics can
be made quantitatively from application of our kinemetric analysis
described by \citet{Sha08}.  This is possible for 15 of the
best-resolved sources with highest quality data; these galaxies are
marked as red and green symbols in most plots of Figures~\ref{fig-Kz} 
to \ref{fig-fbar}.
Our method is adapted from the original technique developed by the SAURON
team for analysis of local galaxies \citep{Kra06} to applications for high
redshift studies.  It provides a measure of the degree of asymmetry in the
observed velocity and velocity dispersion maps, where the lower (higher)
the asymmetry, the more disk-like (merger-like) the object.  
Of the first eleven SINS galaxies classified by kinemetry, eight are
disks and three are mergers \citep[see][]{Sha08}.  This initial set has
now been expanded to include the analysis of four additional sources,
two of which are classified as disk-like and two as merger-like.
The kinemetric classification is reported in Table~\ref{tab-dyn}.
The resulting fractions of disk- and merger-like systems is thus $2/3$ 
and $1/3$, respectively.  The uncertainties of our method are discussed
by \citet{Sha08}, to which we refer for details.  Based on these, we 
expect to correctly classify $\sim 89\%$ of disks and $\sim 80\%$ of
mergers, implying that $\sim 1$ of the ten disks may be misclassified
as merger, and $\sim 1$ of the five mergers may be misclassified as disk.

For the more compact objects or for data sets with lower S/N, kinemetry
is too uncertain or impossible.  In those cases, we sorted the galaxies
based on a qualitative assessment of the asymmetry in the velocity field
and dispersion map (essentially, the same criteria as for our
quantitative kinemetry).  We find in this way similar fractions of
$\sim 2/3$ of the objects that appear to have H$\alpha$ kinematics
consistent with rotation in a single disk, and $\sim 1/3$ with
asymmetric or irregular H$\alpha$ kinematics suggestive of a merger.
We note that for the 15 objects classified quantitatively, our
kinemetry confirmed in all cases our prior qualitative assessment
\citep[see][]{FS06a, Gen06, Gen08, Sha08}.
As noted in \S~\ref{Sect-samples}, the SINS H$\alpha$ sample
includes three pairs of galaxies at approximately the same redshift and
with projected separations of $\rm \approx 15 - 30~kpc$.  The individual
components can in principle be counted and inspected separately (see
\S~\ref{Sub-mergers}) or taken as three merging systems, but this
has little consequences on our overall classification.

Another important characteristic of galaxies is the amount of dynamical
support provided by rotational/orbital motions and by turbulent/random 
motions.  Ideally, the distinction between ``rotation-dominated'' and
``dispersion-dominated'' kinematics would rely on detailed and accurate
dynamical modeling, from which the ratio of circular/orbital velocity
$v_{\rm rot}$ to intrinsic local velocity dispersion $\sigma_{0}$
is derived.  We note that this $\sigma_{0}$ is different from the
source-integrated velocity dispersion $\sigma_{\rm int}$ discussed
so far in this paper.  The $\sigma_{0}$ is a measure of the intrinsic
local random motions of the gas free from contributions from large-scale
velocity gradients, providing dynamical support and related to the
geometrical thickness of rotating disks
\citep{FS06a, Gen08, Sta08, Cre09, Wri09}.
Reliable determination of $v_{\rm rot}/\sigma_{0}$ is
possible for 14 galaxies among our SINS H$\alpha$ sample
(for details about the modeling and the uncertainties, see \citealt{Gen08}
and \citealt{Cre09}; results are given in Table~\ref{tab-dyn}).
Adopting $v_{\rm rot}/\sigma_{0} \sim 1$ as a boundary, we find that
13 sources are rotation-dominated and 1 is dispersion-dominated.
\footnote{
This compares reasonably well with \citet{Epi09}, who determined
the $v_{\rm rot}/\sigma_{0}$ in a similar way for their SINFONI
sample at $1.2 \la z \la 1.6$, and for which 2 of their 9 sources
have a ratio $< 1$.}
Taking into account the uncertainties on the ratios, two of the
rotation-dominated sources as well as the dispersion-dominated
object are within $1\,\sigma$ of the boundary.
This sub-sample with $v_{\rm rot}/\sigma_{0}$ determinations is likely
to be biased towards rotation-dominated systems although, as we note in
\S~\ref{Sub-disks}, the inferred values for our SINS galaxies are
still significantly lower than for present-day spirals.
We preferentially modeled disk-like systems and the $v_{\rm rot}/\sigma_{0}$ 
ratio can be most robustly determined for the larger and brighter ones that
are well-sampled out to large radii, where the rotation curve flattens and
the intrinsic local velocity dispersion is best constrained.

To allow a more general analysis of all of our SINS H$\alpha$ sample
galaxies, we defined a working criterion involving the full observed
velocity gradient $v_{\rm obs}$ (uncorrected for inclination) and the
integrated line width $\sigma_{\rm int}$ as follows.
Based on simulations of disk galaxies with various ratios of
intrinsic circular velocity to local velocity dispersion and a
range of sizes and dynamical masses appropriate for our sample,
the cross-over between rotation- and dispersion-dominated systems
at $v_{\rm rot}/\sigma_{0} \sim 1$ occurs around a ratio of 
$v_{\rm obs} / (2\,\sigma_{\rm int}) \sim 0.4$.
This is the case for the typical spatial resolution achieved with both
AO-assisted as well as seeing-limited observations, with the exception
of very compact sources in seeing-limited data where the small observed
gradients could still be consistent with a rotation-dominated system.
We emphasize that $v_{\rm rot}/\sigma_{0}$ corresponds to an
intrinsic and inclination-corrected property of disks, whereas
$v_{\rm obs} / (2\,\sigma_{\rm int})$ is an observed quantity,
with $v_{\rm obs}$ taken as the maximum projected velocity difference
$v_{\rm max} - v_{\rm min}$ measured from the observed velocity field.

Of the 47 galaxies with sufficient S/N for measuring $v_{\rm obs}$,
14 have $v_{\rm obs} / (2\,\sigma_{\rm int}) < 0.4$ and 33 have
$v_{\rm obs} / (2\,\sigma_{\rm int}) > 0.4$, thus implying that $\sim 1/3$
of the sources are dispersion-dominated systems.  Comparing with the
quantitatively classified systems, all 13 rotation-dominated sources satisfy
$v_{\rm obs} / (2\,\sigma_{\rm int}) > 0.4$ (see Table~\ref{tab-dyn});
the dispersion-dominated source also does but is just $1\,\sigma$ away
from the boundary in both ratios.  Although conceptually devised for
disks, this classification can also be indicative for mergers where
the $v_{\rm rot}$ term then represents the orbital velocity of the
system.  Obviously, the $v_{\rm obs} / (2\,\sigma_{\rm int})$ ratio
is an approximate diagnostic because of its sensitivity to the
inclination of the systems and to the contribution of large-scale
velocity gradients to the integrated line width.  It nevertheless
provides a useful (if approximate) probe of the nature of the
dynamical support in the cases where the data quality prevents
reliable detailed kinematic modeling.

Altogether, combining the above criteria based on the H$\alpha$
kinematics, the SINS H$\alpha$ sample includes $\sim 1/3$ of
clearly identified disk-like galaxies, $\sim 1/3$ of clearly
identified mergers or interacting systems, and $\sim 1/3$
of sources with typically more compact morphologies and kinematics
that appear to be dominated by velocity dispersion as compared to their
velocity gradients.  As we discuss in the following sub-sections, the
proportion of disk-like systems tends to increase at higher masses while
dispersion-dominated systems appear more ubiquitous at lower masses.

The overall classification is unlikely to be significantly
affected by the 10 non-detected sources in our sample.  The disks and
mergers classified quantitatively by kinemetry do not differentiate in
global photometric, stellar, and in integrated H$\alpha$ properties
(see \S~\ref{Sect-representativeness} and \S~\ref{Sub-lineprop}).
Dispersion-dominated objects may possibly be more ubiquitous among
lower-mass galaxies (see \S~\ref{Sub-compact} and \S~\ref{Sub-dynstate}).
However, since the non-detections show no trend with photometric
and stellar properties (\S~\ref{Sect-representativeness}) we do
not expect that they would be biased towards one class or the other.
Because of our observing strategy and sensitivity limits
(\S~\ref{Sub-detlim}), we may be missing more extended sources with
lower averaged H$\alpha$ surface brightnesses, but these could be either
disk- or merger-like systems.  We verified and conclude similarly for the
five galaxies further excluded in the discussion of kinematics because of
too low S/N per resolution element.  Therefore, there is no evidence that
the classification of our SINS H$\alpha$ sample should be biased by the
relatively small fraction ($\approx 25\%$) of undetected and unclassifiable
targets.

\subsection{Rotation-dominated High Redshift Galaxies}
           \label{Sub-disks}

As found in our previous studies of various subsets of the SINS
sample, a majority of those sources exhibit compelling kinematic 
signatures of ordered rotation in a disk-like configuration, including
a smooth and monotonic velocity gradient (in the best cases showing the
classical ``spider-diagram'' of pure disk rotation), alignment of the
morphological and kinematic major axes (see also \S~\ref{Sub-profiles}),
and, in several cases, a global peak in the velocity dispersion map
close to the morphological/kinematic center.
These have been discussed extensively by \citet{FS06a,Gen06,Gen08,Sha08}
and \citet{Cre09} to which we refer for details of individual cases.

Interestingly, even in the largest and most regular massive
disks ($\rm 15 - 20~kpc$ across with rotation velocities of
$v_{\rm rot} \sim 200 - 300~{\rm km\,s^{-1}}$), the inferred
intrinsic local velocity dispersion is quite substantial, with
$\sigma_{0} \sim 30 - 90~{\rm km\,s^{-1}}$.  This suggests the
gas disks have large amounts of random motions/turbulence and
should accordingly be fairly thick.
For the disks where we can carry out reliable dynamical modeling,
we infer $v_{\rm rot}/\sigma_{0} \sim 1 - 7$, with median and mean
of $\approx 4.5$ \citep{Gen08, Cre09}, lower than typical values for
local (late-type) spiral galaxies \citep*[$\sim 10 - 20$; e.g.][]{Dib06}.
Our dynamical modeling accounts for the spatial and spectral resolution
of the data, and so the high inferred levels of intrinsic local velocity
dispersion are not caused by beam-smearing of the inner rotation curve
or of a central unresolved source with broad line width.
The dispersion-dominated systems would have still lower inferred
$v_{\rm rot}/\sigma_{0} \la 1$ ratios, assuming disk dynamics.
However, they are also typically very compact and less well resolved
spatially, so that the final verdict is still out as to what their
intrinsic $v_{\rm rot}/\sigma_{0}$ is.
The large turbulence appears to be a key property of many $z \sim 1 - 3$
star-forming disk-like systems, as inferred also by other groups based on
kinematics \citep{Kas07, Law07b, Law09, Wri07, Wri09, Bour08, Sta08, Epi09}
or indirectly from the determination of large $z$ scale heights of the
stellar light emission \citep[e.g.,][]{Elm05a, Elm06, Elm05b, Elm07}.

Evidently, these high redshift disks are dynamically different from
present-day disks.  In view of the different conditions prevailing at
high redshift, this may not be surprising.  The origin of the inferred
high gas-phase turbulence is still uncertain, but plausible causes include
feedback from intense star formation, heating from the conversion of the
gravitational energy as gas from the halo is accreted onto the galaxies
at high rates, and stirring due to internal dynamical processes
\citep*[e.g.,][]{Aba03, Tho05, FS06a, Gen08, Kho09, Bur09}.
Interestingly, deviations on kpc-scales from pure rotation are seen
in several of the large disks that we observed at higher resolution
with AO, while on large scales the kinematics are consistent with 
disk rotation (e.g., $\rm Q2346-BX482$, $\rm Deep3a-15504$, and
$\rm ZC-782941$; \citealt{Gen06, Gen08}).
These small-scale perturbations could be produced by the presence and 
mutual interactions of the observed luminous/massive star-forming clumps
(as seen in numerical simulations by, e.g., \citealt{Imm04a, Imm04b, Bour08}),
or the proximity of small satellites.

\subsection{Dispersion-dominated High Redshift Galaxies}
            \label{Sub-compact}

The dispersion-dominated objects tend to be the more compact
sources among our SINS sample.  They also tend to have lower
dynamical masses and angular momenta than the rotation-dominated
systems (see \S~\ref{Sub-dynstate}).  In a significant number of
them, we detect velocity gradients that are suggestive of orbital
motions in a disk or a close merger, although the observed amplitude 
is typically much smaller than for the larger massive disks.
The most compact of those sources have $\rm FWHM(H\alpha) \la 4~kpc$,
and so are marginally resolved spatially in our seeing-limited SINFONI
data.

Because of the significant beam smearing effects in small systems,
some fraction could be lower-mass disk-like galaxies with intrinsically
smaller sizes and circular velocities and thus largely unresolved
rotation contributing to the observed velocity dispersion.
Alternatively, some could be nearly face-on disks, where surface 
brightness limitations prevent detection of the emission at larger radii
with our typical integration times.  If so, the limiting sensitivities
of our data sets derived in \S~\ref{Sub-detlim} suggest the surface
brightness of the outer parts would need to be $\ga 10 - 20$ times
fainter than the central detected parts.  Other possibilities include
simply largely unresolved systems whose kinematics are dominated by
random/non-circular motions, late-stage mergers/merger remnants, or
very young systems undergoing their first phases of intense gas
accretion and conversion into stars.
Interestingly, there are two groups among this dispersion-dominated
population in terms of the stellar ages, [\ion{N}{2}]/H$\alpha$ ratio,
and H$\alpha$ equivalent width, suggesting part of them is already fairly
evolved at $z \sim 2$ while others seem to be extremely young systems
(see also \S~\ref{Sub-dynstate}).
The latter may be closely connected with the young and highly gas-rich
objects discussed by \citet[][see also \citealt{Law09}]{Erb06b}.

Several dispersion-dominated systems have been observed
at $\sim 0\farcs 1 - 0\farcs 2$ resolution with Keck/OSIRIS
and AO (\citealt{Law07b, Law09, Wri09}).
In particular, the twelve $z \sim 2 - 2.5$ BX-selected galaxies studied
by \citet{Law09} appear to be mostly comprised of such objects, with at
most five disk-like objects and three resolved multi-component mergers
according to the classification by these authors, and all but five 
satisfying our $v_{\rm obs}/(2\,\sigma_{\rm int}) < 0.4$ criterion.
As we saw in \S~\ref{Sub-sinf_ifus}, an important difference between
the \citet{Law09} sample and our SINS H$\alpha$ sample with \nsinsdet\
detected sources is in terms of intrinsic sizes, with significantly
lower half-light radii for the former; the stellar mass distributions also
indicate the \citet{Law09} sample emphasizes a somewhat lower mass range.
Here we further see kinematic differences, with our sample having
$\sim 1/3$ of dispersion-dominated objects whereas the fraction is
$\sim 60\%$ for the \citet{Law09} sample.
Several factors may play a role in these differences,
from intrinsic properties of the populations probed by the samples
(primary colour and magnitude criteria, stellar mass ranges) to
selection biases and limiting surface brightnesses.
Clearly, larger samples at the highest spatial resolution possible
are needed to better assess the fraction of dispersion-dominated
systems at $z \sim 2$ and their nature.

\subsection{Merger/Interacting High Redshift Systems}
            \label{Sub-mergers}

Our SINS sample also includes a variety of merging and interacting
systems, ranging from well separated galaxies in early stages of
interaction (e.g., the pairs $\rm Q2346-BX404/405$, $\rm GMASS-2113E/W$,
$\rm SA12-8768/8768NW$) to what looks like single systems in our data
but with asymmetric/disturbed kinematics, presumably from later-stage
mergers (e.g., $\rm Q1623-BX528$, $\rm K20-ID7$, $\rm Deep3a-12556$).
These represent roughly $\sim 1/3$ of our SINS sample.
Examination of the well-separated interacting pairs shows a range
of kinematics for the individual components
(see Figures~\ref{fig-vfs}, \ref{fig-maps_bmbx4}, \ref{fig-maps_gmass1},
and \ref{fig-maps_gdds2}).  $\rm Q2346-BX404$ and $\rm BX405$ are both
dispersion-dominated ($v_{\rm obs} / (2\,\sigma_{\rm int}) < 0.4$) but
show kinematic features consistent with disk rotation.  The pair
$\rm GMASS-2113E/W$ appears more dispersion-dominated and with
irregular kinematics although lower S/N data makes the assessment
more uncertain.  $\rm SA12-8768$ is a rotation-dominated 
($v_{\rm obs} / (2\,\sigma_{\rm int}) > 0.4$) disk-like source;
the faint north-western companion has too low S/N to be classified.
We cannot rule out that some of the more compact objects
(or sources with poorer resolution and/or S/N data) are also
mergers, and results from the highest resolution data available to
date indeed suggests $\sim 50\%$ or less are consistent with being
mergers, although the samples are still small
(\citealt{Law07b, Law09, Wri09}).
Again, much larger samples will be needed to assess robustly the
fraction of mergers and other types among these compact, lower-mass
populations, for instance from kinemetry analysis.

It is interesting to note that these merging/interacting systems in
our SINS sample do not appear to differentiate in their integrated
H$\alpha$ or stellar properties (\S~\ref{Sect-representativeness},
\S~\ref{Sub-lineprop}), only in their kinematics.
We are however likely missing the more extreme major mergers
in their most luminous/intensely starbursting phases, which
are more frequent among bright submillimeter-selected samples
\citep[e.g.][]{Tac06, Tac08, Swi06, Bou07}.
These are rarer, but most importantly, more highly dust-obscured,
making studies in the near-IR of their rest-frame optical properties
more difficult.

More generally, with our criterion (\S~\ref{Sub-kinclass}),
with the typical effective field of view of our SINFONI data, and
the fact that we focus on the H$\alpha$ line emission, we are primarily
sensitive to merger stages in which the progenitors have projected
separations of $\rm \la 15 - 20~kpc$
(the central deeper part of the seeing-limited SINFONI data obtained for
all but one targets is $\sim 4^{\prime\prime} - 5^{\prime\prime}$ across, see
\S~\ref{Sect-obsred}), sufficiently elevated star formation rates to
be detected in our data (typically $\rm SFRs \ga 10~M_{\odot}\,yr^{-1}$;
e.g., \S~\ref{Sect-Ha_prop} and Figure~\ref{fig-SFRallMstar}),
and sufficiently perturbed and asymmetric gas kinematics on scales of
$\rm \sim 1 - 5~kpc$.
Based on local interacting/merging systems (e.g., as ubiquitous among 
Ultra-Luminous Infrared Galaxies) as well as on numerical simulations, 
such phases occur on fairly short timescales of a few $\rm \sim 100~Myr$
or less \citep[e.g.,][]{Mih94a, Mih94b, Mih96, San96, Lot08}.
Therefore, we would not expect to find many mergers in these stages
among our SINS H$\alpha$ sample.
We did detect serendipitous star-forming companions at projected distances
of $\rm 15 - 30~kpc$ and within $\rm 100~km\,s^{-1}$ along the line-of-sight
in only two cases among our 60 original targets
($\rm GMASS-2113W$ and $\rm SA12-8768NW$; \S~\ref{Sect-samples}).
It is still possible that we are missing mergers during more quiescent
phases, or for which the companions have too low star formation or are
too obscured to be detected in our H$\alpha$ observations.  It is also
possible that we are missing companions to our main targets at projected
radii $\rm \ga 15 - 20~kpc$, or that would have H$\alpha$ line widths
narrower than $\rm \sim 100~km\,s^{-1}$ (the effective spectral resolution
of our data; \S~\ref{Sect-obsred}) and observed wavelengths coinciding
exactly with strong night sky line residuals.
A detailed assessment would involve complex considerations about mergers
\citep[e.g.,][]{Con08, Lot08} and is well beyond the purpose of this paper,
where we are interested in the evolutionary and dynamical state of the
primary targets of our SINS sample.

\subsection{Dynamical versus Evolutionary State}
            \label{Sub-dynstate}

We explore here whether the dynamical state of our SINS H$\alpha$
sample galaxies can be related to other properties indicative of
their global evolutionary state.  For this purpose, we complement
our SINS sample with the sample studied with OSIRIS by \citet{Law09},
which, as noted above, appears to be distinct in several properties and
thus may probe different evolutionary stages or a different population.
\citet{Law09} discuss exhaustively the differences between their
sample and the SINS galaxies studied in our earlier publications.
These differences remain for the subset for which we have carried out
detailed kinematic modeling and kinemetry but we note that the full SINS
sample presented in this paper extends to lower masses and fainter $K$
band magnitudes, with median $M_{\star} = 3 \times 10^{10}~{\rm M_{\odot}}$
and $K_{\rm s, Vega} = 20.0~{\rm mag}$ or only a factor of two higher and
0.5~mag brighter, respectively, than for the \citeauthor{Law09} sample.
Beyond these differences, which may be driven in part by target selection
and by observational factors, the two samples are complementary in the
following sense.
As we have seen in \S~\ref{Sub-sinf_ifus}, the \citet{Law09} galaxies
have smaller sizes for comparable H$\alpha$ fluxes and luminosities, and
consequently have higher inferred surface brightnesses and star formation 
rates per unit area.  They tend to lie at lower stellar masses compared
to the ensemble of the SINS galaxies (although there is significant
overlap) and show a smaller proportion of rotation-dominated objects,
with overall smaller observed velocity gradients.

For the source in common with \citet{Law09} for which we also
obtained AO-assisted SINFONI observations with resolution FWHM
$\approx 0\farcs 2$, $\rm Q1623-BX502$, the agreement in kinematic
and morphological properties is excellent.  The projected velocity
gradients, half-light radii, and integrated velocity dispersions
are all essentially identical.
The other two sources detected by \citet{Law09} with OSIRIS that
are among our SINS sample were observed in seeing-limited mode
with SINFONI.  For $\rm Q2343-BX513$, the SINFONI kinematics
indicate only a small velocity gradient and are moreover affected
by night sky line residuals on the red side of the H$\alpha$ line,
but the same integrated velocity dispersion is inferred.
Our derived intrinsic half-light radius is nearly twice larger
but uncertain because, for lack of a PSF calibration star for this
data set, we assumed the average seeing of the SINS observations;
however, our H$\alpha$ flux is also about twice higher, suggesting
our SINFONI data may have detected more of the fainter emission
at larger radii.
For $\rm Q1623-BX543$, our observations were taken under strongly
variable seeing conditions and the southern merger component
is not seen in our data (projected distance of $0\farcs 8$).

\subsubsection{Velocity-size relation}
               \label{Sub-velsize}

In view of the size differences and the existence of a
velocity-size relation at $z \sim 2$ \citep{Bou07}, we show in
Figure~\ref{fig-velsize} the SINS and \citet{Law09} galaxies in
the $v_{\rm d}$ versus $r_{1/2}({\rm H\alpha})$ plane.  Here, the
relevant velocity estimate that we denote $v_{\rm d}$ should provide
a measure of the gravitational potential, which we derived using one
of three methods as follows. \\
{\em (1)\/} ``Kinematic modeling'':
for the SINS disk galaxies with kinematic modeling \citep{Gen08, Cre09}, 
we used the circular velocity from the intrinsic, inclination-corrected
rotation curve of the best-fitting model disk.
\\
{\em (2)\/} ``Velocity gradient $+$ width'':
for the SINS galaxies without modeling but with rotation-dominated
kinematics ($v_{\rm obs}/(2\,\sigma_{\rm int}) > 0.4$), we followed
the method described by \citet{FS06a} and computed $v_{\rm d}$ as
the average of the estimate based on the observed H$\alpha$ velocity
gradient, $v_{\rm d}^{\rm vgrad}\sin(i) = 1.3\,v_{\rm obs}({\rm H\alpha})$,
and that based on the integrated H$\alpha$ velocity dispersion, 
$v_{\rm d}^{\rm width}\sin(i) = 0.99\,\sigma_{\rm int}({\rm H\alpha})$.
These relations were obtained from simple disk models with a range
of beam-smearing, sizes, and local isotropic velocity dispersions
appropriate for our SINS galaxies.  We accounted for inclination $i$
using the intrinsic H$\alpha$ morphological axis ratio of each galaxy
whenever possible, otherwise we used the average $<\sin(i)> = \pi / 4$.
\\
{\em (3)\/} ``Velocity width'':
For the SINS systems with dispersion-dominated kinematics
($v_{\rm obs}/(2\,\sigma_{\rm int}) < 0.4$), a virial
approach is more appropriate and we adopted
$v_{\rm d} = \sqrt 3\,\sigma_{\rm int}({\rm H\alpha})$,
where the scaling factor is a representative average for a
variety of realistic three-dimensional isotropic galactic mass
distributions \citep[][]{Bin08}.

The method and resulting value for each galaxy are listed in
Table~\ref{tab-dyn}.
As noted in \S~\ref{Sub-lineprop}, the $\rm H\alpha$ kinematics
of $\rm K20-ID5$ appear importantly affected by the AGN (or shocks),
and we treated its inferred $v_{\rm d}$ as upper limit.
For the \citet{Law09} sample, we applied methods 2 or 3 above
depending on the ratio $v_{\rm obs}/(2\,\sigma_{\rm int})$ (using
the $v_{\rm shear}$ and $\sigma_{\rm net}$ given by these authors),
with an average inclination correction $<\sin(i)> = \pi / 4$ for
all sources.

The galaxies from both the SINS and \citet{Law09} samples follow a
fairly well defined velocity-size relation in Figure~\ref{fig-velsize},
as found previously by \citet{Bou07} with a subset of the SINS galaxies.
A few of the additional galaxies here lie below the relation towards
somewhat higher $v_{\rm d}$ and correspond to lower angular momentum.
Interestingly, the clear merger systems identified by
our kinemetry overlap with the distribution of disks.  Perhaps these
are at earlier stages of merging, before significant loss of angular
momentum occurs in the late merger stages, as more frequently seen
among the luminous dust-rich SMG population \citep{Bou07, Tac08}.
The sample of \citet{Law09}, comprising mostly dispersion-dominated
objects, lies at the low $v_{\rm d}$ -- low size end, suggesting that
these objects may be drawn from the part of the population with lower
angular momentum compared to the ensemble of our SINS galaxies, and
especially the larger and more massive rotating disks.

\subsubsection{Large turbulent velocities}
               \label{Sub-veldisp}

As emphasized above, a key feature of $z \sim 2$ star-forming
galaxies is their large inferred amounts of local random motions.
This is seen not only in the dispersion-dominated objects but also
in the large rotating disks. One of the possible causes for the high
intrinsic local velocity dispersions that we can directly test with
the data available is the effects of feedback from star formation
through supernova explosions, massive stars winds, and radiation
pressure \citep[e.g.][]{Tho05}.  In this case, one would expect
a decrease of $v_{\rm obs}/(2\,\sigma_{\rm int})$ (or of intrinsic
$v_{\rm rot}/\sigma_{0}$) at higher star formation rate surface
densities \citep[e.g.][]{Gen08}. 

In Figure~\ref{fig-vcsig0}{\em a\/}, we plot the observed
$v_{\rm obs}/(2\,\sigma_{\rm int})$ ratio as a function of star
formation rate surface density $\Sigma [{\rm SFR^{00}(H\alpha)}]$,
calculated from the H$\alpha$-derived star formation rates and
half-light radii (Tables~\ref{tab-Hameas} and \ref{tab-SFRs} for
the SINS galaxies) and for the case of extra attenuation towards
the \ion{H}{2} regions with $A_{V,\,{\rm neb}} = A_{V,\,{\rm SED}}/0.44$.
A trend is apparent, although with large scatter; the Spearman's
rank correlation coefficient indicates an anticorrelation with
$\rho = -0.29$ and significance of $2.1\,\sigma$.
The trend remains qualitatively the same when using
$\Sigma [{\rm SFR^{0}(H\alpha)}]$ for the case of no extra
attenuation towards the \ion{H}{2} regions (with $\rho = -0.37$
and correlation significance of $2.7\,\sigma$).
The trend outlined with the SINS and \citet{Law09} samples is
thus consistent with the interpretation that star formation
feedback plays a role in causing the large velocity dispersions
observed in $z \sim 2$ star-forming galaxies.

Figure~\ref{fig-vcsig0}{\em b\/} reveals a clearer
trend of increasing $v_{\rm obs}/(2\,\sigma_{\rm int})$ with
increasing dynamical mass $M_{\rm dyn}$ (derived as explained
in \S~\ref{Sub-massfractions}).  The Spearmans' rank correlation
coefficient is $\rho = 0.45$, and with correlation significance
of $3.4\,\sigma$.  In contrast, there is no clear trend with
stellar mass seen in Figure~\ref{fig-vcsig0}{\em d\/}
($\rho = 0.09$ and $0.6\,\sigma$ significance).  This suggests
that dispersion-dominated objects tend to be more gas-rich.
Figure~\ref{fig-vcsig0}{\em c\/} indicates that stellar age
does not seem to be an important factor ($\rho = -0.05$ and
$0.3\,\sigma$ significance).  In the plot, we used the best-fit
age from the SED modeling, but the same qualitative conclusion
is reached with, e.g., the ratio of $M_{\star} / {\rm SFR}$.
\footnote{
The trends, or lack thereof, seen in Figure~\ref{fig-vcsig0} are
not qualitatively changed for different SED modeling assumptions
as considered in Appendix~\ref{App-sedmod}.}

Thus, dispersion-dominated objects could include genuinely young and
gas-rich lower-mass objects in their earliest evolutionary stages where
the intense star formation activity is fueled by rapid gas accretion
from the halo, as well as more evolved systems where the star formation
activity may have been triggered by a merger event between 
gas-rich progenitors.  In either scenario, star formation
feedback will lead to higher gas-phase turbulence and provide
vertical support against gravity.
In fact, for a marginally (un)stable star-forming disk (with Toomre
parameter $Q = 1$), $\sigma_{0}/v_{\rm rot} = f_{\rm gas}/a$, where
$f_{\rm gas}$ is the gas mass fraction and $a$ is a dimensionless
parameter depending on the distribution of gas and gravitational
potential with typical values $\sim 1.4 - 1.7$ \citep[e.g.][]{FS06a}.
With this, the most gas-rich disks would be expected to approach
an intrinsically dispersion-dominated, spheroidal system in their
dynamical state.

Potential concerns in the above analysis are limitations from surface
brightness sensitivities on one hand, and spatial resolution on the other.
Indeed, the SINS and \citet{Law09} samples segregate significantly in
several of the diagrams.  It is unclear to what extent smaller sizes
and velocity gradients are influenced by the lower instrumental surface
brightness sensitivity of the \citet{Law09} data, affecting the ability
to detect fainter and more diffuse emission at larger radii.  In contrast,
sizes are better constrained with higher spatial resolution, especially
for the more compact objects.  However, some differences exist in properties
measured independently (especially in the somewhat lower stellar masses and
sizes from sensitive broad-band imaging, as pointed out by \citealt{Law09}),
so that differences in H$\alpha$ morphological and kinematic properties may
also reflect (at least in part) real physical differences.  Clearly, it
will be important to expand the samples studied consistently at the highest
spatial resolution with AO to a wide range of galaxy parameters to confirm
the trends outlined here.

\subsection{Mass Fractions and Constraints on Dark Matter Contribution}
            \label{Sub-massfractions}

With the data at hand, we can constrain the baryonic mass fraction
$f_{\rm baryons} = (M_{\rm gas} + M_{\star}) / M_{\rm dyn}$ among
our SINS H$\alpha$ sample galaxies.
The total stellar masses were obtained from our SED modeling,
which assumed a \citet{Cha03} IMF (Appendix~\ref{App-sedmod}).
For the gas masses, we relied on our H$\alpha$-derived star formation
rates normalized to unit area within the intrinsic half-light radius
and applied the Schmidt-Kennicutt relation between star formation rate
and gas mass surface density.  This relation has been established for
local star-forming galaxies \citep[e.g.,][]{Ken98} and its validity
has recently been tested at high redshift from direct measurements
of CO molecular line emission of bright SMGs
\citep[][see also \citealt{Tac06, Tac08}]{Bou07}.  These results
as well as very recent CO line detections in several rest-UV/optically
selected star-forming galaxies (BX and $sBzK$ objects) at $z \sim 1 - 2$
\citep{Dad08, Tac09} all show that both low and high redshift star-forming
galaxies lie approximately along a universal relation.
We used the relation derived by \citet{Bou07}, which implies:
\begin{equation}
 M_{\rm gas}~[{\rm M_{\odot}}] = 3.66 \times 10^{8}\,
 ({\rm SFR}~[{\rm M_{\odot}\,yr^{-1}}])^{0.58}\,
 ({r_{1/2}}~[{\rm kpc}])^{0.83}.
\label{Eq-SK}
\end{equation}
In applying equation~\ref{Eq-SK}, we took half of the inferred H$\alpha$
star formation rate for the area enclosed within $r_{1/2}{\rm (H\alpha)}$,
and multiplied by two to get the total gas mass.  
We considered again the two cases without and with extra attenuation
towards the \ion{H}{2} regions relative to the stars, with the \sfrha's
and \sfrhaoo's from Table~\ref{tab-SFRs}, giving $M_{\rm gas}^{0}$
and $M_{\rm gas}^{00}$ (differing by about a factor of two on average).

For the dynamical masses, we again followed one of the methods used
in \S~\ref{Sub-dynstate} to compute $v_{\rm d}$.
For the 18 disk galaxies with detailed kinematic modeling, we adopted
the total dynamical masses (i.e., within $r < 10~{\rm kpc}$) derived
by \citet{Gen08} and \citet{Cre09}.
For the rotation-dominated systems, we assumed disk rotation and
calculated the enclosed dynamical mass as
$M_{\rm dyn}(r < r_{1/2}) = (v_{\rm d}^2\,r_{1/2}) / G$,
where $G$ is the gravitational constant.
We averaged the masses obtained with $v_{\rm d}^{\rm vgrad}$ and
$v_{\rm d}^{\rm width}$ calculated from the observed velocity
gradient and from the integrated velocity dispersion, respectively
\citep[as described by][]{FS06a}, and corrected for inclination.
Here, the radius we used is half of the major axis $\rm FWHM(H\alpha)$
(given in Table~\ref{tab-Hameas}), which is more appropriate as measure
of the intrinsic deprojected radius of inclined disks.  We then
multiplied the resulting mass by two to obtain the total dynamical mass.
For the dispersion-dominated objects, we applied the isotropic virial
estimator with
$M_{\rm dyn} = (6.7\,\sigma_{\rm int}^2\,r_{1/2}) / G$,
appropriate for a variety of galactic mass distributions \citep{Bin08}.
For this case, $M_{\rm dyn}$ represents the total dynamical mass
and we used $r_{1/2}({\rm H\alpha})$ as measure of the intrinsic
half-light radius of dispersion-dominated systems.
As for $v_{\rm d}$ above, we considered the dynamical mass derived
for $\rm K20-ID5$ as upper limit since its kinematics are likely
affected by AGN and/or shocks.
The gas and dynamical masses are listed in Table~\ref{tab-dyn}
(stellar masses are given in Table~\ref{tab-sedprop}).

Figure~\ref{fig-fbar} shows the derived baryonic mass fractions for our
SINS galaxies.  The median value is $f_{\rm baryons} \sim 70\% - 80\%$,
depending on which gas mass estimate is adopted, with scatter of 0.3~dex.
For different SED modeling assumptions (see Appendix~\ref{App-sedmod}),
the median values are in the range $\sim 60\% - 80\%$; variations
in stellar mass fractions are typically partly compensated by opposite
variations in gas mass fractions because of the changes in best-fit
$A_{V}$ used to correct the H$\alpha$ SFRs on which our $M_{\rm gas}$ 
estimates are based (Equation~\ref{Eq-SK}).
The results are not strongly sensitive to the extinction correction
assumed in computing the SFRs from H$\alpha$ because overall the
stellar mass dominates the baryonic mass budget.  For our SINS sample,
the median gas mass fraction is $\sim 15\% - 30\%$, depending on the
H$\alpha$ extinction correction adopted.  This is somewhat lower than
the first estimates from millimeter CO line emission obtained to date
in several similarly selected galaxies at $z \sim 1 - 2.5$
\citep[$\sim 20\% - 50\%$][]{Dad08, Tac09} but may be consistent in view
of the large scatter of 0.35~dex in our data and the still small samples
with CO measurements available.
Our results suggest that the dark matter contribution within a radius
of $\rm \sim 10~kpc$ is $\sim 20\% - 30\%$ for our SINS H$\alpha$ sample.
We assumed a \citet{Cha03} in deriving the stellar masses; for more
``bottom-light'' IMFs at high redshift, as have been discussed in
recent literature \citep[e.g.][]{Dok08, Dave08}, the inferred baryonic
mass fraction would be lower and the dark matter contribution
correspondingly higher.

\section{SUMMARY}
         \label{Sect-conclu}

We have presented the SINS survey of star-forming galaxies at
$z \sim 1 - 3$ carried out with SINFONI at the VLT.
With a total of 80 objects observed, this is the largest
survey of near-IR integral field spectroscopy to date.
The largest subset, the SINS H$\alpha$ sample, includes \nsins\
optically- and near-/mid-IR selected galaxies at $1.3 < z < 2.6$.
Although with some bias towards the bluer part of the galaxy population
compared to purely $K$-selected samples at similar redshifts (due to
the requirement of an optical spectroscopic redshift), the SINS H$\alpha$
sample provides a reasonable representation of massive actively star-forming
galaxies at $z \sim 2$, in the range
$M_{\star} \sim 3 \times 10^{9} - 3 \times 10^{11}~{\rm M_{\odot}}$,
with median $M_{\star} = 2.7 \times 10^{10}~{\rm M_{\odot}}$ and
$\rm SFR(SED) = 72~M_{\odot}\,yr^{-1}$.

We discussed the ensemble integrated H$\alpha$ properties,
and demonstrated that our deep SINFONI data provide reliable
measurements of the total line fluxes, kinematics, and sizes.
The typical surface brightness sensitivities ($3\,\sigma$ per
resolution element) of our data sets imply limiting star
formation rates per unit intrinsic area of
$\rm \sim 0.03~M_{\odot}\,yr^{-1}\,kpc^{-2}$
for the average integration time of 3.4\,hr, or
$\rm \sim 0.1~M_{\odot}\,yr^{-1}\,kpc^{-2}$ in 1\,hr.
We showed quantitatively how observational strategies possibly 
affect trends of galaxy sizes with line fluxes and luminosities,
and stellar masses, emphasizing the importance of taking these
effects into account in comparing samples and assessing whether
observed trends reflect true physical relationships.

The main scientific conclusions of this paper and of our SINS
survey can be summarized as follows:
\\
$\bullet$
Analysis of the H$\alpha$ luminosities and equivalent widths
provides evidence for differential extinction between the
\ion{H}{2} regions and the stars by roughly a factor of
$\sim 2$, similar to what is inferred in local star-forming
and starburst galaxies.
\\
$\bullet$
With extra attenuation by a factor of $\sim 2$ towards the \ion{H}{2}
regions, the H$\alpha$ star formation rates are in good agreement with
those derived from the broad-band SED modeling for our SINS H$\alpha$
sample.
The data support that our SINS galaxies have had, on the whole,
roughly constant star formation rates over their lifetimes.
\\
$\bullet$
We find that many of the massive $z \sim 2$ star-forming galaxies
studied typically exhibit a large component of intrinsic local random
motions.  Inferred intrinsic velocity dispersions range from $\sim 30$
to $\rm 90~km\,s^{-1}$.
\\
$\bullet$
The observed morphologies of the H$\alpha$ line emission
and rest-UV/optical continuum emission are generally irregular
and asymmetric.  Large star-forming clumps of size $\rm \sim 1~kpc$
often dominate the appearance.  Despite these irregular and clumpy
morphologies of the nebular line emission tracing star-forming regions
and young stellar populations, the kinematics of the gas is often
surprisingly ordered.  Well defined velocity gradients are apparent
in about $80\%$ of the cases, where such measurements were possible
given sufficient resolution and S/N.  Two-dimensional ``spider-diagram''
patterns characteristic of ordered disk rotation are seen in the velocity
fields of several of the galaxies with highest quality SINFONI data.
\\
$\bullet$
Taking the SINS H$\alpha$ sample as a whole, $\sim 1/3$ of the galaxies
appear to have rotation-dominated kinematics, $\sim 1/3$ are interacting
or merging systems, and $\sim 1/3$ appear to have kinematics dominated
by large amounts of random motions and are thus ``dispersion-dominated.''
The fraction of rotation-dominated systems increases among the more
massive and evolved part of the SINS sample.
\\
$\bullet$
The rotation-dominated systems follow a velocity-size relation similar
to local disk galaxies.
\\
$\bullet$
The dispersion-dominated objects tend to be compact and have a lower
mass and lower angular momentum than the rotation-dominated systems.
The dispersion-dominated objects exhibit a wide range of ages but
include a population of young and probably very gas-rich galaxies in
the first stages of formation.  Other dispersion-dominated objects may
be late stage mergers.

\acknowledgments

We wish to thank the ESO staff, and in particular at Paranal Observatory,
for their helpful and enthusiastic support during the many observing runs
and several years over which SINFONI GTO were carried out.  We also thank
the SINFONI and PARSEC teams for their hard work on the instrument and
the laser, which allowed our observational program to be carried out so
successfully.
We thank the referee for useful comments and suggestions that helped
improve the quality and presentation of the paper.
This paper and the SINS survey have benefitted from many constructive,
insightful, and enthusiastic discussions with many colleagues whom we
are very grateful to, especially Marijn Franx for numerous inspiring
discussions, as well as Andi Burkert, Thorsten Naab, Peter Johansson, 
Ortwin Gerhard, Avishai Dekel, Pieter van Dokkum, Guinevere Kauffmann,
Simon White, Hans-Walter Rix, St\'ephane Courteau, Martin Bureau, 
Claudia Maraston, among many others.
We are grateful to Ian Smail, Scott Chapman, and Rob Ivison for providing
the necessary information and imaging data of SMGs targeted as part of
our SINS survey.
N. M. F. S. acknowledges support by the Schwerpunkt Programm SPP1177
of the Deutsche Forschungsgemeinschaft and by the Minerva Program of
the Max-Planck-Gesellschaft.
N. A. is supported by a Grant-in-Aid for Science Research (No. 19549245)
by the Japanese Ministry of Education, Culture, Sports, Science and Technology.

\appendix

\section{SED MODELING}
         \label{App-sedmod}

For the purpose of investigating the ensemble properties of our SINS
H$\alpha$ sample, we complemented our SINFONI data of the line emission
with properties derived from modeling of their optical- to near-/mid-IR
emission.  Parameters such as stellar mass and age, interstellar
extinction, absolute and specific star formation rates are available
from several of the surveys from which we drew our SINS targets.
However, the details of the modeling (assumptions, model ingredients,
and modeling techniques) are different from one survey to the other.
In order to allow for consistent comparisons among the SINS galaxies
as well as with the $K$-selected reference sample from the FIREWORKS
catalogue in CDFS \citep[][see \S~\ref{Sect-representativeness}]{Wuy08},
we remodeled all of the SINS galaxies in the same manner.
One limitation remains, due to the different wavelength coverage of the
different surveys, ranging from 4 up to over 10 bands and some including
IRAC data at $\rm 3 - 8~\mu m$.  However, this will mostly have an impact
on the uncertainties of the best-fit parameters
\citep[e.g.][]{Sha05a, Wuy07}.

For the optically-selected \bxbm\ objects, we used the 
$U_{n}G{\mathcal R}JK_{\rm s}$ photometry published by
\citet[][see also \citealt{Ste04}]{Erb06b}.
Targets in the Q2346 field have no $J$ band photometry.
$\rm Q2346-BX482$ lies in an area not covered with the $K_{\rm s}$-band
imaging and no near-IR photometry was available for $\rm SSA22a-MD41$.
For the latter two sources, we used the total $H_{160}$ magnitudes
measured from the deep HST/NICMOS imaging presented by \citet{FS09a}.
We further complemented the photometry of $\rm SSA22a-MD41$ with the total
$K$-band magnitude measured from publicly available imaging obtained with
the SOFI instrument at the ESO NTT (under program 071.A-0639, P.I.: Lehnert).
These data were reduced following procedures described by \citet{FS06b}.
The original photometric data for the K20 targets are described by
\citet[][see also \citealt{Cim02c}]{Dad04a}.  Since deeper, higher
resolution data with wider wavelength coverage are now available in
CDFS, we cross-identified our K20 targets in, and used the data from,
the FIREWORKS catalogue of \citet{Wuy08}.
For the Deep3a sources, we used the $U - K$ catalogue based on the
Subaru/SuprimeCam $BR_{c}I_{c}z^{\prime}$ and NTT/SOFI $JK_{\rm s}$
data described by \citet{Kon06}, supplemented with photometry through
the $U_{841}$ and $V_{843}$ filters from NTT/WFI \citep{Dad09}.
For the GMASS targets, we used the $B - 8~{\rm \mu m}$ photometry
from the catalogue generated and kindly provided by the GMASS team
based on the HST/ACS $BVIZ$, VLT/ISAAC $JHK_{\rm s}$, and Spitzer/IRAC
3.6, 4.5, 5.8, $\rm 8.0~\mu m$ deep imaging
\citep[][see also, e.g., \citealt{Cim08}]{Kur09}.
For the zCOSMOS sources, we collected photometry taken in the
Subaru/SuprimeCam $Bi^{\prime}z^{\prime}$, UKIRT/WIRCAM $J$,
and CFHT/WIRCAM $K_{\rm s}$ filters from the imaging data
presented by \citet{Cap07} and \citet{McC09}
\footnote{The WIRCAM $K_{\rm s}$ photometry was kindly
made available to us in advance of publication.}.
For the targets taken from the GDDS survey, we retrieved the seven-band
$BVRIz^{\prime}HK_{\rm s}$ photometric catalogue available through the
GDDS web site\,\footnote{See http://www.ociw.edu/lcirs/gdds.html} and
described by \citet{Abr04} and \citet{Che02}.

Photometric uncertainties were either as explicitly given in the
publications or databases or, if unavailable from those references,
were inferred from the depths of the imaging data.  In addition,
a minimum uncertainty was adopted (typically $\rm 0.08 - 0.1~mag$
depending on the depth and quality of the data sets) to account for
absolute calibration uncertainties and PSF-/aperture-matching across
the bands.  We used in all cases estimates of the ``total'' photometric
fluxes.  The input photometry for the SED modeling was further corrected
for the Galactic extinction towards the various fields, based on the
dust maps of \citet*{Sch98}.

The modeling was carried out following the procedures
described by \citet[][see also \citealt{Wuy07, Wuy08}]{FS04}.
In summary, we generated the synthetic spectra using the synthesis code
of \citet{BC03}, for a range of ages and a set of star formation histories.
We employed the set of ``Padova 1994'' evolutionary tracks and the
lower resolution but wider wavelength coverage set of stellar libraries
based on the BaSeL 3.1 library.  We adopted a fixed solar metallicity,
a \citet{Cha03} IMF, and the \citet{Cal00} reddening law (applied for
a uniform foreground screen of obscuring dust).  The attenuation due
to intergalactic H opacity was accounted for following the prescriptions
of \citet{Mad95}, and Lyman continuum absorption was approximated by
setting the flux of the synthetic templates equal to zero at
$\lambda_{\rm rest} < 912$~\AA.  We considered three combinations of
star formation history (SFH) and dust content: constant star formation
(CSF) and dust, single stellar population (SSP) with instantaneous
star formation at $t = 0$ and no dust, and an intermediate case of
an exponentially declining star formation rate with timescale
$\rm \tau = 300~Myr$.
These choices are admittedly simplistic, and likely to somewhat
bias the overall results (for instance in terms of absolute ages).
For consistency, comparisons of derived stellar properties for our SINS
H$\alpha$ sample should thus be limited to those for other samples at
similar redshifts obtained with similar SFHs (or families thereof), as
we do in the context of this paper.

The model SEDs were obtained by convolving the synthetic spectra
with the filter curves, which account properly for the full system
throughput for each of the photometric bandpasses considered.
The redshift was fixed at the H$\alpha$ redshift of the sources
(or the optical redshift for sources undetected in our H$\alpha$ data).
The age, extinction, and luminosity scaling of the model SEDs to the
observed SEDs were the free parameters in the fitting, which is based
on chi-squared minimization.  We restricted the ages considered between
a minimum of 50~Myr (to avoid implausibly young and extremely obscured
best-fits to the reddest galaxies) and a maximum corresponding to the age
of the universe at the redshift of each source.  The stellar mass we use
corresponds to the mass of stars still alive and stellar remnants.
The adopted results were then taken as the best-fit among the three
combinations of SFH$+$dust.
The three choices of SFH$+$dust are obviously very simplistic.
However, we opted for this as some of the targets have only $4 - 5$
photometric bands, limiting the number of free parameters possible
to keep the number of degrees of freedom $\ge 1$ in the fits.
Formal (random) uncertainties on the best-fitting parameters were
obtained from 200 Monte Carlo realizations (randomly varying the observed
SEDs assuming Gaussian photometric uncertainties), and taking the 68\%
confidence intervals of the distributions of best-fit results.
For five of our SINS H$\alpha$ sample galaxies, lower confidence
intervals of 0 are derived on their best-fit ages of 50~Myr, i.e. the
minimum allowed.  This is obviously artificial and due to the discrete
age grid and strict lower age limit in our modeling procedure.

We did not attempt to correct for emission line contribution in our
SED modeling.  Again, the main reason is consistency with the SED
modeling of the reference $K$-selected sample, for which this 
contribution is unknown for the very large majority of sources.
In addition, while it would in principle be possible to correct for
H$\alpha$ and [\ion{N}{2}] emission for all our sources, it is not
always possible to account for other potentially bright emission lines
in other near-IR and optical bands (e.g., Ly$\alpha$, [\ion{O}{3}])
for lack of measurements.
This is not expected to affect our results in a major way, as the line
fluxes --- at least for H$\alpha$ --- contribute on average $\sim 10\%$
of the broad-band emission (median of 7\%, with first and third quartile
of the distribution at 5\% and 13\%; see Table~\ref{tab-Hameas}).
Other lines together will not make significantly larger contribution to
the broad-band fluxes.  The main effect of correcting for H$\alpha$ is
generally to reduce the derived stellar masses, ages, and extinction,
and would be largest for our $K$-faintest targets with highest H$\alpha$
equivalent width and specific star formation rate
\citep[see also, e.g.,][]{Erb06b, Kri08a}.
By far, the largest H$\alpha$ contribution is inferred for $\rm Q1623-BX502$
(57\%) and this likely drives the best-fit towards a higher stellar mass,
an older age, and a higher extinction.
Nevertheless, the trends in the ensemble properties discussed in
\S~\ref{Sect-Ha_prop} and \ref{Sect-dust_sf} are not qualitatively
altered because few galaxies have contributions in excess of $10\%$.
Comparison of our SED modeling results to those from the studies of the 
respective surveys (when available) indicate overall good agreement, with 
differences generally attributable to the different model assumptions and 
ingredients, or to our not accounting for line emission contribution.

The formal fitting uncertainties derived from our Monte Carlo
realizations do not take into account the impact of our choice of
model ingredients and assumptions.  In particular, the metallicity,
reddening law, and IMF as well as the adopted synthesis code and set
of star formation histories can all have important effects and lead
to systematic variations in derived properties.  These are still
poorly constrained from observations for $z \sim 2$ galaxies.  
For our SINS H$\alpha$ sample, the number of bands available to construct
the SEDs is limited for many of the galaxies.  Together with the non-uniform
depth and wavelength coverage of the photometry from the various parent
surveys and the known degeneracies among model parameters, this prevents
a meaningful attempt at constraining these parameters in our modeling.
In addition, other independent empirical constraints either do not
exist or are insufficient (except for very few sources).
Since our main purpose is to investigate relative trends and ensemble
properties within our SINS H$\alpha$ sample and with respect to the
general population of $z \sim 2$ galaxies, an exhaustive discussion
of the effects of variations in SED modeling parameters is beyond the
scope of this paper.  It is nevertheless worth assessing the possible
impact of different choices in order to estimate the systematic
uncertainties and verify the robustness of our conclusions.

We tested the impact of models with a different treatment of stellar
evolutionary phases.
Specifically, we used the \citet{Mar05} models with the \citet{Kro01}
IMF (the differences between the \citealt{Cha03} and \citealt{Kro01}
IMFs have a negligible effect compared to that of the different models).
To gauge the possible impact of changes in assumed metallicity and
reddening law, we ran additional suites of \citet{BC03} models with
metallicity of $1/5$ and 2.5 times solar, and using alternatively
extinction laws for the Milky Way \citep{All76} and for the Small
Magellanic Cloud \citep{Pre84, Bou85}.  The metallicities explored
bracket the range inferred for $z \sim 2$ star-forming galaxies in
a similar stellar mass range as our SINS H$\alpha$ sample
\citep[e.g.,][]{Dok04, Erb06a, Hal08}.
Based on the variations in best-fit properties of individual objects
with these different models, we infer typical (median) systematic 
uncertainties of $\pm 30\%$ for the stellar masses, of $\rm \pm 0.3~mag$
for the visual extinctions $A_{V}$, and of factors of $\sim 2 - 3$ for
the stellar ages as well as for the absolute and specific star formation
rates (SFRs).

The main impact on the SINS H$\alpha$ sample properties of using the
\citet{Mar05} instead of the \citet{BC03} models is for the stellar ages
(the median becomes about twice younger) and for the absolute and specific
SFRs (median values higher by factors of $\approx 2$ and 3, respectively).
The median best-fit extinction increases a little by $\rm 0.2~mag$.  The
effects on the stellar masses are moderate, with the median decreasing by
$\approx 25\%$.  The variations in ensemble properties for the reference
$K_{\rm Vega} < 22~{\rm mag}$, $1.3 < z < 2.6$ sample from the CDFS FIREWORKS
catalogue  \citep{Wuy08} considered in \S~\ref{Sect-representativeness}
are similar (within $\approx 10\%$ for the median values).
As for the \citet{Mar05} models, the effects of changes in the assumed
metallicity and extinction law are most important for the stellar ages
and for the absolute and specific SFRs.  Compared to the results for
solar metallicity and the \citet{Cal00} reddening law, the variations
in median values for the SINS H$\alpha$ sample are by factors of
$\approx 0.4 - 4.5$ for the age, $\approx 0.3 - 1.5$ for the SFR, and
$\approx 0.2 - 3.6$ for the specific SFR.  The variations in median
$A_{V}$ is within $-0.6$ to $\rm +0.2~mag$.  The stellar masses are
least affected, with the median varying by factors of $0.7 - 1.2$.
We further computed models with a \citet{Sal55} IMF; this affects
essentially only the stellar masses and absolute SFRs, which increase
by a nearly identical factor of $\approx 1.7$.
In all cases, variations in ensemble properties for the reference CDFS
sample are comparable and in the same sense as for the SINS H$\alpha$
sample.  The relative comparisons between the two samples in
\S~\ref{Sect-representativeness} are therefore unaffected.

We verified the consequences of the variations in model ingredients
and assumptions considered above on all other results of this paper
that depend on properties derived from our SED modeling.
While this leads to systematic shifts in ensemble properties,
none of the main conclusion is significantly altered (see
\S~\ref{Sect-Ha_prop}, \S~\ref{Sect-dust_sf}, and \S~\ref{Sect-kinematics}).

\section{CHARACTERISTICS OF THE EFFECTIVE POINT-SPREAD FUNCTIONS}
         \label{App-psfs}

All effective PSFs were constructed, as described in \S~\ref{Sub-datared},
from the stars used for acquisition (which are also the AO reference stars
for the AO-assisted data sets) and observed at the start and in between
science OBs.
For the seeing-limited data, the PSF is very close to Gaussian.
The shape sometimes shows a noticeable elongation but, from
two-dimensional elliptical Gaussian fits, the median and mean
ellipticity is $\approx 0.1$ (or an axis ratio of $\approx 0.8$).
For AO-assisted data, the effective PSF shape is also very close to
Gaussian.  For the purpose of characterizing the achieved resolution
the assumption of a Gaussian provides a satisfactory estimate of the
spatial resolution element of the AO data as well (and the median and
average ellipticity is 0.06, or an axis ratio of 0.89).

To examine possible systematic structure in our PSFs, we constructed
higher S/N profiles by averaging the effective PSF images associated
with H$\alpha$ data sets of the SINS H$\alpha$ sample galaxies.
One PSF was created for the $\rm 125~mas\,pixel^{-1}$ scale, including
both seeing-limited and AO-assisted data (there are only three PSFs
obtained with AO at this scale, and excluding them does not change the
averaged profile), and one for the $\rm 50~mas\,pixel^{-1}$ scale with AO.
These are plotted in Figure~\ref{fig-psfs}.
To obtain a representative average PSF for the $\rm 125~mas\,pixel^{-1}$
data, we excluded the six PSFs with $\rm FWHM > 0\farcs 8$ (four of
which are for undetected sources), so that 46 PSF images were combined.
For the AO PSF at $\rm 50~mas\,pixel^{-1}$, we included all but one
of the five effective PSFs; the PSF of $\rm Deep3a-15504$ was excluded
because it is a double star resolved in our high resolution data.
The PSFs were normalized to a common peak value of unity before
combination, and an additional $5\,\sigma$-clipping was applied
to exclude residual bad pixels present in some cases.
The combined PSFs reveal more clearly extended wings and, as expected,
somewhat more prominently in the AO-assisted $\rm 50~mas\,pixel^{-1}$
data.  The profiles are best fit by a narrow core and a broad underlying
component, both elliptical Gaussian in shape.
The relative peak intensities of the narrow core to the broad component
are 3.7 and 2.7 for the averaged PSFs at 125 and $\rm 50~mas\,pixel^{-1}$,
respectively, and their relative FWHMs are approximately 0.5 and 0.4.

To quantify the effects of uncertainties in the PSF FWHM and shape
on kinematic modeling, we performed the following simulations.
We used model thin disks generated with DYSMAL (the same code as
used by \citealt{Cre09} and previously by \citealt{FS06a, Gen06, Gen08}).
We chose a fiducial model with parameters representative of the SINS
disk-like galaxies, and we varied the inclination between 10 and 80 degrees.
The models were binned spatially and spectrally to the SINFONI pixel size.
We convolved the models spectrally as appropriate for $K$-band SINFONI data,
and spatially using different PSFs: the real effective PSFs constructed
from the acquisition stars, and various model PSFs obtained by fitting
the real PSFs with a two-dimensional single circular Gaussian, single
elliptical Gaussian, and double elliptical Gaussian (with narrow core
and broad wings).
We also considered a simple circular Gaussian of FWHM $0\farcs 5$ for
the seeing-limited mode data and $0\farcs 15$ for the AO data at the
$\rm 50~mas\,pixel^{-1}$ scale.  We used 18 sets of such real and model
PSFs, corresponding to the 18 galaxies modeled by \citet{Cre09}.
We then extracted the velocity field and velocity dispersion maps
from the convolved model disks as described in (\S~\ref{Sub-extrmaps}).
For a given disk inclination and set of PSFs, we compared the differences
in extracted velocity fields and dispersion maps.

The maximum differences in relative velocities amount to $10\%$ or less
across the velocity fields.  They are smaller for the velocity widths,
$\la 7\%$ for the AO cases and $\la 3\%$ for the seeing-limited cases
across the dispersion maps.  The results are little sensitive to galaxy
inclination.  These maximum differences are comparable to or smaller
than our typical formal measurements uncertainties.
We conclude from these simulations that the typical uncertainties on
the PSF size and shape of our SINFONI data, including the presence or
not of possible extended wings or the assumption of a common PSF with
representative average FWHM of the data sets, have only a small impact
on the interpretation of the extracted kinematics and on the modeling,
and are not significant in view of other uncertainties such as the
intrinsic mass distribution or deviations from pure disk kinematics.

\section{NOISE PROPERTIES OF THE SINFONI DATA CUBES}
         \label{App-noise}

The data reduction procedure described in \S~\ref{Sub-datared} produces a
noise cube by taking the standard deviation of all values that are averaged
for a given pixel in the final combined 3D cube (after clipping outliers) 
and normalizing by the squared root of the number of pixels used.
Due to various factors, including the slitlet projection onto the detector
(with two pixels sampling a resolution element along one spatial axis) and,
most importantly, the data reduction, the resulting noise is not expected 
to scale linearly with aperture size as for pure uncorrelated Gaussian
noise.  This is analogous to what is seen in broad-band imaging data
\citep[e.g.][]{Lab03, FS06b}, where the effective noise $\sigma_{\rm real}$
increases faster with linear size of aperture $N \equiv \sqrt{A}$ than the
Gaussian scaling $\propto N \times \sigma_{\rm pix}$, where $\sigma_{\rm pix}$
is the pixel-to-pixel rms, even if for any given aperture size the noise has
a Gaussian behaviour.

To investigate the noise properties in our reduced SINFONI data,
we carried out a similar analysis as described by \citet{Lab03} and
\citet{FS06b}.  For every spectral channel, we measured the flux in
non-overlapping square apertures of equal size placed at random over the
area of deepest integration (i.e. the region of overlap of all exposures
for data taken with the on-source dithering pattern).  Aperture sizes $N$
from 1 to 8 pixels were considered in turn.  For the spectral channels that
include line emission from the galaxies (and for all channels for those that
are brightest in continuum emission), we excluded apertures that overlap with
the source.  For each $N$ and spectral channel, the distribution of the
measurement fluctuations in the ``empty apertures'' is well approximated
by a Gaussian, indicating Gaussian behaviour for a given aperture size
and spectral channel.

The effective noise $\sigma_{\rm real}(N, \lambda)$ was taken as
the dispersion of the best-fit Gaussian to the distribution of
empty aperture fluxes for each $N$ and channel.  The resulting
function normalized by the relation for uncorrelated noise,
$\sigma_{\rm real}(N, \lambda) / [N\,\times\,\sigma_{\rm pix}(\lambda)]$,
first rises rapidly with $N$ and then flattens and varies much more
slowly.  This transition in effective noise properties occurs around
a characteristic spatial scale of 4~pixels, corresponding to twice a
slitlet width.  It thus most plausibly reflects differences in the
contribution from correlated noise within and across slitlets.
Analytically, a logarithmic function of the form
\begin{equation}
\sigma_{\rm real}(N, \lambda) / [N\,\times\,\sigma_{\rm pix}(\lambda)]
  = a(\lambda) + b(\lambda)\,\log(N)
\label{Eq-noisemodellam}
\end{equation}
provides a good description of the observed noise behaviour.

Obviously, the number of non-overlapping apertures is rather
limited for the largest $N$ values because of the small FOV and so
the $\sigma_{\rm real}(N, \lambda)$ are less tightly constrained.
Moreover, we are ultimately also interested in the noise properties
of the channels including the emission lines to be fitted, where we
can only measure reliably the fluctuations in empty sky regions around
the source for small aperture sizes.  However, the measured values of
$\sigma_{\rm real}(N, \lambda) / [N\,\times\,\sigma_{\rm pix}(\lambda)]$
across all wavelength channels show a typical rms scatter around the
median by $\sim 25\%$, and the $a$ and $b$ coefficients show fairly
narrow distributions.  This suggests that a single set of $a$ and $b$
values can provide a reasonably accurate estimate of the noise
properties throughout all spectral channels of a given data cube.

We thus repeated the analysis considering all empty apertures flux
measurements over all spectral channels to derive a global
$\sigma_{\rm real}(N) / [N\,\times\,\sigma_{\rm pix}]$.
The corresponding relation follows closely the median of 
$\sigma_{\rm real}(N, \lambda) / [N\,\times\,\sigma_{\rm pix}(\lambda)]$,
and we can fit the same function as in Eq.~\ref{Eq-noisemodellam}
to derive global values of $a$ and $b$.  In general, the median of the
$a$ and $b$ values for the individual channels is close to the $a_{med}$
and $b_{med}$ obtained by a fit to the median of
$\sigma_{\rm real}(N) / [N\,\times\,\sigma_{\rm pix}]$
taken over the individual channels.  Also, the fit to the global
$\sigma_{\rm real}(N) / [N\,\times\,\sigma_{\rm pix}]$ from the analysis
carried out over all channels together leads to parameters close to
$a_{med}$ and $b_{med}$ (though sometimes with significantly lower $a$),
confirming the functional form of our noise model out to the larger
apertures considered.  For our applications, we adopted the set of
$a_{med}$ and $b_{med}$.

To compute the noise spectrum to be used in the emission line fitting to
spectra integrated over apertures of equivalent linear size $N = \sqrt{A}$,
we applied
\begin{equation}
\sigma_{\rm real}(N, \lambda) =
   [N\,\times\,\sigma_{\rm pix}(\lambda)] \times
    (a_{\rm med} + b_{\rm med}\,\log(N)),
\label{Eq-noisemodel}
\end{equation}
where the first factor accounts for the wavelength dependence of the
noise level and the second factor provides a global description of the
non-Gaussian, correlated nature of the noise properties in the reduced
SINFONI data cubes.  While this is not an exact measurement since the
wavelength dependence of the $a$ and $b$ parameters is ignored, the
analysis shows that this approximates the effective noise in $N > 1$
apertures across all wavelengths on average to $\sim 25\%$ in our
H$\alpha$ data sets (with a range from $\sim 5\%$ up to $\sim 50\%$).

As a quantitative example of the application of our noise model,
we consider the apertures used to extract the integrated spectrum.
The noise spectrum derived from this empirical model for each source
is on average a factor of 2 higher than what would be inferred assuming
pure Gaussian noise propagation (with a range from 1.6 to 2.7 among the
data sets).
We also determined the spectral pixel-to-pixel rms directly from the
integrated spectra, in regions free of night sky lines and of galaxy
line emission out to $\rm \pm 10000~km\,s^{-1}$ around H$\alpha$.
The spectral rms is typically $5\% - 15\%$ lower than the $1\,\sigma$
noise in the same wavelength interval obtained from application of
our noise model.
Only one galaxy appears to deviate significantly in that respect:
for $\rm Q1623-BX455$, which has the largest aperture size scaling from
the noise model (factor of 2.7 higher than for Gaussian noise propagation),
the noise spectrum gives $1\,\sigma$ uncertainties a factor of 1.8 higher
than the spectral rms in intervals free of night sky and galaxy emission
lines.
While the noise properties for such data sets as obtained with
SINFONI are complex, the results above indicate that overall
our method is able to constrain them to within $\sim 10\%$.
The advantage is that it allows to derive the noise at each wavelength
(thus preserving the variations across the full spectrum) and takes
into account the ``redistribution'' of the noise on different spatial
scales resulting notably from the reduction procedure (e.g., from
interpolations applied at different stages).

\section{H$\alpha$ MAPS, POSITION-VELOCITY DIAGRAMS, AND INTEGRATED SPECTRA}
         \label{App-allmaps}

Figures~\ref{fig-maps_bmbx1} to \ref{fig-maps_gdds2} present the
velocity-integrated H$\alpha$ linemaps, the position-velocity diagrams,
and the integrated spectra of all detected sources from our SINS H$\alpha$
sample.
The position-velocity diagrams were extracted from the data cubes, without
additional smoothing from median-filtering, in a synthetic slit 6~pixels
wide along the major axis of the galaxies, indicated by the rectangle on
the H$\alpha$ maps.  This width corresponds to $0\farcs 75$ or $1.3\,\times$
the median PSF FWHM for the data sets at the $\rm 125~mas\,pixel^{-1}$ scale,
and $0\farcs 30$ or $1.75\,\times$ the median PSF FWHM for those at the
$\rm 50~mas\,pixel^{-1}$ scale.
The integrated spectra were extracted from the unsmoothed data cubes in
circular apertures, with the radii adopted so as to enclose $> 90\%$ of
the total flux based on the curve-of-growth analysis.  These apertures
for each galaxy and instrument setup are also shown on the H$\alpha$ maps,
with the radii listed in Table~\ref{tab-Hameas}.
For galaxies obtained at both seeing-limited and AO pixel scales, we
show the results for each setup.


\clearpage


\setcounter{figure}{0}
\setcounter{section}{0}

\LongTables

\tabletypesize{\small}
\begin{deluxetable}{llllll}
\tablecolumns{6}
\tablewidth{0pt}
\tablecaption{SINS survey: galaxies observed
              \label{tab-samples}}
\tablehead{
   \colhead{Source} & 
   \colhead{$z_{\rm sp}$\,\tablenotemark{a}} & 
   \colhead{Class\,\tablenotemark{b}} & 
   \colhead{$K_{\rm Vega}$} & 
   \colhead{Parent survey or field} & 
   \colhead{References} \\
   \colhead{} & 
   \colhead{} & 
   \colhead{} & 
   \colhead{(mag)} & 
   \colhead{} & 
   \colhead{}
}
\startdata
$\rm Q1307-BM1163$  & 1.4105 & BM          & \ldots  &  BX/BM NIRSPEC  &  1, 2 \\
$\rm Q1623-BX376$   & 2.4085 & BX          &  20.84  &  BX/BM NIRSPEC  &  1, 2, 3 \\
$\rm Q1623-BX447$   & 2.1481 & BX          &  20.55  &  BX/BM NIRSPEC  &  1, 2, 3 \\
$\rm Q1623-BX455$   & 2.4074 & BX          &  21.56  &  BX/BM NIRSPEC  &  1, 2 \\
$\rm Q1623-BX502$   & 2.1550 & BX          &  22.04  &  BX/BM NIRSPEC  &  1, 2 \\
$\rm Q1623-BX528$   & 2.2682 & BX          &  19.75  &  BX/BM NIRSPEC  &  1, 2 \\
$\rm Q1623-BX543$   & 2.5211 & BX          &  20.54  &  BX/BM NIRSPEC  &  1, 2 \\
$\rm Q1623-BX599$   & 2.3304 & BX          &  19.93  &  BX/BM NIRSPEC  &  1, 2 \\
\enddata
\end{deluxetable}

\clearpage

\tabletypesize{\footnotesize}
\begin{deluxetable}{lccccccccc}
\tablecolumns{10}
\tablewidth{0pt}
\tablecaption{Photometric properties of the SINS H$\alpha$ sample galaxies
              \label{tab-photprop}}
\tablehead{
   \colhead{Source} & 
   \colhead{$B_{\rm AB}$} & 
   \colhead{$G_{\rm AB}$} & 
   \colhead{$H_{\rm Vega}$} & 
   \colhead{$K_{\rm s,\,Vega}$} & 
   \colhead{$B_{\rm AB} - K_{\rm s,\,Vega}$} & 
   \colhead{$G_{\rm AB} - K_{\rm s,\,Vega}$} & 
   \colhead{$B_{\rm AB} - z_{\rm AB}$} & 
   \colhead{$z_{\rm AB} - K_{\rm s,\,Vega}$} & 
   \colhead{$J_{\rm Vega} - K_{\rm s,\,Vega}$} \\
   \colhead{} & 
   \colhead{(mag)} & 
   \colhead{(mag)} & 
   \colhead{(mag)} & 
   \colhead{(mag)} & 
   \colhead{(mag)} & 
   \colhead{(mag)} & 
   \colhead{(mag)} & 
   \colhead{(mag)} & 
   \colhead{(mag)}
}
\startdata
$\rm Q1307-BM1163$  & \ldots           & $21.83 \pm 0.01$ & \ldots           & \ldots           &
           \ldots           & \ldots           & \ldots           & \ldots           & \ldots           \\
$\rm Q1623-BX376$   & \ldots           & $23.43 \pm 0.01$ & \ldots           & $20.83 \pm 0.09$ &
           \ldots           & $2.60 \pm 0.09$  & \ldots           & \ldots           & $1.59 \pm 0.14$  \\
$\rm Q1623-BX447$   & \ldots           & $24.53 \pm 0.03$ & \ldots           & $20.54 \pm 0.07$ &
           \ldots           & $3.99 \pm 0.08$  & \ldots           & \ldots           & $1.64 \pm 0.11$  \\
$\rm Q1623-BX455$   & \ldots           & $25.03 \pm 0.05$ & \ldots           & $21.55 \pm 0.18$ &
           \ldots           & $3.48 \pm 0.19$  & \ldots           & \ldots           & $1.81 \pm 0.31$  \\
$\rm Q1623-BX502$   & \ldots           & $24.45 \pm 0.03$ & \ldots           & $22.03 \pm 0.28$ &
           \ldots           & $2.42 \pm 0.28$  & \ldots           & \ldots           & $0.97 \pm 0.33$  \\
$\rm Q1623-BX528$   & \ldots           & $23.69 \pm 0.01$ & $20.97 \pm 0.06$ & $19.74 \pm 0.03$ &
           \ldots           & $3.95 \pm 0.03$  & \ldots           & \ldots           & $1.77 \pm 0.06$  \\
$\rm Q1623-BX543$   & \ldots           & $23.43 \pm 0.01$ & \ldots           & $20.53 \pm 0.07$ &
           \ldots           & $2.90 \pm 0.07$  & \ldots           & \ldots           & $1.29 \pm 0.09$  \\
$\rm Q1623-BX599$   & \ldots           & $23.54 \pm 0.01$ & \ldots           & $19.92 \pm 0.04$ &
           \ldots           & $3.62 \pm 0.04$  & \ldots           & \ldots           & $2.07 \pm 0.08$  \\
\enddata
\tablecomments
{
All photometry has been corrected for Galactic extinction based on
the dust maps and extinction curve of \citet{Sch98}.
The following $E(B-V)$ values were used for the various fields:
0.007~mag for Q1307,
0.033~mag for Q1623,
0.036~mag for Q2343,
0.029~mag for Q2346,
0.063~mag for SSA22a,
0.008~mag for K20 and GMASS,
0.043~mag for Deep3a,
0.019~mag for zCOSMOS,
0.029~mag for SA12, and
0.059~mag for SA15.
}
\tablerefs
{
The references or photometric catalogues for sources in the
various fields are:
\citet{Erb06b} for Q1623, Q2343, Q2346, and SSA22a;
\citet{Kon06} and \citet{Dad09} for Deep3a;
\citet{Kur09} for GMASS;
\citet{Cap07} and \citet{McC09} for zCOSMOS;
\citet{Abr04} and \citet{Che02} for GDDS (SA12 and SA15);
\citet{FS09a} for the NICMOS $H_{160}$ photometry of
$\rm Q1623-BX528$, $\rm Q1623-BX663$, $\rm SSA22a-MD41$,
$\rm Q2343-BX389$, $\rm Q2343-BX610$, and $\rm Q2346-BX482$.
The original K20 catalogue is presented by \citet{Dad04a}
but we used the CDFS FIREWORKS catalogue of \citet{Wuy08},
based on more recent, deeper imaging with larger wavelength
coverage.
$\rm GMASS-2113E$ is not in the GMASS catalogue but
we identified it in the CDFS FIREWORKS catalogue from
which we adopt the photometry.
}
\tablenotetext{a}
{
$K$ band photometry is not available from \citet{Erb06b}
in the SSA22a field; we used publicly available archival
$K_{\rm s}$ imaging from SOFI at the ESO NTT, obtained
under program 071.A-0639 (P.I. Lehnert).
}
\end{deluxetable}

\clearpage

\tabletypesize{\small}
\begin{deluxetable}{lccccccc}
\tablecolumns{8}
\tablewidth{0pt}
\tablecaption{Properties derived from SED modeling of the
              SINS H$\alpha$ sample galaxies
              \label{tab-sedprop}}
\tablehead{
   \colhead{Source} & 
   \colhead{SFH\,\tablenotemark{a}} & 
   \colhead{Age} & 
   \colhead{$A_{V}$} & 
   \colhead{$M_{\star}$} & 
   \colhead{$M_{V,\,{\rm AB}}$\,\tablenotemark{b}} & 
   \colhead{SFR} & 
   \colhead{sSFR\,\tablenotemark{c}} \\ 
   \colhead{} & 
   \colhead{} & 
   \colhead{(Myr)} & 
   \colhead{(mag)} & 
   \colhead{($\rm 10^{10}~M_{\odot}$)} & 
   \colhead{(mag)} & 
   \colhead{($\rm M_{\odot}\,yr^{-1}$)} & 
   \colhead{($\rm Gyr^{-1}$)}
}
\startdata
$\rm Q1307-BM1163$  & \ldots       & \ldots                    & \ldots                & \ldots                  &
         \ldots                    & \ldots                    & \ldots                    \\
$\rm Q1623-BX376$   & CSF          & $286^{+520}_{-31}$        & $0.4 \pm 0.2$         & $0.84^{+0.46}_{-0.10}$  &
         $-22.26^{+0.11}_{-0.06}$  & $40^{+1}_{-17}$           & $4.8^{+1.1}_{-3.0}$       \\
$\rm Q1623-BX447$   & $\tau 300$   & $509^{+132}_{-55}$        & $0.8 \pm 0.2$         & $2.12^{+0.36}_{-0.04}$  &
         $-22.17^{+0.08}_{-0.02}$  & $24^{+12}_{-8}$           & $1.1^{+0.3}_{-0.4}$       \\
$\rm Q1623-BX455$   & CSF          & $1015^{+885}_{-611}$      & $0.6 \pm 0.2$         & $1.03^{+0.52}_{-0.39}$  &
         $-21.51^{+0.20}_{-0.12}$  & $15^{+10}_{-1}$           & $1.5^{+2.0}_{-0.6}$       \\
$\rm Q1623-BX502$   & CSF          & $227^{+177}_{-137}$       & $0.4 \pm 0.2$         & $0.23^{+0.14}_{-0.10}$  &
         $-20.98^{+0.19}_{-0.17}$  & $14^{+9}_{-1}$            & $6.0^{+8.0}_{-2.5}$       \\
$\rm Q1623-BX528$   & CSF          & $2300^{+450}_{-200}$      & $0.6 \pm 0.2$         & $6.54^{+1.10}_{-0.58}$  &
         $-22.97^{+0.04}_{-0.04}$  & $46 \pm 1$            & $0.7 \pm 0.1$             \\
$\rm Q1623-BX543$   & CSF          & $81^{+21}_{-9}$           & $0.8 \pm 0.2$         & $0.94^{+0.20}_{-0.09}$  &
         $-22.72^{+0.08}_{-0.04}$  & $150^{+2}_{-5}$           & $15^{+2}_{-3}$            \\
$\rm Q1623-BX599$   & CSF          & $2750^{+18}_{-1141}$      & $0.4 \pm 0.2$         & $5.66^{+0.07}_{-0.24}$  &
         $-22.88^{+0.15}_{-0.02}$  & $34^{+19}_{-1}$           & $0.6^{+0.4}_{-0.1}$       \\
\enddata
\tablecomments
{
The formal (random) fitting uncertainties are given, derived from
the 68\% confidence intervals based on 200 Monte Carlo simulations
for the default set of \citealt{BC03} models with solar metallicity,
the \citealt{Cha03} IMF, and the \citealt{Cal00} reddening law;
systematic uncertainties (from SED modeling assumptions) are estimated
to be typically $\pm 30\%$ for the stellar masses, $\rm \pm 0.3~mag$
for the extinctions, and factors of $\sim 2 - 3$ for the ages as well
as for the absolute and specific star formation rates (see
\S~\ref{Sect-representativeness} and Appendix~\ref{App-sedmod}).
}
\tablenotetext{a}
{
The best-fitting star formation history:
``CSF'': constant star formation rate;
``$\tau 300$'': exponentially declining star formation rate with $e$-folding
timescale of $\rm \tau = 300~Myr$.
}
\tablenotetext{b}
{
Rest-frame absolute $V$-band magnitude, uncorrected for extinction.
}
\tablenotetext{c}
{
Specific star formation rate,
i.e. the ratio of star formation rate over stellar mass.
}
\end{deluxetable}

\clearpage

\tabletypesize{\small}
\begin{deluxetable}{lllc}
\tablecolumns{3}
\tablecaption{Observing runs for SINS SINFONI observations
              \label{tab-runs}}
\tablehead{
   \colhead{Run ID} & 
   \colhead{Dates} & 
   \colhead{Program ID}
}
\startdata
Mar03  & 2003 March 22 to April 11      &  070.A-0229, 070.B-0545  \\
Jul04  & 2004 July 8 to 22              &  073.B-9018  \\
Aug04  & 2004 August 13 to 22           &  073.B-9018  \\
Nov04  & 2004 November 29 to 30         &  074.A-9011  \\ 
Dec04  & 2004 December 20 to 21         &  074.A-9011  \\
Mar05  & 2005 March 13 to 23            &  074.A-9011  \\
Apr05  & 2005 April 4 to 8              &  075.A-0466  \\
Jun05  & 2005 June 15 to 17             &  075.A-0466  \\
Aug05  & 2005 August 27 to September 4  &  075.A-0466  \\
Oct05  & 2005 October 2 to 12           &  076.A-0527  \\
Mar06  & 2006 March 16 to 20            &  076.A-0527  \\
Apr06  & 2006 April 20 to 22            &  077.A-0576  \\
Jun06  & 2006 June 6 to 7               &  077.A-0576  \\
Aug06  & 2006 August 16 to 19           &  077.A-0576  \\
Nov06  & 2006 November 24 to 28         &  078.A-0600  \\
Dec06  & 2006 November 29 to December 3 &  078.A-0055\,\tablenotemark{a} \\
Mar07  & 2007 March 25 to 28            &  078.A-0600  \\
Apr07  & 2007 April 16 to 23            &  079.A-0341  \\
Aug07  & 2007 August 18 to 20           &  079.A-0341  \\
Oct07  & 2007 October 27 to 29          &  080.A-0330  \\
Nov07  & 2007 November 13 to 15         &  080.A-0635\,\tablenotemark{a} \\
Mar08  & 2008 March 25 to 27            &  080.A-0330  \\
Apr08  & 2008 April 4 to 9              &  080.A-0330 \\
Jul08  & 2008 July 27 to 31             &  080.A-0339
\enddata
\tablenotetext{a}
{
These SINFONI observing runs were carried out as part of the
collaboration between the SINS and GMASS teams, under normal
ESO Open Time programs (P.I.: A. Cimatti).
}
\end{deluxetable}

\clearpage

\tabletypesize{\small}
\begin{deluxetable}{llcclrcl}
\tablecolumns{8}
\tablewidth{0pt}
\tablecaption{Summary of the SINFONI observations
              \label{tab-obs}}
\tablehead{
   \colhead{Source} & 
   \colhead{$z_{\rm sp}$\,\tablenotemark{a}} & 
   \colhead{Band} & 
   \colhead{Scale\,\tablenotemark{b}} & 
   \colhead{Mode} & 
   \colhead{$t_{\rm int}$\,\tablenotemark{c}} & 
   \colhead{PSF FWHM\,\tablenotemark{d}} & 
   \colhead{Run ID\,\tablenotemark{e}} \\
   \colhead{} & 
   \colhead{} & 
   \colhead{} & 
   \colhead{(mas)} & 
   \colhead{} & 
   \colhead{(s)} & 
   \colhead{} & 
   \colhead{}
}
\startdata
$\rm Q1307-BM1163$  & 1.4105 & $H$  & 125 & Seeing-limited & 14400 & $0\farcs 61$ & Mar05 \\
                    &        & $J$  & 125 & Seeing-limited &  7200 & $0\farcs 77$ & Mar05 \\
$\rm Q1623-BX376$   & 2.4085 & $K$  & 125 & Seeing-limited & 15600 & $0\farcs 49$ & Mar05, Apr05 \\
$\rm Q1623-BX447$   & 2.1481 & $K$  & 125 & Seeing-limited & 14400 & $0\farcs 56$ & Mar06, Aug06 \\ 
                    &        & $H$  & 125 & Seeing-limited &  1800 & $0\farcs 83$ & Apr07 \\
$\rm Q1623-BX455$   & 2.4074 & $K$  & 125 & Seeing-limited & 12000 & $0\farcs 57$ & Mar05 \\
$\rm Q1623-BX502$   & 2.1550 & $K$  &  50 & NGS/LGS-AO     & 22800 & $0\farcs 24$ & Apr05, Mar08, Apr08 \\
$\rm Q1623-BX528$   & 2.2682 & $K$  & 125 & Seeing-limited & 24300 & $0\farcs 63$ & Jul04, (Aug05), Mar07 \\
\enddata
\end{deluxetable}

\clearpage

\tabletypesize{\small}
\begin{deluxetable}{lllcccccccc}
\tablecolumns{8}
\tablewidth{0pt}
\tablecaption{H$\alpha$ properties of the SINS H$\alpha$ sample
              \label{tab-Hameas}}
\tablehead{
   \colhead{Source} & 
   \colhead{$r_{\rm aper}$\,\tablenotemark{a}} & 
   \colhead{$z_{\rm H\alpha}$\,\tablenotemark{b}} & 
   \colhead{$F({\rm H\alpha})$\,\tablenotemark{b}} & 
   \colhead{$\rm \sigma_{int}(H\alpha)$\,\tablenotemark{b}} & 
   \colhead{$r_{\rm 1/2}({\rm H\alpha})$\,\tablenotemark{c}} & 
   \colhead{$\rm FWHM(H\alpha)$\,\tablenotemark{d}} &
   \colhead{$f_{\rm BB}({\rm H\alpha})$\,\tablenotemark{e}} \\
   \colhead{} & 
   \colhead{(arcsec)} & 
   \colhead{} & 
   \colhead{($\rm 10^{-17}~erg\,s^{-1}\,cm^{-2}$)} & 
   \colhead{($\rm km\,s^{-1}$)} &
   \colhead{(kpc)} &
   \colhead{(kpc)} &
   \colhead{}
}
\startdata
$\rm Q1307-BM1163$  & $1.00$  & $1.4104$  & $64.2 \pm 1.7$            &
         $153^{+6}_{-5}$       & $2.4 \pm 1.1$         & $6.5 \pm 2.2$         & \ldots   \\
$\rm Q1623-BX376$   & $1.00$  & $2.4088$  & $8.5^{+0.8}_{-0.6}$       &
         $99^{+11}_{-9}$       & $2.4 \pm 0.9$         & $4.6 \pm 1.8$         & $0.15$   \\
$\rm Q1623-BX447$   & $1.00$  & $2.1473$  & $10.2^{+1.0}_{-0.9}$      &
         $144 \pm 17$          & $3.7 \pm 1.0$         & $10.0 \pm 2.4$        & $0.12$   \\
$\rm Q1623-BX455$   & $1.00$  & $2.4072$  & $10.5^{+1.5}_{-1.2}$      &
         $130 \pm 28$          & $3.0 \pm 1.0$         & $7.5 \pm 2.1$         & $0.36$   \\
$\rm Q1623-BX502$   & $0.50$  & $2.1556$  & $12.3^{+0.7}_{-0.6}$      &
         $72^{+7}_{-5}$        & $1.7 \pm 0.5$         & $4.3 \pm 1.0$         & $0.57$   \\
$\rm Q1623-BX528$   & $1.25$  & $2.2683$  & $11.4 \pm 0.4$            &
         $141 \pm 8$           & $4.6 \pm 1.3$         & $11.7 \pm 2.9$        & $0.07$   \\
$\rm Q1623-BX543$   & $1.00$  & $2.5209$  & $19.5^{+1.3}_{-1.4}$      &
         $149^{+22}_{-23}$     & $2.8 \pm 1.0$         & $8.6 \pm 2.3$         & $0.28$   \\
$\rm Q1623-BX599$   & $1.00$  & $2.3313$  & $30.6^{+0.9}_{-0.8}$      &
         $153 \pm 9$           & $2.8 \pm 1.0$         & $4.7 \pm 2.1$         & $0.23$   \\
\enddata
\tablenotetext{a}
{
Radius of the circular aperture used to extract the
spatially-integrated spectrum of each source.
}
\tablenotetext{b}
{
Redshift (vacuum), flux, and velocity dispersion of H$\alpha$
derived by fitting a Gaussian profile to the line emission in
the spatially-integrated spectrum.  
The velocity dispersion is corrected for spectral instrumental resolution.
The uncertainties correspond to the formal 68\% confidence intervals
derived from 100 Monte Carlo simulations.
For undetected sources, the optical redshift is given in
parenthesis and the $3\,\sigma$ upper limit on the flux is given.
The limits are computed based on the noise spectrum within an
aperture of radius 1\arcsec, assuming H$\alpha$ at the wavelength
expected for the optical redshift and an intrinsic width corresponding
$\rm \sigma_{\rm int} = 130~km\,s^{-1}$, the average for the
detected sources.
}
\tablenotetext{c}
{
Intrinsic (corrected for spatial resolution) H$\alpha$
half-light radius derived from curve-of-growth analysis from
spectra integrated in circular apertures of increasing radius.
The measurement errors account for typical seeing variations
during the observations and uncertainties from the PSF shape.
}
\tablenotetext{d}
{
Intrinsic FWHM size along the major axis, derived from fitting a
1-D Gaussian to the H$\alpha$ light profile taken in 6-pixels-wide
circular apertures along the major axis of each galaxy.
For some of the sources with faintest H$\alpha$ surface brightness,
major axis profiles are too unreliable for a FWHM determination.
The measurement errors account for typical seeing variations
during the observations and uncertainties from the PSF shape.
}
\tablenotetext{e}
{
Fractional contribution of the H$\alpha$ emission line to the
observed broad-band flux density ($K$ for sources at $2 < z < 2.6$
and $H$ for those at $1.3 < z < 2$).
}
\end{deluxetable}

\clearpage

\tabletypesize{\small}
\begin{deluxetable}{lcccccccc}
\tablecolumns{9}
\tablewidth{0pt}
\tablecaption{H$\alpha$ luminosities and equivalent widths
              of the SINS H$\alpha$ sample
              \label{tab-LHaWHa}}
\tablehead{
   \colhead{Source} & 
   \colhead{$L^{\rm obs}({\rm H\alpha})$\,\tablenotemark{a}} & 
   \colhead{$L^{0}({\rm H\alpha})$\,\tablenotemark{a}} & 
   \colhead{$L^{00}({\rm H\alpha})$\,\tablenotemark{a}} & 
   \colhead{$L^{0}_{\rm pred.}({\rm H\alpha})$\,\tablenotemark{b}} & 
   \colhead{$W^{\rm rest}_{\rm SINF}({\rm H\alpha})$\,\tablenotemark{c}} & 
   \colhead{$W^{\rm rest}_{\rm BB}({\rm H\alpha})$\,\tablenotemark{d}} & 
   \colhead{$W^{\rm rest,00}_{\rm BB}({\rm H\alpha})$\,\tablenotemark{d}} & 
   \colhead{$W^{\rm rest}_{\rm pred.}({\rm H\alpha})$\,\tablenotemark{e}} \\
   \colhead{} & 
   \colhead{($\rm 10^{42}~erg\,s^{-1}$)} & 
   \colhead{($\rm 10^{42}~erg\,s^{-1}$)} & 
   \colhead{($\rm 10^{42}~erg\,s^{-1}$)} & 
   \colhead{($\rm 10^{42}~erg\,s^{-1}$)} & 
   \colhead{(\AA)} & 
   \colhead{(\AA)} & 
   \colhead{(\AA)} & 
   \colhead{(\AA)}
}
\startdata
$\rm Q1307-BM1163$  & $7.9 \pm 0.2$             & \ldots                    & \ldots                    & \ldots                    &
         $182^{+8}_{-7}$           & \ldots                    & \ldots                    & \ldots                    \\
$\rm Q1623-BX376$   & $3.9^{+0.4}_{-0.3}$       & $5.3 \pm 0.9$             & $7.7^{+2.8}_{-2.7}$       & $8.7^{+0.2}_{-3.6}$       &
         $24 \pm 2$                & $174^{+18}_{-16}$         & $256^{+56}_{-55}$         & $254^{+13}_{-87}$         \\
$\rm Q1623-BX447$   & $3.5^{+0.4}_{-0.3}$       & $6.4^{+1.2}_{-1.1}$       & $13.9^{+5.0}_{-4.9}$      & $5.1^{+2.5}_{-1.6}$       &
         $78^{+10}_{-9}$           & $142^{+15}_{-13}$         & $306^{+67}_{-65}$         & $108^{+15}_{-28}$         \\
$\rm Q1623-BX455$   & $4.8^{+0.7}_{-0.6}$       & $7.5^{+1.6}_{-1.4}$       & $13.4^{+5.0}_{-4.9}$      & $3.3^{+2.1}_{-0.1}$       &
         $>633$                    & $554^{+77}_{-71}$         & $987^{+235}_{-228}$       & $153^{+67}_{-32}$         \\
$\rm Q1623-BX502$   & $4.3 \pm 0.2$             & $5.8 \pm 0.9$             & $8.5 \pm 3.0$             & $2.9^{+1.9}_{-0.1}$       &
         $>1136$                   & $1380^{+157}_{-156}$      & $2029^{+454}_{-453}$      & $281^{+128}_{-60}$        \\
$\rm Q1623-BX528$   & $4.5 \pm 0.2$             & $7.1 \pm 1.1$             & $12.6 \pm 4.4$            & $9.9 \pm 0.1$             &
         $111^{+13}_{-12}$         & $74 \pm 3$                & $132 \pm 26$              & $114^{+3}_{-6}$           \\
$\rm Q1623-BX543$   & $9.9 \pm 0.7$             & $18.2 \pm 3.0$            & $39.3 \pm 13.8$           & $31.4^{+0.5}_{-1.1}$      &
         $>693$                    & $368 \pm 25$              & $795 \pm 162$             & $428^{+20}_{-39}$         \\
\enddata
\end{deluxetable}

\clearpage

\tabletypesize{\small}
\begin{deluxetable}{lcccccc}
\tablecolumns{7}
\tablewidth{0pt}
\tablecaption{Star formation rate estimates of the SINS H$\alpha$ sample
              \label{tab-SFRs}}
\tablehead{
   \colhead{Source} & 
   \colhead{$\rm SFR(SED)$\,\tablenotemark{a}} & 
   \colhead{$\rm SFR^{0}(H\alpha)$\,\tablenotemark{b}} & 
   \colhead{$\rm SFR^{00}(H\alpha)$\,\tablenotemark{b}} & 
   \colhead{$L^{\rm obs}({\rm UV})$\,\tablenotemark{c}} &
   \colhead{$L^{0}({\rm UV})$\,\tablenotemark{c}} &
   \colhead{$\rm SFR^{0}(UV)$\,\tablenotemark{d}} \\
   \colhead{} & 
   \colhead{($\rm M_{\odot}\,yr^{-1}$)} & 
   \colhead{($\rm M_{\odot}\,yr^{-1}$)} & 
   \colhead{($\rm M_{\odot}\,yr^{-1}$)} & 
   \colhead{($\rm 10^{28}~erg\,s^{-1}\,Hz^{-1}$)} & 
   \colhead{($\rm 10^{28}~erg\,s^{-1}\,Hz^{-1}$)} & 
   \colhead{($\rm M_{\odot}\,yr^{-1}$)}
}
\startdata
$\rm Q1307-BM1163$  & \ldots                    & \ldots                    & \ldots                    &
         \ldots                    & \ldots                    & \ldots                    \\
$\rm Q1623-BX376$   & $40^{+1}_{-17}$           & $24 \pm 4$                & $36 \pm 13$               &
         $19.2 \pm 0.2$            & $49 \pm 23$               & $40 \pm 19$               \\
$\rm Q1623-BX447$   & $24^{+12}_{-8}$           & $30 \pm 5$                & $65 \pm 23$               &
         $5.7 \pm 0.2$             & $37 \pm 18$               & $31 \pm 14$               \\
$\rm Q1623-BX455$   & $15^{+10}_{-1}$           & $35 \pm 7$                & $62 \pm 23$               &
         $4.4 \pm 0.2$             & $18 \pm 8$                & $15 \pm 7$                \\
$\rm Q1623-BX502$   & $14^{+9}_{-1}$            & $27 \pm 4$                & $40 \pm 14$               &
         $6.2 \pm 0.2$             & $16 \pm 7$                & $13 \pm 6$                \\
$\rm Q1623-BX528$   & $46 \pm 1$                & $33 \pm 5$                & $59 \pm 20$               &
         $13.6 \pm 0.1$            & $56 \pm 26$               & $46 \pm 22$               \\
$\rm Q1623-BX543$   & $150^{+2}_{-5}$           & $85 \pm 14$               & $180 \pm 64$              &
         $20.7 \pm 0.2$            & $136 \pm 64$              & $110 \pm 53$              \\
$\rm Q1623-BX599$   & $34^{+19}_{-1}$           & $81 \pm 12$               & $120 \pm 41$              &
         $16.4 \pm 0.2$            & $42 \pm 20$               & $35 \pm 16$               \\
\enddata
\tablenotetext{a}
{
Intrinsic star formation rate and corresponding 68\% confidence
intervals derived from the SED modeling (Table~\ref{tab-sedprop}).
}
\tablenotetext{b}
{
Intrinsic star formation rate derived from the H$\alpha$ line
luminosity (see Table~\ref{tab-LHaWHa}), applying the conversion
of \citet{Ken98} adjusted to a \citet{Cha03} IMF.
The \sfrha\ are computed assuming the best-fit extinction
$A_{V,{\rm SED}}$ from the SED modeling and the \citet{Cal00}
reddening law, and the \sfrhaoo\ are computed assuming extra
attenuation towards the \ion{H}{2} regions with
$A_{V,{\rm neb}}$ = $A_{V,{\rm SED}}/0.44$.
Uncertainties are propagated from those of the luminosities;
$3\,\sigma$ upper limits are given for sources undetected in H$\alpha$.
}
\tablenotetext{c}
{
Observed and intrinsic rest-frame UV luminosity densities,
derived from either the $B$ or $G$ band magnitude
(see Table~\ref{tab-photprop}) as described in
\S~\ref{Sub-intrinsic_prop}.
The extinction correction uses the best-fit extinction
$A_{V,{\rm SED}}$ from the SED modeling and is computed
for a rest-frame wavelength of 1500~\AA\ assuming the
\citet{Cal00} reddening law.
Uncertainties are computed from those of the observed
magnitudes, as well as of the best-fit $A_{V}$ from the
SED modeling for the intrinsic luminosity;
the $3\,\sigma$ upper limit is given for $\rm SA15-7353$,
undetected in the $B$ band.
}
\tablenotetext{d}
{
Intrinsic star formation rate derived from the rest-frame UV
luminosity density, applying the conversion of \citet{Ken98}
adjusted to a \citet{Cha03} IMF and corrected for extinction
using the best-fit extinction $A_{V,{\rm SED}}$ from the SED
modeling as for the intrinsic rest-frame UV luminosity.
Uncertainties are propagated from those of the luminosities;
the $3\,\sigma$ upper limit is given for $\rm SA15-7353$,
undetected in the $B$ band.
}
\end{deluxetable}

\clearpage

\tabletypesize{\footnotesize}
\begin{deluxetable}{lllccccccc}
\tablecolumns{10}
\tablewidth{0pt}
\tablecaption{Dynamical properties and mass estimates of the SINS H$\alpha$ sample
              \label{tab-dyn}}
\tablehead{
   \colhead{Source} & 
   \colhead{Method\,\tablenotemark{a}} & 
   \colhead{Kinemetry\,\tablenotemark{b}} & 
   \colhead{$v_{\rm obs}/2$\,\tablenotemark{c}} & 
   \colhead{$v_{\rm obs}/(2\,\sigma_{\rm int})$\,\tablenotemark{d}} & 
   \colhead{$v_{\rm rot}/\sigma_{0}$\,\tablenotemark{e}} & 
   \colhead{$v_{\rm d}$\,\tablenotemark{f}} & 
   \colhead{$M_{\rm gas}^{0}$\,\tablenotemark{g}} &
   \colhead{$M_{\rm gas}^{00}$\,\tablenotemark{g}} &
   \colhead{$M_{\rm dyn}$\,\tablenotemark{h}} \\
   \colhead{} & 
   \colhead{} & 
   \colhead{} & 
   \colhead{($\rm km\,s^{-1}$)} & 
   \colhead{} & 
   \colhead{} & 
   \colhead{($\rm km\,s^{-1}$)} & 
   \colhead{($\rm 10^{10}~M_{\odot}$)} & 
   \colhead{($\rm 10^{10}~M_{\odot}$)} & 
   \colhead{($\rm 10^{10}~M_{\odot}$)}
}
\startdata
$\rm Q1307-BM1163$  & Velocity width               & \ldots & $60 \pm 18$               & $0.39 \pm 0.12$       & \ldots                &
         $264^{+10}_{-9}$          & \ldots                    & \ldots                    & $8.8 \pm 4.1$             \\
$\rm Q1623-BX376$   & Velocity gradient $+$ width  & \ldots & $60 \pm 18$               & $0.60 \pm 0.19$       & \ldots                &
         $112 \pm 29$              & $0.64 \pm 0.22$           & $0.80 \pm 0.30$           & $1.4 \pm 0.6$             \\
$\rm Q1623-BX447$   & Kinematic modeling           & Disk   & $100 \pm 30$              & $0.69 \pm 0.22$       & \ldots                &
         $229 \pm 15$              & $1.1 \pm 0.3$             & $1.6 \pm 0.5$             & $12 \pm 1$                \\
$\rm Q1623-BX455$   & Velocity gradient $+$ width  & \ldots & $55 \pm 17$               & $0.42 \pm 0.16$       & \ldots                &
         $119 \pm 27$              & $0.95 \pm 0.28$           & $1.3 \pm 0.5$             & $2.7 \pm 1.1$             \\
$\rm Q1623-BX502$   & Velocity gradient $+$ width  & \ldots & $45 \pm 14$               & $0.62^{+0.19}_{-0.20}$& $0.8 \pm 0.2$         &
         $91^{+26}_{-25}$          & $0.52^{+0.13}_{-0.12}$    & $0.65^{+0.20}_{-0.19}$    & $0.85 \pm 0.36$           \\
$\rm Q1623-BX528$   & Velocity gradient $+$ width  & Merger & $67 \pm 20$               & $0.48^{+0.14}_{-0.15}$& \ldots                &
         $145 \pm 30$              & $1.3 \pm 0.3$             & $1.8 \pm 0.6$             & $6.0 \pm 2.1$             \\
$\rm Q1623-BX543$   & Velocity width               & \ldots & $55 \pm 17$               & $0.37 \pm 0.12$       & \ldots                &
         $257^{+38}_{-39}$         & $1.5 \pm 0.5$             & $2.3 \pm 0.9$             & $9.5^{+4.0}_{-4.1}$       \\
\enddata
\end{deluxetable}

\clearpage


\setcounter{figure}{0}

\begin{figure}[t]
\figurenum{1}
\epsscale{1.20}
\plotone{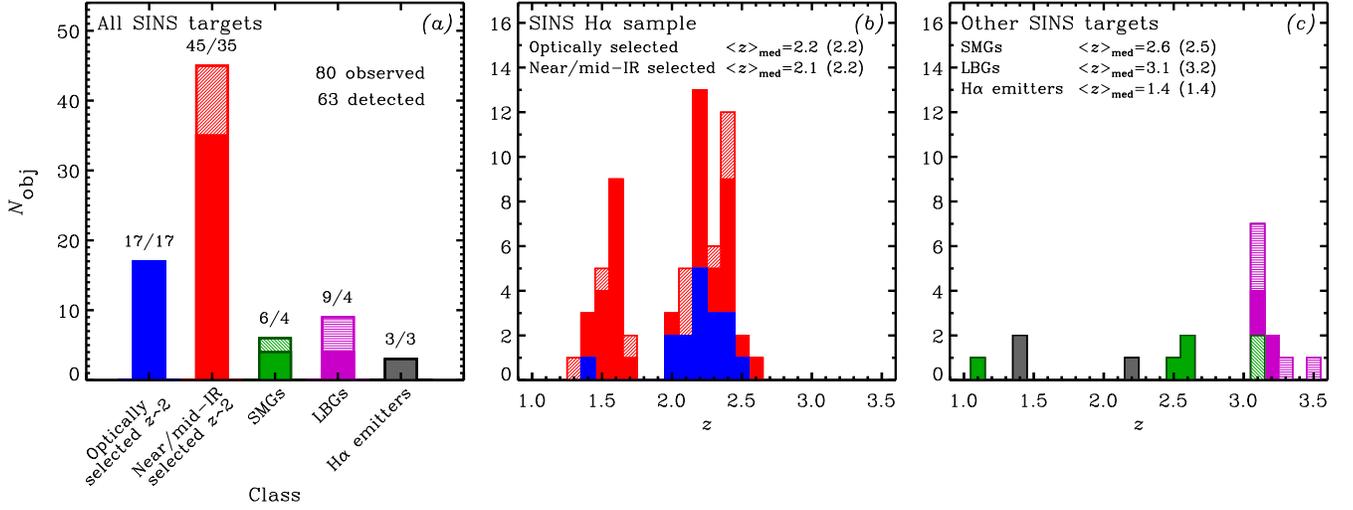}
\vspace{-0.8cm}
\caption{
\small
Distribution of the SINS galaxies as a function of class and redshift.
{\em (a)\/}
Number of sources observed (hatched histograms) and detected
(superposed solid filled histograms) for each of the galaxy
classes considered.
{\em (b)\/}
Redshift distribution of the \nsins\ optically-selected and
near-/mid-IR-selected sources spanning the range $1.3 < z < 2.6$, 
which form the ``SINS H$\alpha$ sample'' that is the focus of this paper.
{\em (c)\/}
Redshift distribution of the other sub-samples observed as part of SINS.
In panels {\em (b)\/} and {\em (c)\/}, cumulative histograms are plotted,
and different galaxy classes are shown with different colours as in
panel {\em (a)\/}; the median redshift per class is given for the
observed targets (hatched histograms) and for the detected subsets
(in parenthesis, solid-filled histograms).
The redshift distributions reflect the primary photometric selection
criteria, but are also importantly affected by the observability of
the target emission lines (H$\alpha$ or [\ion{O}{3}]\,$\lambda\,5007$)
in the near-IR atmospheric bands and between the night sky lines.
\label{fig-samples}
}
\end{figure}

\clearpage

\begin{figure}[p]
\figurenum{2}
\epsscale{0.70}
\plotone{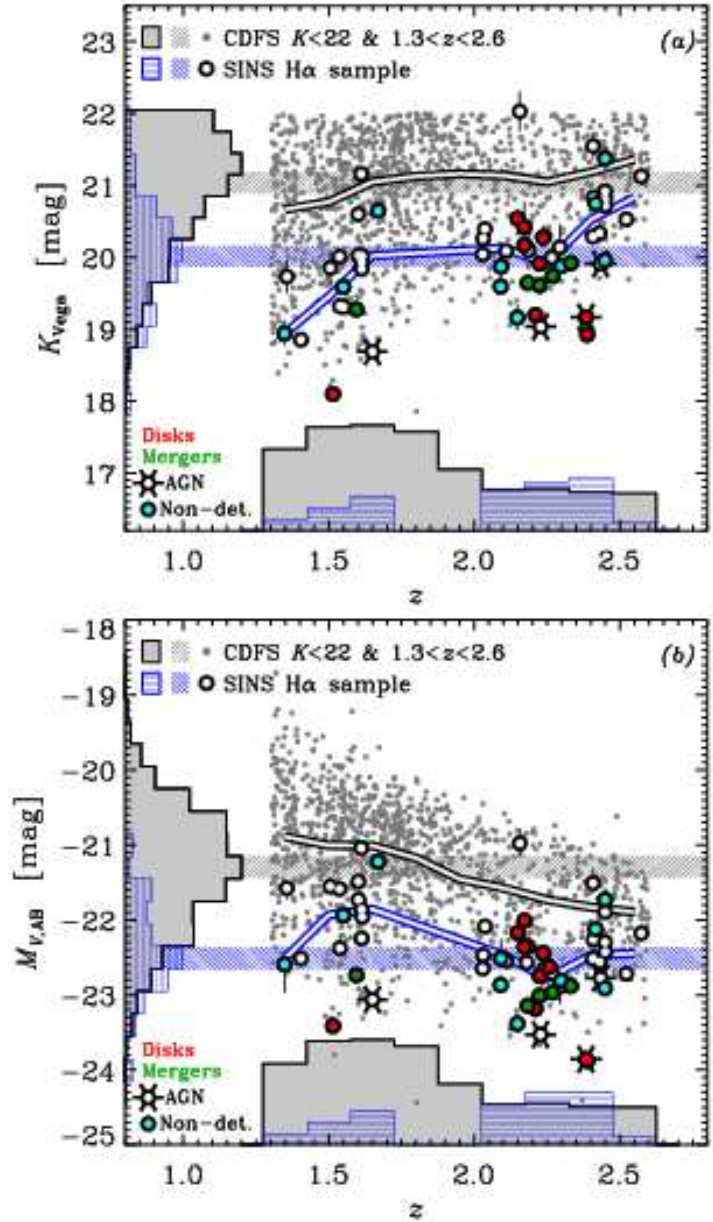}
\vspace{-0.5cm}
\caption{
\small
Redshift and magnitude distributions for the SINS H$\alpha$ sample
at $1.3 < z < 2.6$.
The properties of the SINS galaxies are compared to those of $K$-selected
galaxies from the CDFS \citep{Wuy08} in the same redshift interval and at
$K_{\rm s} < 22.0~{\rm mag}$, i.e. the magnitude of the faintest of the
SINS H$\alpha$ sample galaxies in the $K$ band.
The SINS data points are shown with large filled dots, their projected
distribution onto each axis with blue-hatched histograms, and their
median magnitudes as blue-hatched horizontal bars.
The CDFS data are plotted with small grey dots, grey filled histograms,
and grey-hatched bars.  The histograms are arbitrarily normalized.
In addition, the running median through the SINS and CDFS magnitude
distributions are overplotted as thick blue-white and black-white
lines, respectively.
The galaxies classified as disk-like and merger-like by our kinemetry
\citep{Sha08} are plotted as red- and green-filled circles.  Sources
that were known to host an AGN based on optical (rest-UV) or previous
long-slit near-IR (rest-frame optical) spectroscopy are indicated with
a 6-pointed skeletal star.  Targets that were not detected in H$\alpha$
line emission in our SINFONI data are marked as cyan-filled circles.
{\em (a)} Apparent observed $K$-band magnitude versus redshift.
{\em (b)} Absolute rest-frame $V$-band magnitude versus redshift.
The SINS sample redshift distribution is strongly bi-modal as a result
of the requirement of H$\alpha$ line observability between the near-IR
night sky lines and in spectral regions with high atmospheric transmission.
\label{fig-Kz}
}
\end{figure}

\clearpage

\begin{figure}[p]
\figurenum{3}
\epsscale{0.70}
\plotone{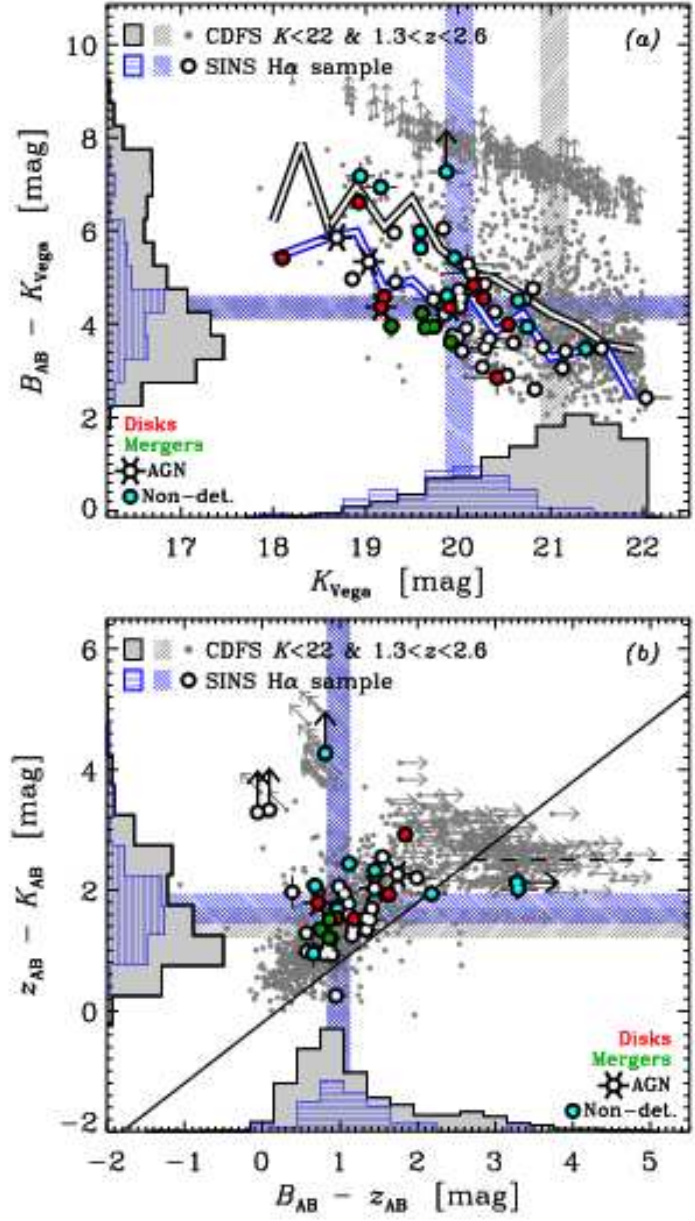}
\vspace{-0.5cm}
\caption{
\small
Colour and magnitude distributions for the SINS H$\alpha$ sample
at $1.3 < z < 2.6$ compared to those of $K$-selected galaxies from
the CDFS \citep{Wuy08} in the same redshift interval and at
$K_{\rm s} < 22.0~{\rm mag}$.
The samples, symbols used, histograms, hatched bars, and thick
lines are the same as for Figure~\ref{fig-Kz}, and as indicated
by the labels in each plot.  The histograms are arbitrarily
normalized.  Arrows correspond to $1\,\sigma$ limits from the
photometric measurements.
{\em (a)} $B - K$ versus $K$ colour-magnitude diagram, where we
have used here $G$ band as proxy for the $B$ band for the 17 \bxbm\
galaxies.
{\em (b)} $z - K$ versus $B - z$ colour diagram, where the 17 \bxbm\
galaxies are excluded because they do not have $z$-band or equivalent
photometry.
The solid diagonal line indicates the
$BzK \equiv (z-K)_{\rm AB} - (B-z)_{\rm AB} > -0.2~{\rm mag}$ colour
criterion for selecting star-forming BzK galaxies ($sBzK$),
and the dashed line indicates the $BzK < -0.2~{\rm mag}$ and
$(z-K)_{\rm AB} > 4.0~{\rm mag}$ criteria for passive BzK galaxies ($pBzK$).
The typically bluer optical to near-IR colours of the SINS sample most
likely results from the bias introduced by the mandatory optical 
spectroscopic redshift for our targets.
The distributions of the reference CDFS sample include a large contribution
from faint $z < 1.9$ galaxies, as can be seen from Figure~\ref{fig-Kz}.
\label{fig-magcol}
}
\end{figure}

\clearpage

\begin{figure}[p]
\figurenum{4}
\epsscale{1.20}
\plotone{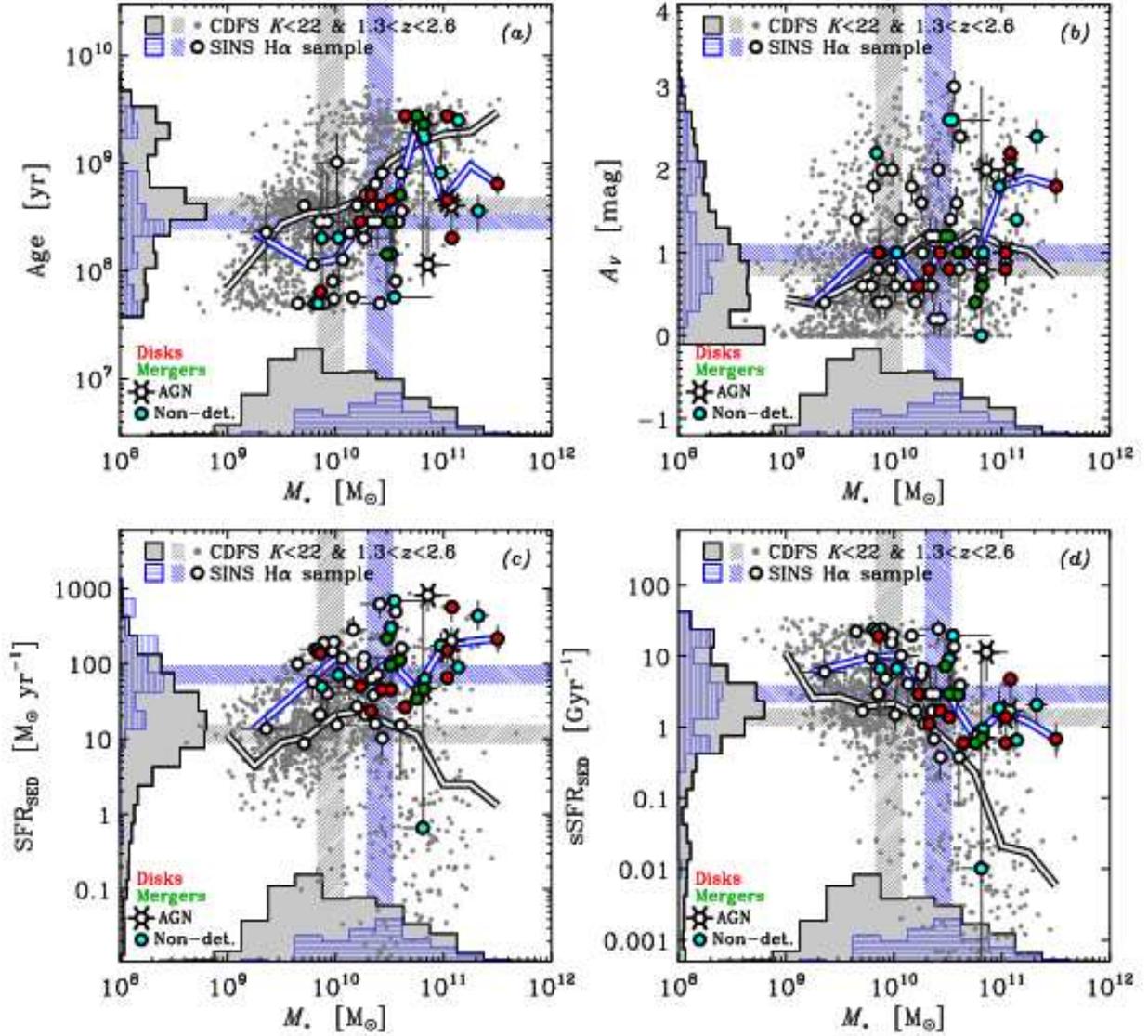}
\vspace{-2.0cm}
\caption{
\small
Properties derived from SED modeling of the SINS H$\alpha$ sample
at $1.3 < z < 2.6$ compared to those of $K$-selected galaxies from
the CDFS \citep{Wuy08} in the same redshift interval and at
$K_{\rm s} < 22.0~{\rm mag}$.
The symbols, histograms, hatched bars, and thick lines are the
same as for Figure~\ref{fig-Kz}, and as indicated by the labels
in each plot.  The histograms are arbitrarily normalized.
{\em (a)} Stellar age,
{\em (b)} visual extinction,
{\em (c)} star formation rate, and
{\em (d)} specific star formation rate, i.e., ratio of star formation
rate and stellar mass, all plotted as a function of stellar mass.
The modeling results correspond to the best fit among three
possible combinations of star formation history $+$ dust considered
(CSF$+$dust, $\tau_{\rm 300\,Myr}$$+$dust, and SSP$+$no-dust models;
see text).
The error bars (shown for the SINS galaxies) correspond to the formal
fitting 68\% confidence intervals listed in Table~\ref{tab-sedprop} (see
\S~\ref{Sect-representativeness} and Appendix~\ref{App-sedmod}).
For a given mass, the SINS galaxies probe the younger part of the
population, with higher absolute and specific star formation rates
as a result of our observational sensitivity limits for H$\alpha$,
of the $K$-brightness distribution, and of the mandatory optical
spectroscopic redshift implying a bias towards bluer galaxies.
Nevertheless, the SINS galaxies span a wide range in all properties,
and significantly larger than the differences in the median values
for the SINS and the reference $K_{\rm s} < 22.0~{\rm mag}$ CDFS
sample.
\label{fig-sedprop}
}
\end{figure}

\clearpage

\begin{figure}[t]
\figurenum{5}
\epsscale{0.65}
\plotone{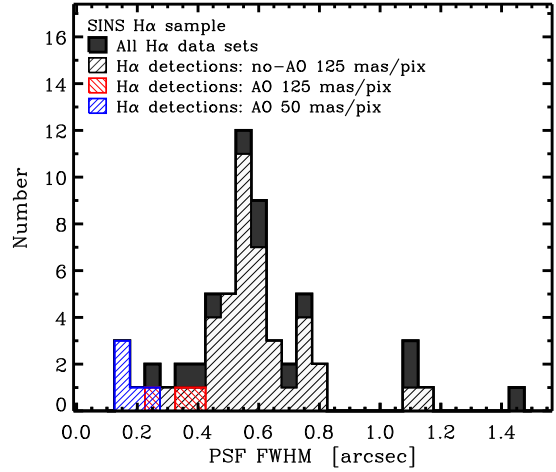}
\vspace{-0.5cm}
\caption{
\small
Distribution of the PSF FWHMs for the H$\alpha$ data sets
of the SINS H$\alpha$ sample at $1.3 < z < 2.6$).
The filled histogram shows the distribution for all data sets for which
a PSF measurement is available (including sets for undetected sources).
The hatched histograms correspond to the data sets of detected sources,
with the black-, red-, and blue-hatched ones for seeing-limited data, 
AO-assisted data at $\rm 125~mas$ pixel scale, and AO-assisted data
at $\rm 50~mas$ pixel scale, respectively.
\label{fig-psfdet}
}
\end{figure}


\begin{figure}[p]
\figurenum{6}
\epsscale{1.20}
\plotone{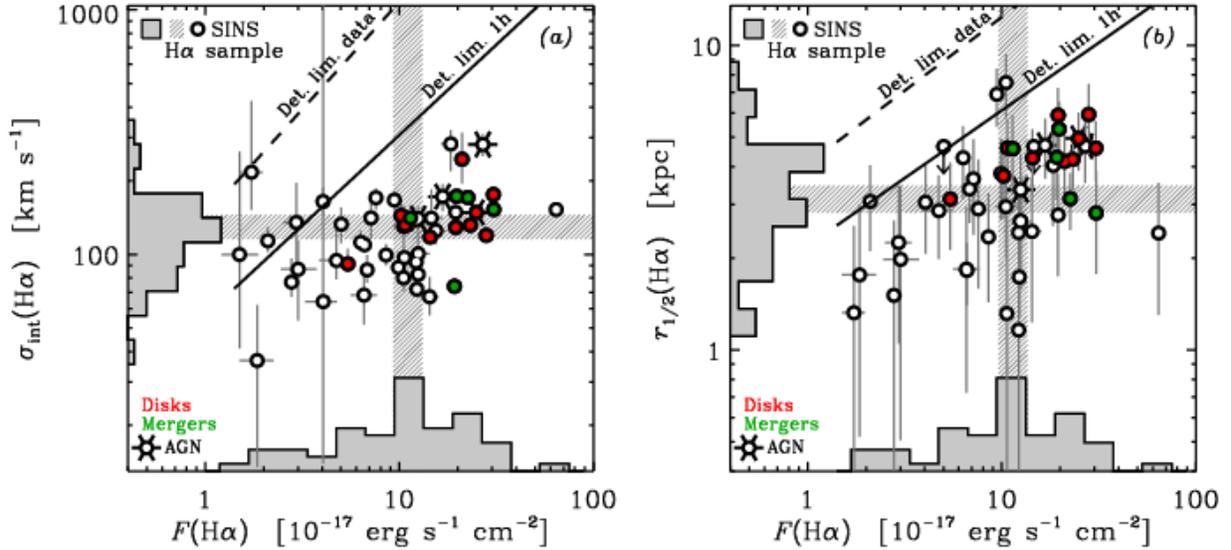}
\vspace{-0.5cm}
\caption{
\small
Integrated H$\alpha$ properties of all detected galaxies from the
SINS H$\alpha$ sample at $1.3 < z < 2.6$.
{\em (a)} Source-integrated velocity dispersion versus H$\alpha$ line flux,
derived from Gaussian profile fitting to the integrated spectrum of each
galaxy. The velocity dispersion is corrected for instrumental spectral
resolution.
{\em (b)} Half-light radius versus integrated H$\alpha$ line flux;
the half-light radius is inferred from the curve-of-growth analysis of
the H$\alpha$ flux from Gaussian profile fitting to spectra integrated
over circular apertures of increasing radius, and is corrected for the
spatial resolution of the data based on estimates of the PSF FWHM.
Error bars for the H$\alpha$ line fluxes and velocity dispersions
represent the formal best-fit uncertainties corresponding to the
68\% confidence intervals computed from Monte Carlo simulations
in the line fitting procedure.
Uncertainties on the sizes are estimated taking into account
typical variations of the effective resolution during the 
observations of the galaxies and errors from the PSF shape,
as described in \S~\ref{Sub-extrsizes}.
Upper limits on the size correspond to the observed half-light
radii when these were smaller than half the resolution element.
The grey histograms in each panel show the projected distributions along
the horizontal and vertical axes of the respective H$\alpha$ properties
(excluding limits).  The histograms are arbitrarily normalized.
The hatched bars indicate the median of the distributions (excluding limits).
The black lines indicate the line widths and sizes above which the
galaxies would be undetected (i.e., $\rm S/N < 3$ per spectral or
spatial resolution element, respectively) in the data sets with full
integration times ({\em dashed lines\/}) or normalized to an integration
time of 1\,hr ({\em solid lines\/}), keeping all other properties constant
(see \S~\ref{Sub-detlim}).
The galaxies classified as disk-like and merger-like by our kinemetry
\citep{Sha08} are plotted as red- and green-filled circles.  Sources
that were known to host an AGN based on optical (rest-UV) or previous
long-slit near-IR (rest-frame optical) spectroscopy are indicated with
a 6-pointed skeletal star.
\label{fig-HaSigmaRHa}
}
\end{figure}

\clearpage

\begin{figure}[p]
\figurenum{7}
\epsscale{1.20}
\plotone{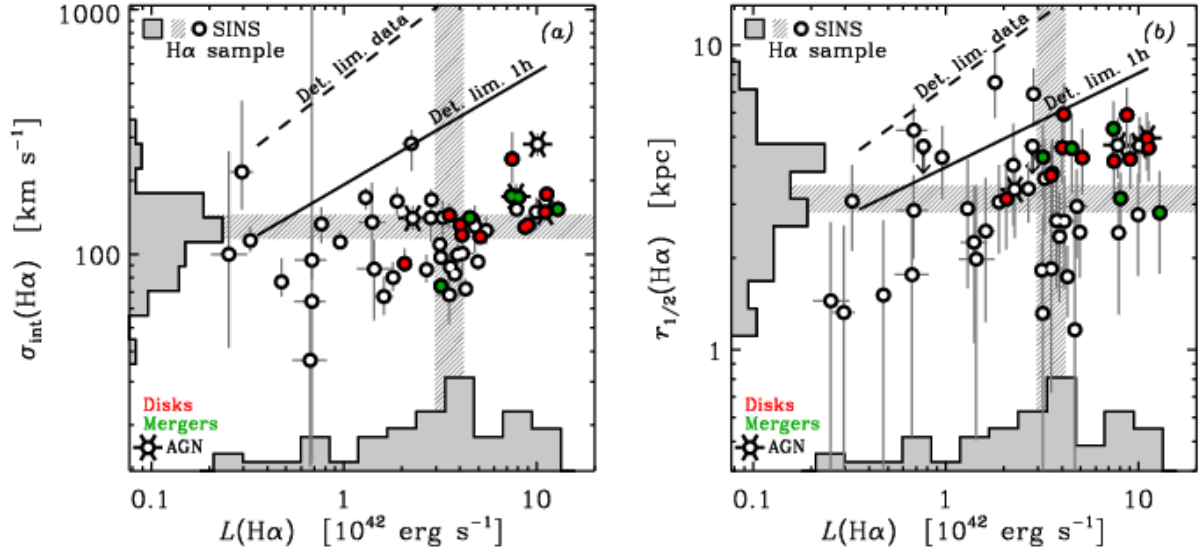}
\vspace{0.0cm}
\caption{
\small
Same as Figure~\ref{fig-HaSigmaRHa} but plotting the integrated H$\alpha$
velocity dispersion and half-light radius as a function of the integrated
H$\alpha$ luminosity instead of flux to remove the effects of redshift.
The $L({\rm H\alpha})$ is uncorrected for extinction in these plots.
\label{fig-LHaSigmaRHa}
}
\end{figure}


\begin{figure}[p]
\figurenum{8}
\epsscale{1.20}
\plotone{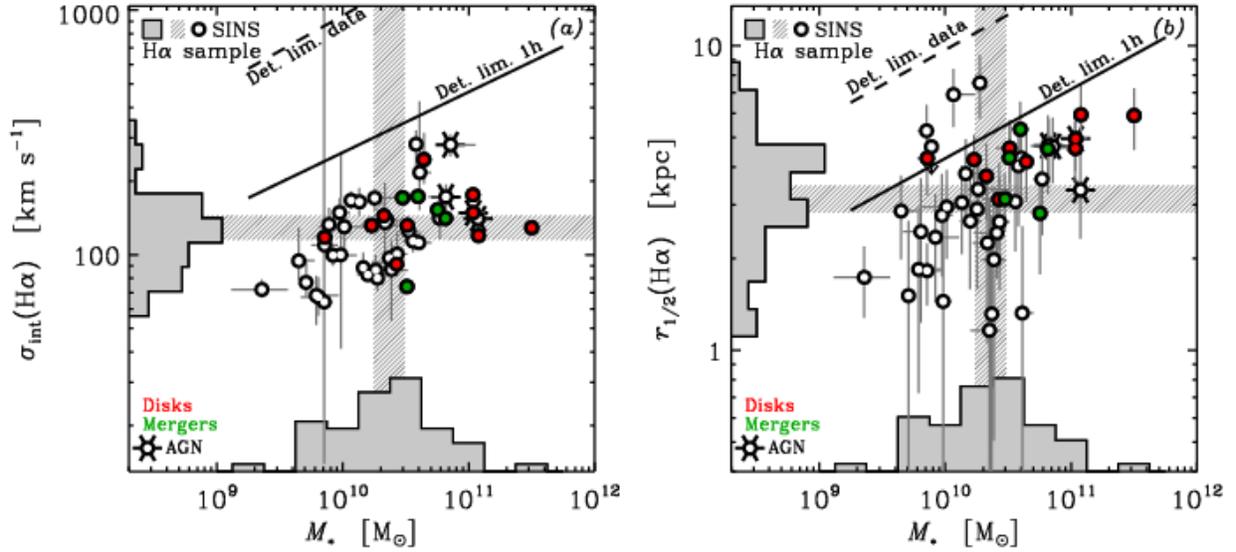}
\vspace{-0.5cm}
\caption{
\small
Same as Figure~\ref{fig-HaSigmaRHa} but plotting the integrated H$\alpha$
velocity dispersion and half-light radius as a function of the stellar
mass from the SED modeling instead of H$\alpha$ flux.
\label{fig-MstarSigmaRHa}
}
\end{figure}

\clearpage

\begin{figure}[p]
\figurenum{9}
\epsscale{0.60}
\plotone{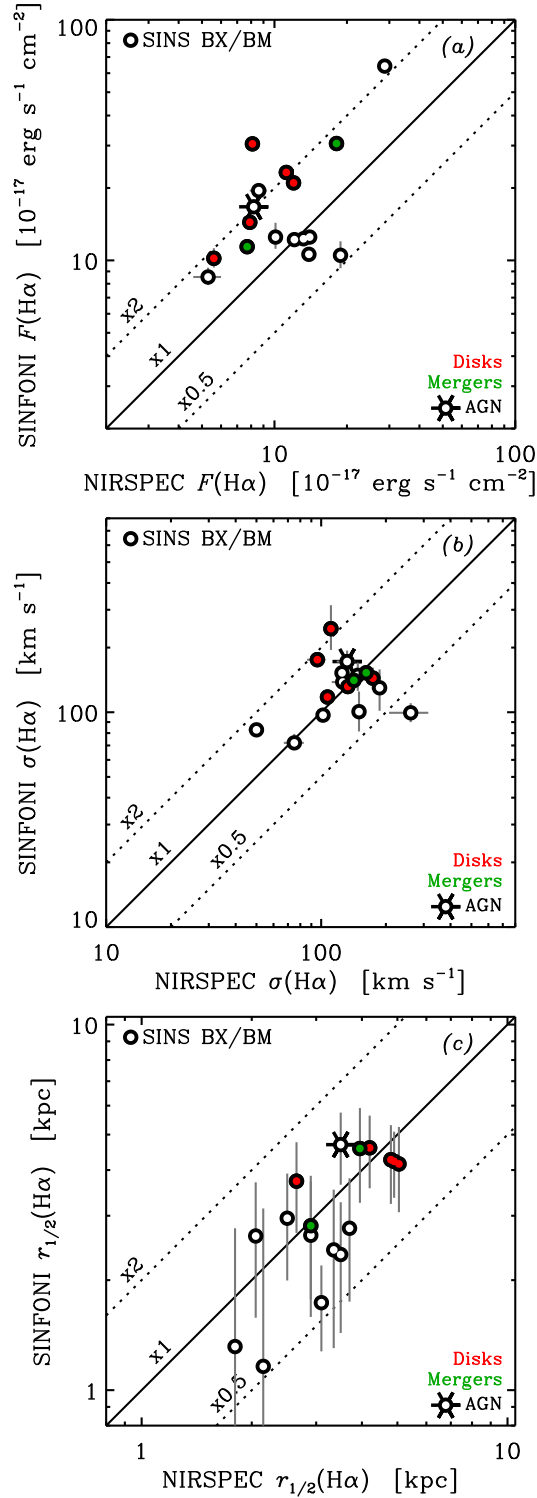}
\vspace{-0.5cm}
\caption{
\small
Comparison of the H$\alpha$ properties of the SINS \bxbm-selected
galaxies derived from our SINFONI integral field data and from the
NIRSPEC long-slit spectroscopy of \citet{Erb06b}.
{\em (a)} H$\alpha$ line flux, where the values from \citet{Erb06b}
are their observed fluxes as reported in their Table~4 and thus do
not include any correction for flux missing due to the long-slit
aperture.
{\em (b)} H$\alpha$ velocity dispersion, from the integrated spectra
and corrected for the instrumental spectral resolution.
{\em (c)} Half-light radius of the H$\alpha$ emitting regions,
corrected for the spatial resolution of the data.
The solid and dotted lines show proportionality factors of 1, 0.5, and
2 as labeled in each panel.
The SINS rotation-dominated, merger, and AGN systems are plotted with
symbols as in Figure~\ref{fig-HaSigmaRHa}.
\label{fig-sins_nirspec_comp}
}
\end{figure}

\clearpage

\begin{figure}[p]
\figurenum{10}
\epsscale{1.20}
\plotone{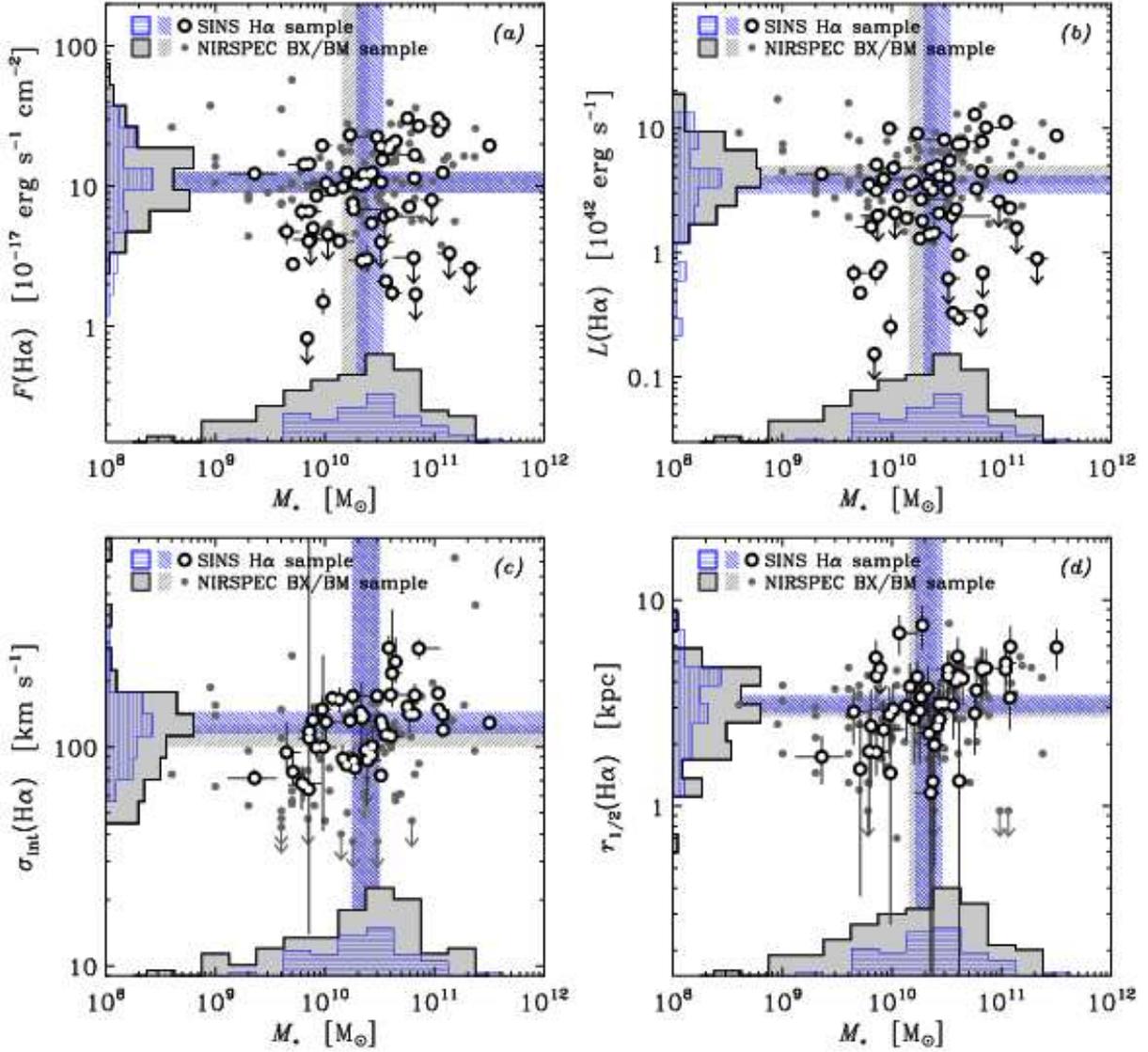}
\vspace{-2.0cm}
\caption{
\small
Integrated H$\alpha$ properties of all detected galaxies of the
SINS H$\alpha$ sample at $1.3 < z < 2.6$ compared to those of the
NIRSPEC long-slit spectroscopic sample of \bxbm\ objects at similar
redshifts by \citet{Erb06b}.  All properties are shown as a function of
stellar mass (computed for the same \citealt{Cha03} IMF for both samples).
{\em (a)} Observed H$\alpha$ line flux.
{\em (b)} Corresponding H$\alpha$ line luminosity
(uncorrected for extinction).
{\em (c)} Velocity dispersion from the integrated H$\alpha$ line width,
corrected for instrumental spectral resolution.
{\em (d)} H$\alpha$ half-light radius, corrected for the spatial
resolution of the data.
The SINS H$\alpha$ sample data are plotted with large dots, with the
distribution and median values for each quantity shown by the blue
histograms and hatched bars.
The NIRSPEC \bxbm\ sample data are plotted with the small grey dots,
grey histograms and grey hatched bars.
The histograms (arbitrarily normalized) and median values exclude
the upper limits.
Error bars correspond to $1\,\sigma$ uncertainties (not shown for
the NIRSPEC sample).  Fluxes and luminosities of undetected sources are
plotted at their $3\,\sigma$ limits in panels {\em (a)\/} and {\em (b)\/}.
Sources for which the H$\alpha$ line emission is spectrally or spatially
unresolved are shown as upper limits in panels {\em (c)\/} and {\em (d)\/}.
The H$\alpha$ fluxes for the NIRSPEC sample are taken from Table~4 of
\citet{Erb06b} and multiplied by the factor of two aperture correction
(for these long-slit data) estimated by \citet{Erb06c}.
For the SINS galaxies, the half-light radius is derived from the H$\alpha$
curve-of-growth analysis.
For the NIRSPEC sample, the half-light radius reported by \citet{Erb06b}
corresponds to half the full spatial extent of the H$\alpha$ emission
in their long-slit spectra.
\label{fig-sins_nirspec}
}
\end{figure}

\clearpage

\begin{figure}[p]
\figurenum{11}
\epsscale{1.20}
\plotone{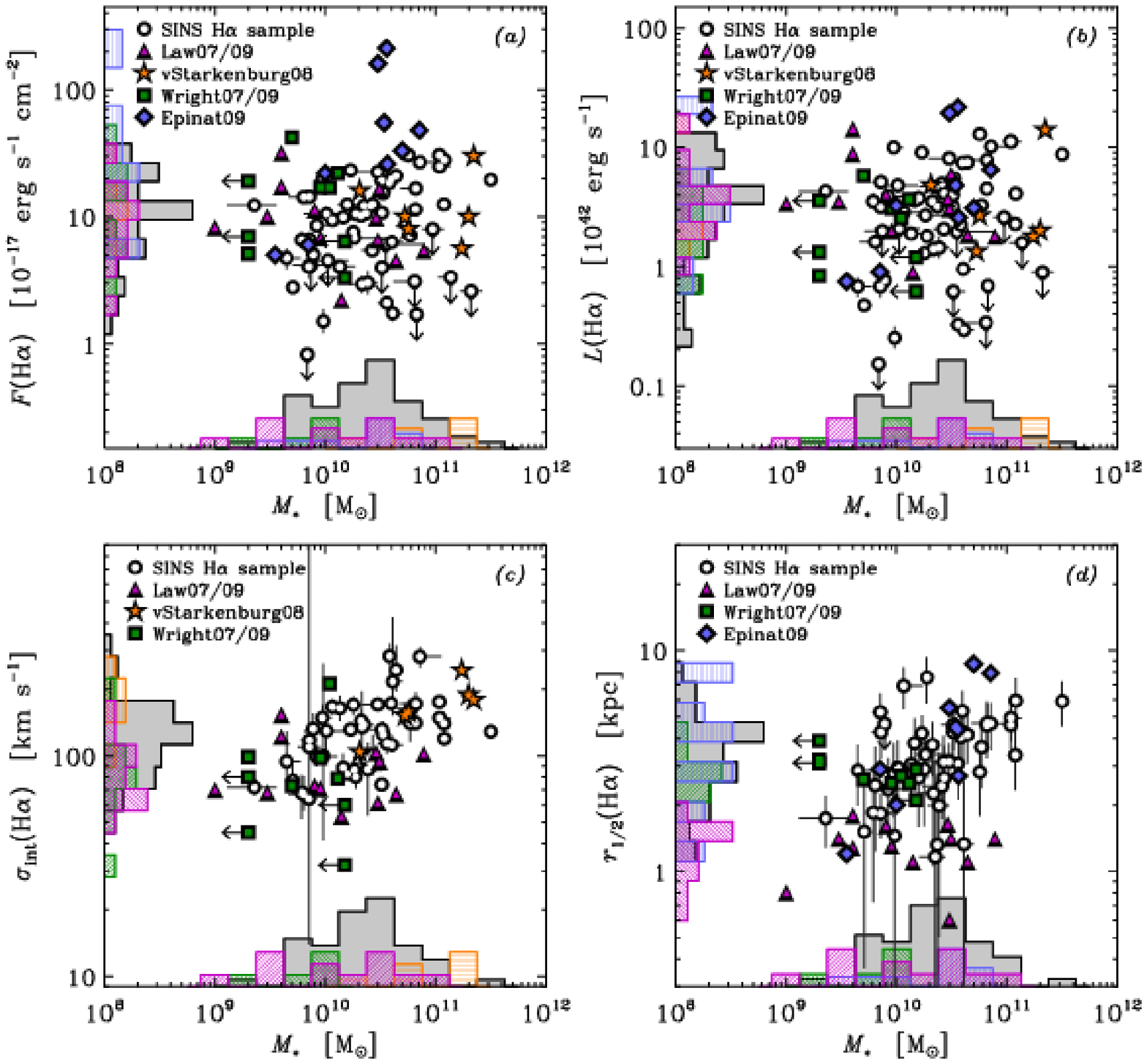}
\vspace{-2.0cm}
\caption{
\small
Same as Figure~\ref{fig-sins_nirspec} but comparing the SINS H$\alpha$
sample at $1.3 < z < 2.6$ to other samples in the same redshift interval
with published near-IR integral field spectroscopy.
The large black/white dots and grey-shaded histograms indicate the data
for the SINS H$\alpha$ sample.
Purple triangles show the objects from \citet{Law07b, Law09},
green squares those from
\citet[][with separated merger components plotted individually]{Wri07, Wri09},
observed with OSIRIS.
Orange stars show the sources from
\citet[][size measurements not available]{Sta08},
and blue lozenges those from
\citet[][integrated velocity dispersions not available]{Epi09},
observed with SINFONI.
Histograms for those samples follow the same colour scheme.
For consistent comparison with our SINS sample, the total system 
properties are used for all sources of \citet{Law07b, Law09},
including the resolved mergers.
For their two separated mergers, \citet{Wri09} give the stellar
masses of both components together but H$\alpha$ properties for
the individual components; the data of the individual components
are thus plotted here using the total $M_{\star}$ as upper limit.
Stellar masses given by \citet{Sta08} and \citet{Epi09} are
for a \citet{Sal55} IMF and have been corrected (dividing by 1.7)
to the \citet{Cha03} IMF used in the other studies.
\label{fig-sins_ifus}
}
\end{figure}

\clearpage

\begin{figure}[p]
\figurenum{12}
\epsscale{0.60}
\plotone{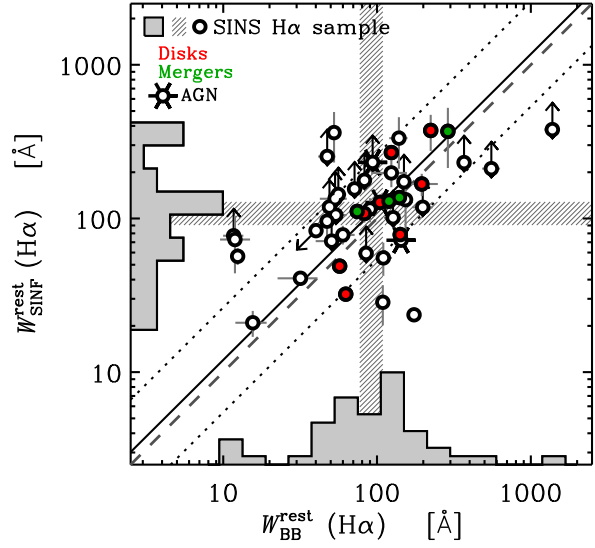}
\vspace{-0.5cm}
\caption{
\small
Comparison of H$\alpha$ equivalent width measurements
for the SINS H$\alpha$ sample.
The $W^{\rm rest}({\rm H\alpha})$ derived using the continuum flux
density estimated from the broad-band magnitudes ($K$ band for sources
at $2 < z < 2.6$ and $H$ band for those at $1.3 < z < 2$) and corrected
for the H$\alpha$ line contribution are plotted along the horizontal
axis, and those derived using the continuum flux density estimated
around H$\alpha$ in the SINFONI integrated spectra are plotted along
the vertical axis.
The dashed line indicates a one-to-one relation.  The best robust
linear bisector fit to the data (exluding limits) and the standard
deviation of the residuals are shown with the solid and dotted lines,
respectively.
The symbols used, histograms, and hatched bars are the same as for
Figure~\ref{fig-HaSigmaRHa}, and as indicated by the labels.
The histograms are arbitrarily normalized.
Error bars represent the $1\,\sigma$ uncertainties,
and non-detections are plotted at their $3\,\sigma$ limits.
\label{fig-ews_comp}
}
\end{figure}

\clearpage

\begin{figure}[p]
\figurenum{13}
\epsscale{1.20}
\plotone{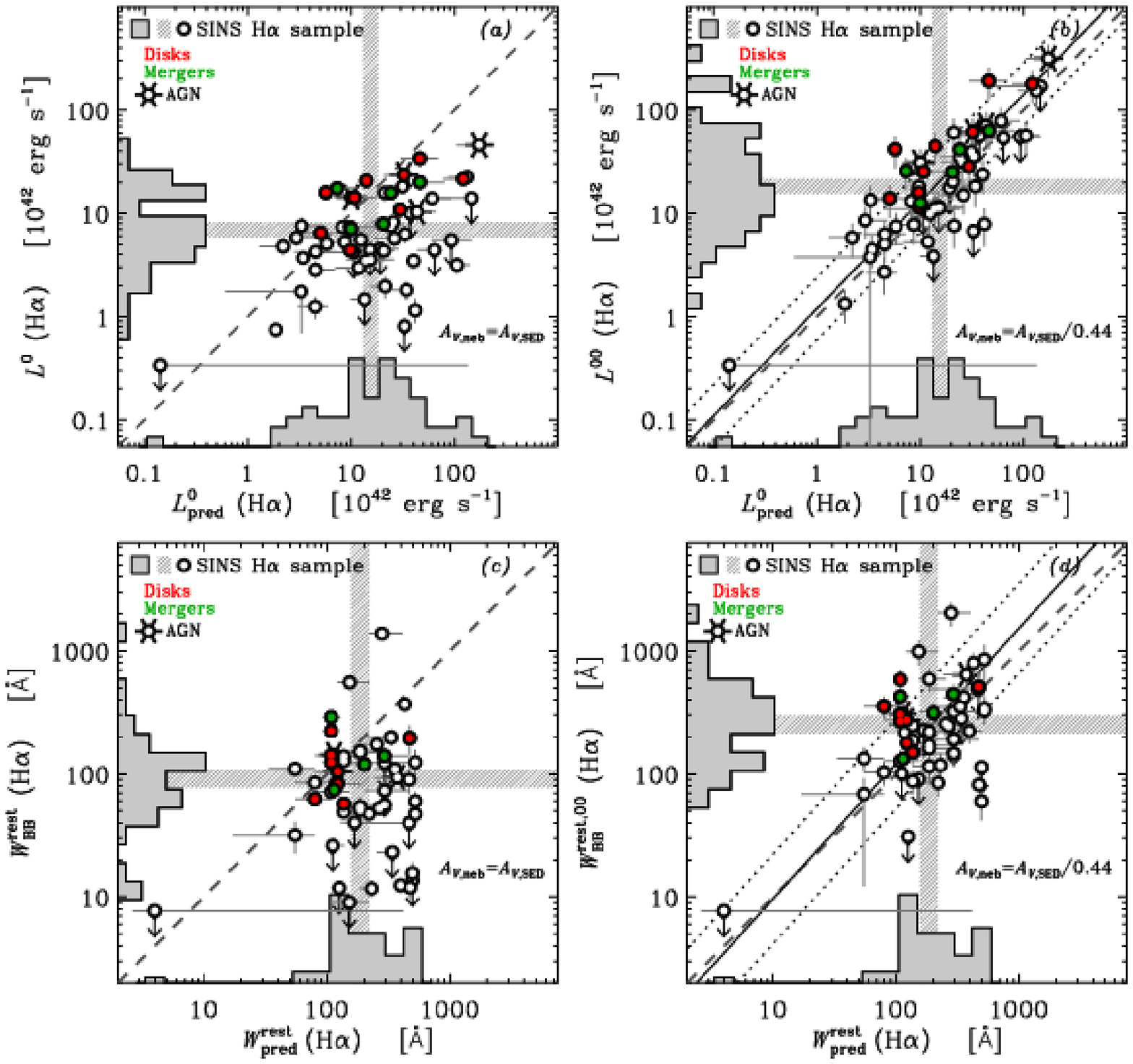}
\vspace{-2.0cm}
\caption{
\small
Comparison of the measured and predicted H$\alpha$
luminosities $L{\rm (H\alpha)}$ and rest-frame equivalent
widths $W^{\rm rest}{\rm (H\alpha)}$ for the SINS H$\alpha$ sample.
The $W^{\rm rest}_{\rm BB}({\rm H\alpha})$ are calculated from the
H$\alpha$ line flux and the broad-band magnitude, assuming a flat
$f_{\nu}$ continuum and correcting for the H$\alpha$ line contribution.
Predictions are calculated from \citet{BC03} models for the best-fit
parameters of each galaxy, assuming solar metallicity and the
\citet{Cal00} reddening law.
The symbols, histograms, and hatched bars are as for
Figure~\ref{fig-HaSigmaRHa}.
Panels {\em (a)\/} and {\em (c)\/} assume the same extinction applies
for the \ion{H}{2} regions and the stars dominating the underlying
continuum emission.
Panels {\em (b)\/} and {\em (d)\/} assume extra attenuation towards the
\ion{H}{2} regions, in an extinction-dependent manner and following the
prescription proposed by \citet{Cal00}.
The dashed line in all panels shows a one-to-one relation. 
The solid and dotted lines in panels {\em (b)\/} and {\em (d)\/}
show the robust linear bisector fit to the data and the standard
deviation of the fit residuals (excluding limits).
\label{fig-LHaEWsVSpred}
}
\end{figure}

\clearpage

\begin{figure}[p]
\figurenum{14}
\epsscale{1.20}
\plotone{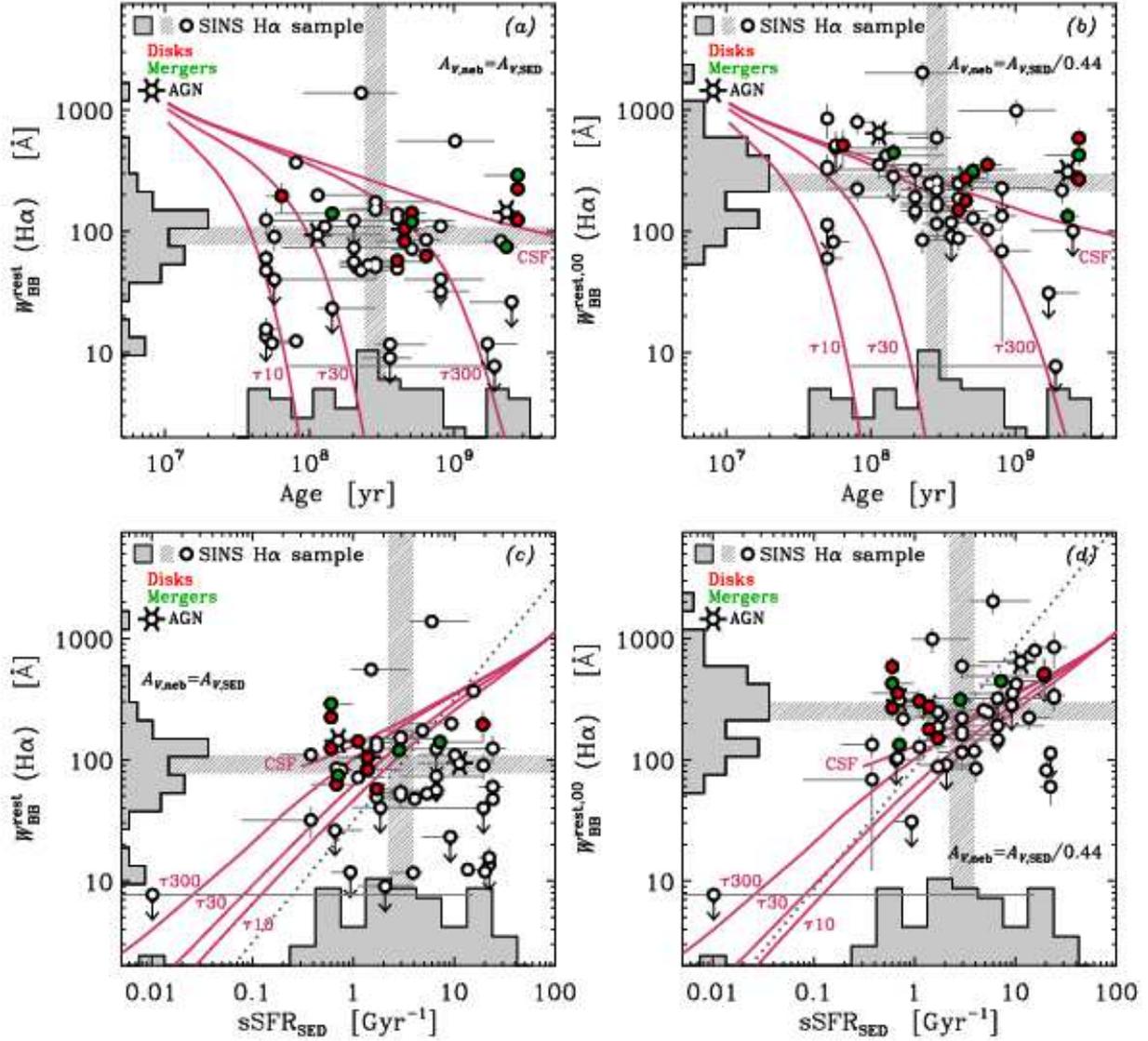}
\vspace{-2.0cm}
\caption{
\small
Rest-frame H$\alpha$ equivalent widths as a function of SED-derived
properties for the SINS H$\alpha$ sample.
{\em (a)\/} Equivalent widths uncorrected for extra dust attenuation
towards the \ion{H}{2} regions relative to the stars, as a function
of best-fit derived stellar age.
{\em (b)\/} Same as {\em a\/} but with correction for extra dust
attenuation towards the \ion{H}{2} regions.
{\em (c), (d)\/} Same as {\em (a), (b)\/} but as a function of
specific star formation rate.
The symbols, histograms, and hatched bars are as for
Figure~\ref{fig-HaSigmaRHa}.
The dotted line in panels {\em (c)\/} and {\em (d)\/} shows direct
proportionality, passing through the median values of the quantities
plotted on each axis.
Purple curves show models computed for \citet{BC03} models with solar
metallicity for different star formation histories as labeled: CSF for
constant star formation rate, $\tau\,300$, $\tau\,30$, and $\tau\,10$
for exponentially declining star formation rates with $e$-folding
timescales of 300, 30, and 10~Myr.
\label{fig-EWsSEDprop}
}
\end{figure}

\clearpage

\begin{figure}[p]
\figurenum{15}
\epsscale{1.20}
\plotone{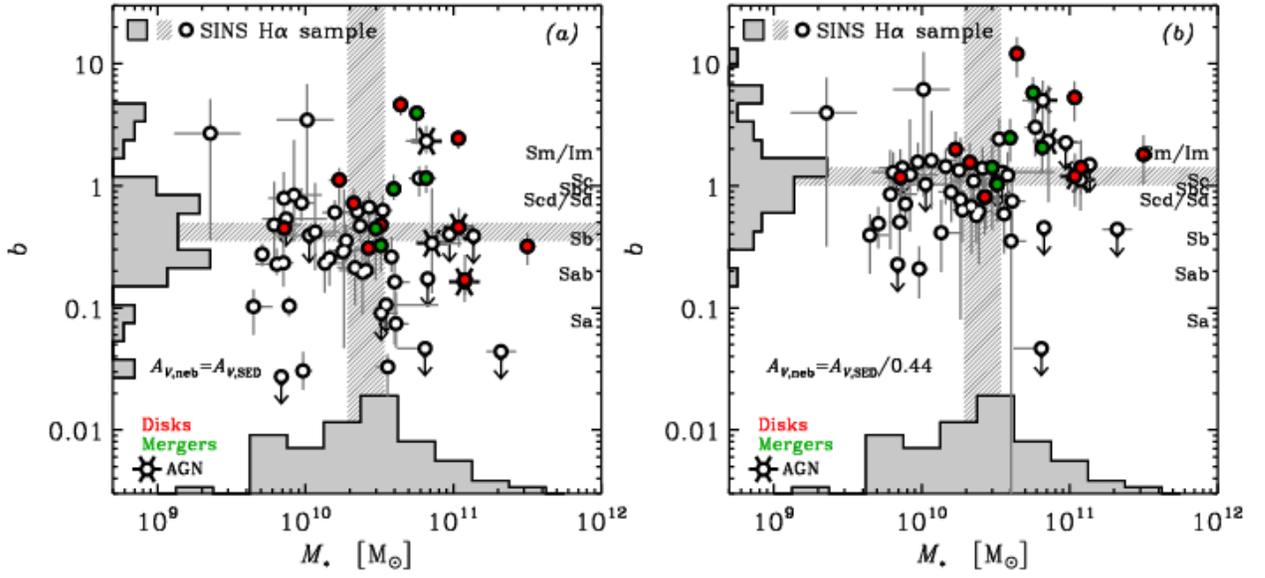}
\vspace{-0.5cm}
\caption{
\small
Scalo birthrate parameter $b$ for our SINS H$\alpha$ sample galaxies
as a function of stellar mass $M_{\star}$.
The $b$ parameter represents the ratio of current to past-averaged
star formation rate, which we computed using the SFR from H$\alpha$
with the conversion from \citet{Ken98} adjusted to a \citet{Cha03} IMF,
and the ratio of stellar mass to stellar age from the SED modeling,
respectively.
{\em (a)\/} $b$ parameter calculated using the $\rm SFR(H\alpha)$
with correction assuming the same extinction applies for the
\ion{H}{2} regions as for the stars.
{\em (b)\/} Same as {\em (a)\/} but with correction for extra dust
attenuation towards the \ion{H}{2} regions.
The symbols, histograms, and hatched bars are as for
Figure~\ref{fig-HaSigmaRHa}.
Median values inferred for a sample of nearby disk galaxies by \citet{Ken94},
as a function of Hubble type, are labeled on the right side of the plots.
\label{fig-bMstar}
}
\end{figure}

\clearpage

\begin{figure}[p]
\figurenum{16}
\epsscale{1.20}
\plotone{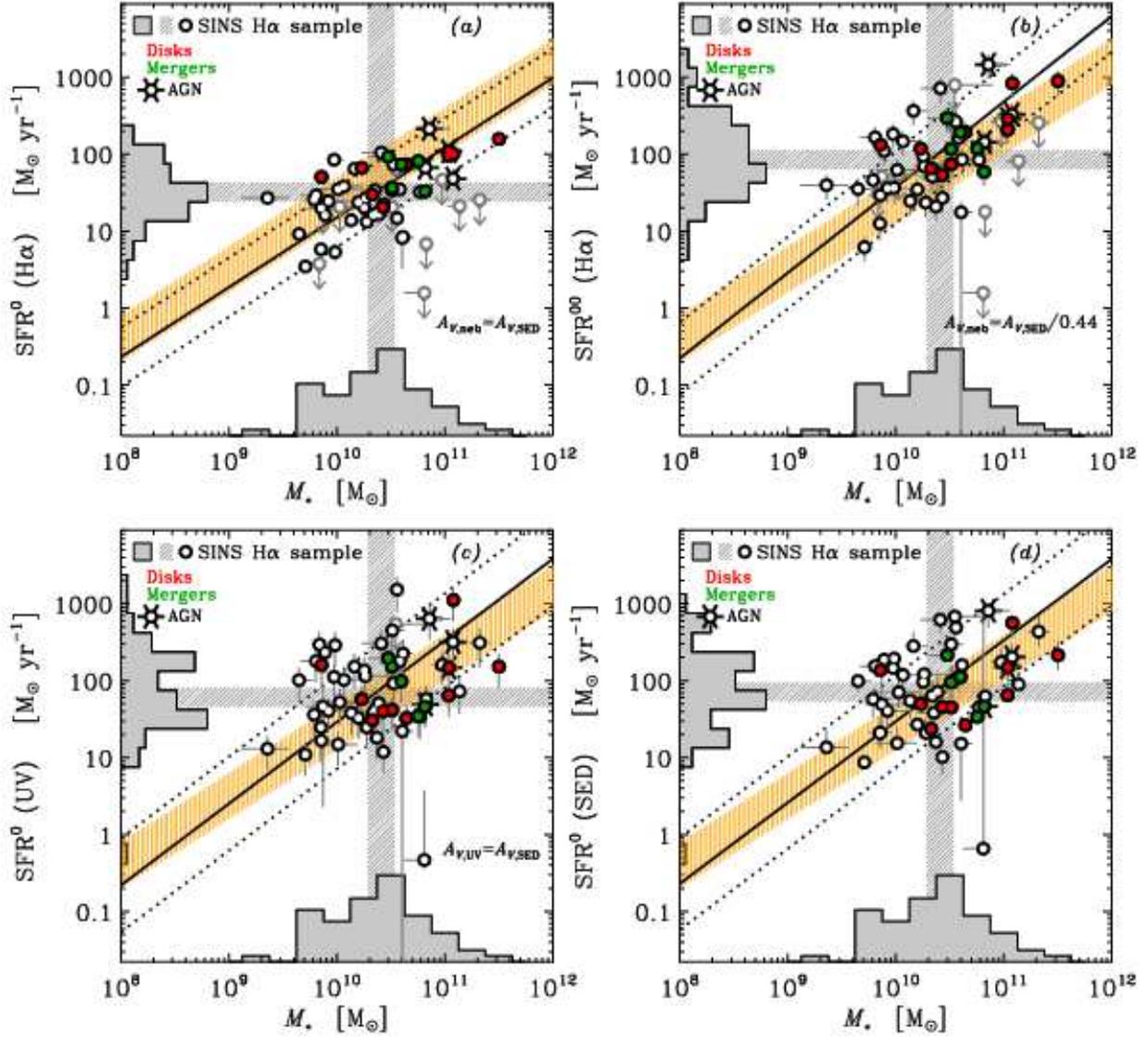}
\vspace{-2.0cm}
\caption{
\small
Comparison of stellar mass -- star formation rate relations
obtained from different star formation rate indicators for
the SINS H$\alpha$ sample.
{\em (a)\/} SFR from H$\alpha$ line luminosity, corrected for
the best-fit extinction derived from the SED modeling and using
the conversion from \citet{Ken98} adjusted to a \citet{Cha03} IMF,
versus $M_{\star}$.
{\em (b)\/} Same as {\em (a)\/} but applying the extra attenuation
correction towards the \ion{H}{2} regions following \citet{Cal00}.
{\em (c)\/} SFR from the rest-frame UV luminosity, corrected for
the best-fit extinction derived from the SED modeling and using
the conversion from \citet{Ken98} adjusted to a \citet{Cha03} IMF,
versus $M_{\star}$.  The rest-frame UV luminosity is computed from the
$G$ (all \bxbm\ sources) or $B$ band photometry (all other sources),
probing the rest-frame $\sim 1200 - 2100$~\AA\ continuum emission for
the redshift interval spanned by our sample.
{\em (d)\/} SFR derived from the SED modeling versus $M_{\star}$.
Solid and dotted lines show the robust linear bisector fit to
the data and the standard deviation of the fit residuals.
The orange hatched bar indicates the slope and standard deviation
of the relation derived at $z \sim 2$ by \citet{Dad07}, for
reference.
\label{fig-SFRallMstar}
}
\end{figure}

\clearpage

\hphantom{hhh}
\vspace{3.0cm}

\begin{figure}[!h]
\figurenum{17}
\epsscale{1.12}
\plotone{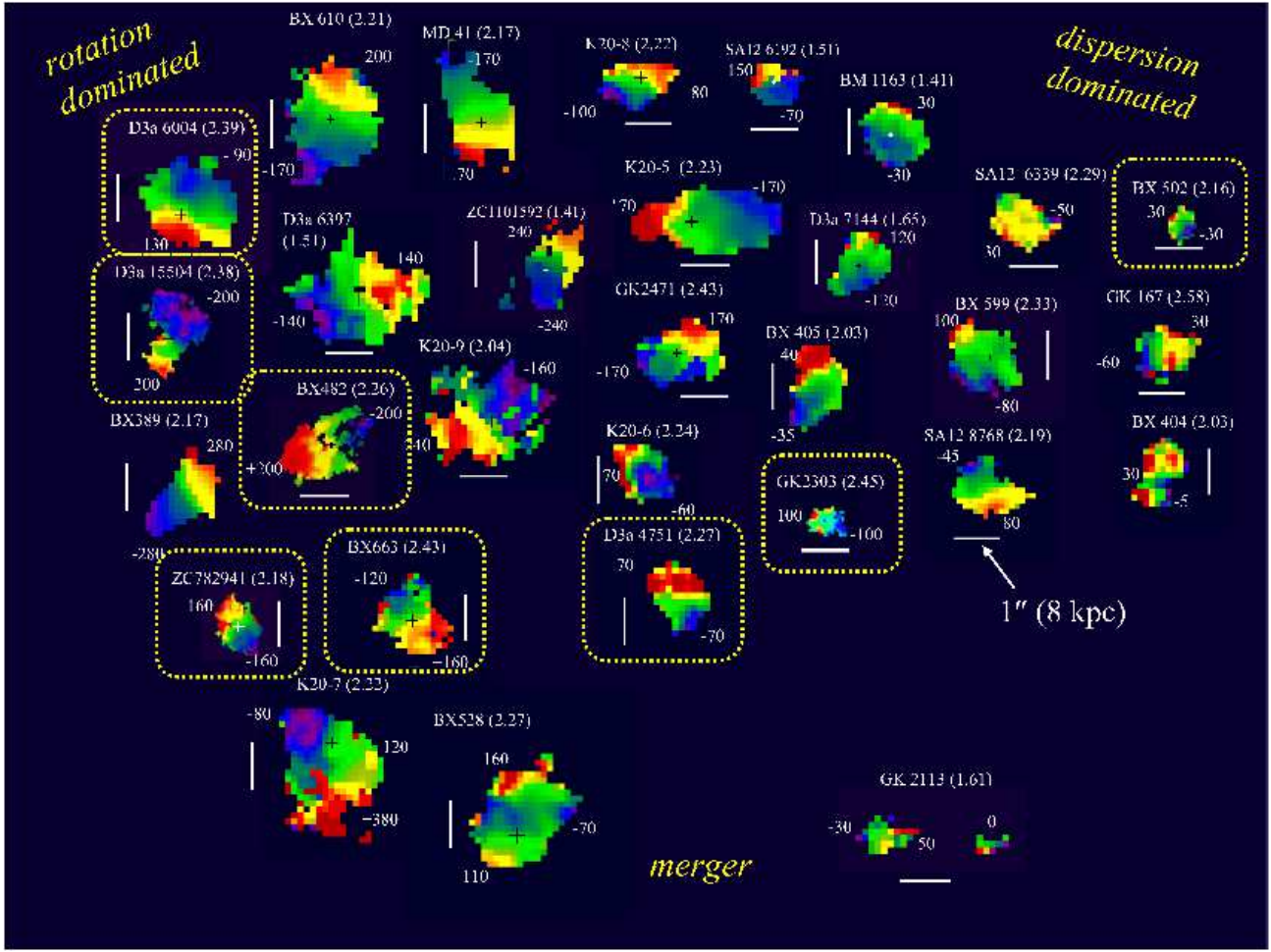}
\vspace{1.0cm}
\caption{
\small
Velocity fields for 30 of the \nsins\ galaxies of the SINS H$\alpha$
sample.  The velocity fields correspond to that derived from the
H$\alpha$ line emission as described in \S~\ref{Sub-extrmaps}
(the exception is $\rm K20-ID5$ for which it was obtained from
the [\ion{O}{3}]\,$\lambda\,5007$ line instead).  
The colour-coding is such that blue to red colours correspond to the
blueshifted to redshifted line emission with respect to the systemic
velocity.  The minimum and maximum relative velocities are labeled
for each galaxy (in $\rm km\,s^{-1}$).
All sources are shown on the same angular scale; the white bars
correspond to $1^{\prime\prime}$, or about $\rm 8~kpc$ at $z = 2$.
The galaxies are approximately sorted from left to right according to
whether their kinematics are rotation-dominated or dispersion-dominated,
and from top to bottom according to whether they are disk-like
or merger-like as quantified by our kinemetry \citep[][]{Sha08}.
Galaxies observed with the aid of adaptive optics (both at the
50 and $\rm 125~mas\,pixel^{-1}$ scales) are indicated by the
yellow rounded rectangles.
\label{fig-vfs}
}
\end{figure}

\clearpage

\begin{figure}[p]
\figurenum{18}
\epsscale{0.60}
\plotone{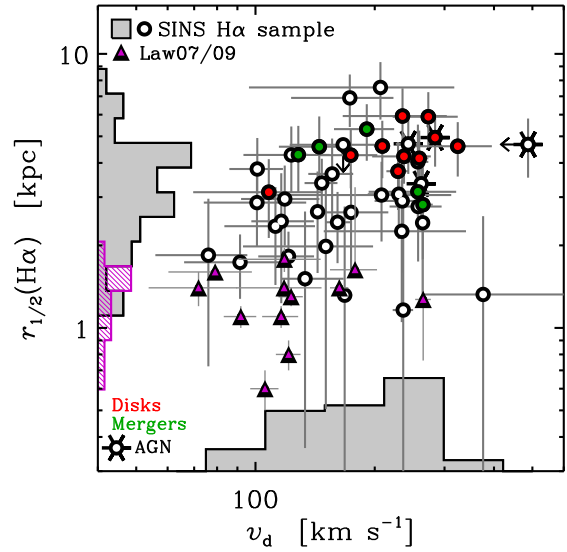}
\vspace{-0.5cm}
\caption{
\small
Velocity-size diagram for $z \sim 2$ star-forming galaxies.
The SINS H$\alpha$ sample galaxies observed with SINFONI
(large circles) are combined with the galaxies observed with OSIRIS by
\citet[][purple triangles, excluding the one at $z = 3.32$]{Law09}.
The circular velocity $v_{\rm d}$ plotted along the horizontal axis
is derived as explained in \S~\ref{Sub-dynstate}, and corrrected
for inclination where appropriate.
The size is taken as the H$\alpha$ half-light radius.
Error bars represent $1\,\sigma$ uncertainties, propagated
analytically from the primary measurements.
Upper limits on the size correspond to the observed half-light
radii when these were smaller than half the resolution element.
Grey and purple histograms (arbitrarily normalized) show the
projected distributions along each axis of the SINS and
\citeauthor{Law09} samples, respectively.
The galaxies classified as disk-like and merger-like by our kinemetry
\citep{Sha08} are plotted as red- and green-filled circles.  Sources
that were known to host an AGN based on optical (rest-UV) or previous
long-slit near-IR (rest-frame optical) spectroscopy are indicated with
a 6-pointed skeletal star.
\label{fig-velsize}
}
\end{figure}

\clearpage

\begin{figure}[p]
\figurenum{19}
\epsscale{1.20}
\plotone{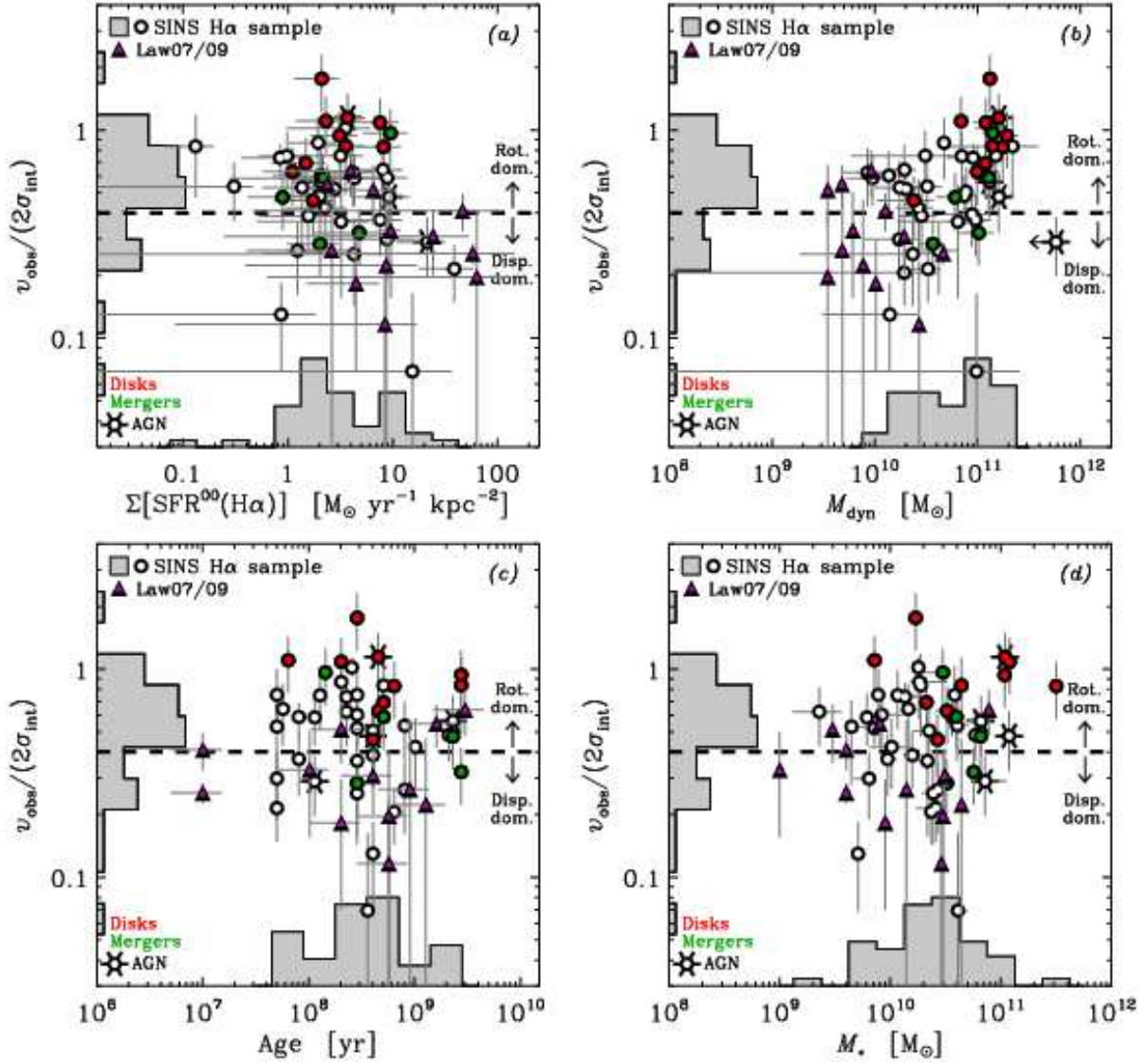}
\vspace{-2.0cm}
\caption{
\small
Kinematic ratio of half the observed velocity gradient to the
integrated velocity line width $v_{\rm obs}/(2\,\sigma_{\rm int})$ of
$z \sim 2$ star-forming galaxies.  As for Figure~\ref{fig-velsize},
the SINS H$\alpha$ sample observed with SINFONI is combined with
that of \citet{Law09} observed with OSIRIS, and the same symbols
and colour-coding are used for the data points and histograms.
{\em (a)\/} $v_{\rm obs}/(2\,\sigma_{\rm int})$ as a function of star
formation rate per unit area, taking $\rm SFR^{00}(H\alpha)$ within the
half-light radius $r_{1/2}({\rm H\alpha})$.  The extinction correction
involved in deriving the SFR from H$\alpha$ used here assumes extra
attenuation towards the \ion{H}{2} regions with
$A_{V,\,{\rm neb}} = A_{V,\,{\rm SED}} / 0.44$
but the trend remains qualitatively the same without this
extra attenuation.
{\em (b)\/} $v_{\rm obs}/(2\,\sigma_{\rm int})$ as a function of
dynamical mass derived as explained in \S~\ref{Sub-massfractions}.
{\em (c)\/} $v_{\rm obs}/(2\,\sigma_{\rm int})$ as a function of
stellar age from the SED modeling.
{\em (d)\/} $v_{\rm obs}/(2\,\sigma_{\rm int})$ as a function of
stellar mass from the SED modeling.
All error bars represent $1\,\sigma$ uncertainties, propagated
analytically from the primary measurements as appropriate.
Our working criterion to discriminate between sources with
rotation- and dispersion-dominated kinematics at
$v_{\rm obs}/(2\,\sigma_{\rm int}) = 0.4$
(see \S~\ref{Sub-dynstate}) is shown by the
dashed horizontal line.
\label{fig-vcsig0}
}
\end{figure}

\clearpage

\begin{figure}[p]
\figurenum{20}
\epsscale{0.60}
\plotone{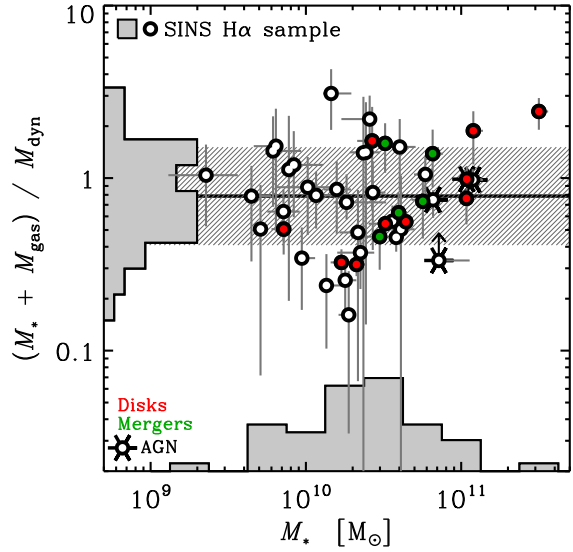}
\vspace{-0.9cm}
\caption{
\small
Baryonic mass fraction from our SINS H$\alpha$ sample galaxies.
Symbols and histograms are as for Figure~\ref{fig-velsize};
the thick solid line shows the median value and the hatched
horizontal bar shows the standard deviation of the data about
the median.
The stellar masses are derived from the SED modeling,
the gas masses are computed from the H$\alpha$ star formation rates
per unit area within the half-light radius and the Schmidt-Kennicutt
relation as obtained by \citet{Bou07}, and the dynamical masses are
inferred from the observed kinematics, as explained in the text
(\S~\ref{Sub-massfractions}).
The data shown in the plot use gas mass estimates based on the
\sfrhaoo's computed assuming extra attenuation towards the \ion{H}{2}
regions relative to the stars; without this extra attenuation, the
baryonic mass fractions decrease by $\sim 10\%$.
\label{fig-fbar}
}
\end{figure}


\begin{figure}[p]
\figurenum{21}
\epsscale{0.90}
\plotone{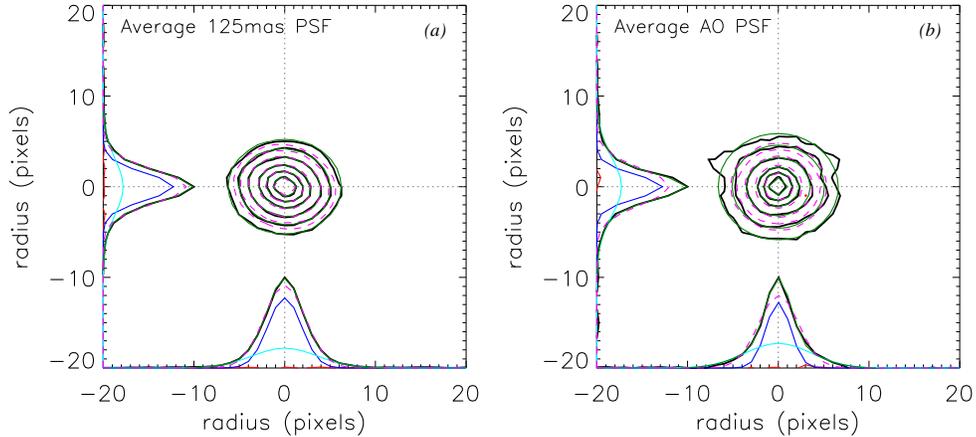}
\vspace{-0.1cm}
\caption{
\small
Profiles of the averaged PSFs for the SINFONI H$\alpha$ data sets
of the SINS H$\alpha$ sample.
{\em (a)} Average PSF for the seeing-limited data at the
$\rm 125~mas\,pixel^{-1}$ scale.
{\em (b)} Average PSF for the AO-assisted observations at the
$\rm 50~mas\,pixel^{-1}$ scale.
The images of the effective PSFs for the reduced and combined OBs
for each galaxy have been averaged together after normalizing to a
peak value of unity.
The PSFs are shown as contour plots and the profiles are projected
onto the vertical and horizontal axes (i.e., in declination and 
right-ascension for reduced SINFONI cubes and extracted images). 
The contours are at 5\%, 10\%, 20\%, 40\%, 60\%, and 80\% of the
peak flux of the PSF.
The data from the average PSFs are plotted with the black solid line.
The green line corresponds to the best two-component Gaussian fits,
with profiles of the narrow and broad components plotted individually
as blue and cyan lines.  The red line shows the residuals from this
two-component Gaussian fit.  The dashed pink line is the best fit 
with a single Gaussian profile.  Both single- and two-components
fits have elliptical Gaussian profiles.
\label{fig-psfs}
}
\end{figure}

\clearpage

\begin{figure}[p]
\figurenum{22}
\epsscale{1.15}
\plotone{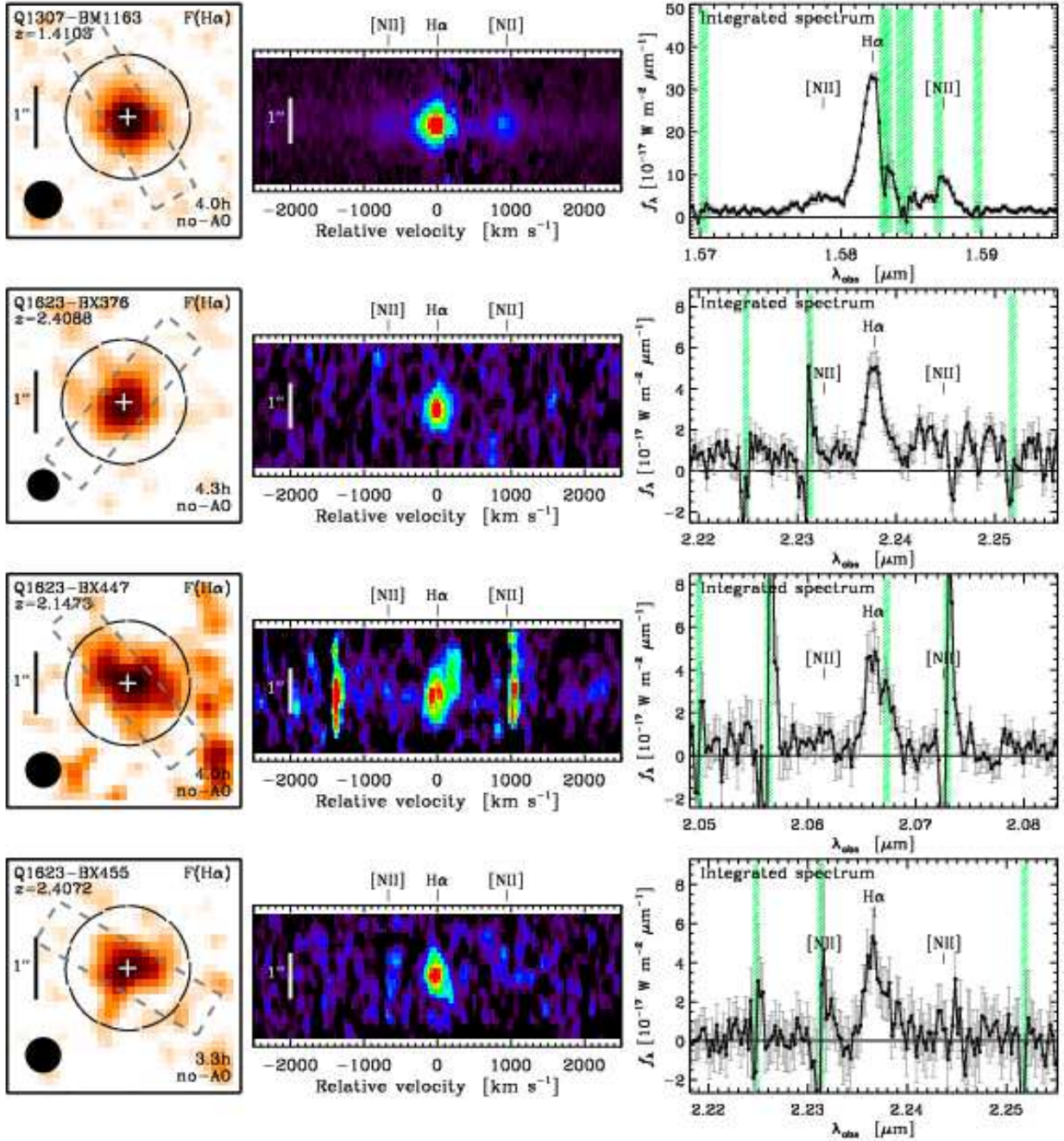}
\vspace{-5.0cm}
\caption{
\small
H$\alpha$ line maps, position-velocity diagrams, and integrated spectra
of the SINS H$\alpha$ sample.
In this figure, four of the \bxbm\ sources taken from the NIRSPEC
long-slit sample of \citet{Erb06b} are shown.  The remaining SINS
galaxies are presented in the following Figures~\ref{fig-maps_bmbx2}
through \ref{fig-maps_gdds2}.
For sources that were observed in seeing-limited and AO modes, the
maps, diagrams, and spectra from both data sets are shown successively.
{\em Left panels:\/}
Velocity-integrated flux extracted at each pixel position.
The total on-source integration time and whether the observations were
carried out in seeing-limited mode (``no-AO'') or with adaptive optics
(``NGS-AO'' and ``LGS-AO'' for Natural or Laser Guide Star) are given
at the bottom right of each panel.
The colour coding scales linearly with flux from white to black for
the minimum to maximum levels displayed (varying for each galaxy).
The spatial resolution is represented by the filled circle at the bottom
left (with diameter corresponding to the FWHM of the effective PSF, which
includes the spatial 2-3 pixel median filtering applied in extracting the
maps).
The angular scale is indicated by the vertical bars on the left.
The dashed rectangle and solid circle overlaid on each map show the
synthetic slit used to extract the position-velocity diagram and the
aperture used to extract the integrated spectrum, respectively.
In all maps, north is up and east is to the left.
{\em Middle panels:\/}
Position-velocity diagrams, obtained by integrating the flux spatially
perpendicular to the synthetic slit shown on the H$\alpha$ maps.
The horizontal axis corresponds to the velocity relative to the systemic
velocity, taken as the redshift derived from the integrated spectrum.
The vertical axis corresponding to the spatial position along the synthetic
slit, with bottom to top running from the south to the north end of the slit
and the angular scale indicated by the vertical bars on the left.
The colours scale linearly from dark blue to red with increasing flux
(for each galaxy, the same minimum and maximum levels are used as for
the line maps).
{\em Right panels:\/}
Integrated spectrum taken in the circular aperture shown on the maps.
The wavelength range corresponds to the same velocity range as for the
position-velocity diagrams ($\rm \pm 2500~km\,s^{-1}$ around H$\alpha$).
The error bars show the $1\sigma$ uncertainties derived from the noise
properties of each data set, and include the scaling with aperture size
following the model described in Appendix~\ref{App-noise}, which accounts
for the fact that the effective noise is not purely Gaussian.
Vertical green hatched bars show the locations of bright night sky lines
that can lead to significant residuals, with width of the bars corresponding
to the FWHM of the effective spectral resolution of the data.
\label{fig-maps_bmbx1}
}
\end{figure}

\clearpage

\begin{figure}[p]
\figurenum{23}
\epsscale{1.15}
\plotone{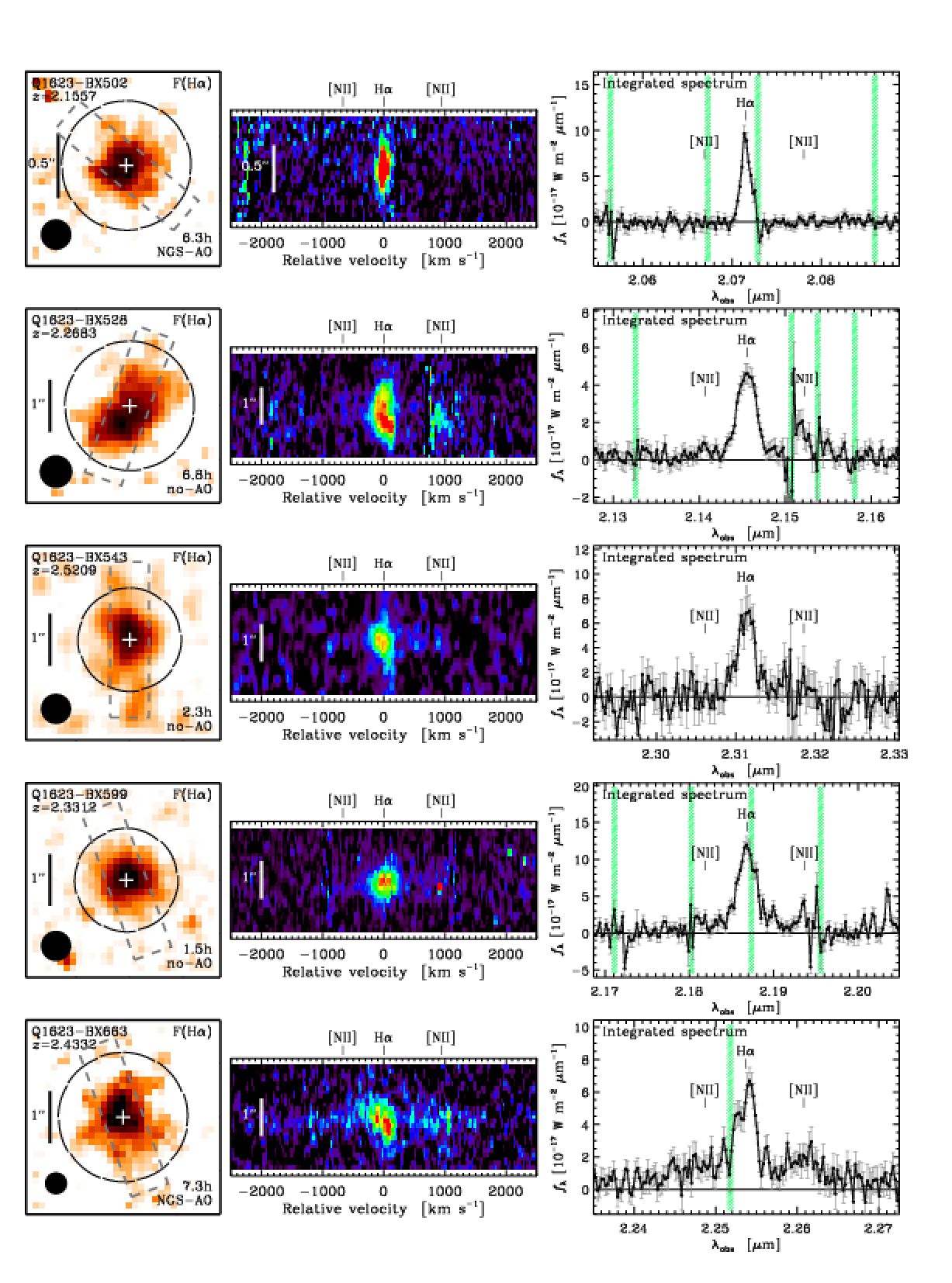}
\vspace{-0.0cm}
\caption{
\small
Same as Figure~\ref{fig-maps_bmbx1} for \bxbm\ galaxies
of the SINS H$\alpha$ sample taken from the NIRSPEC long-slit
sample of \citet{Erb06b}.
\label{fig-maps_bmbx2}
}
\end{figure}

\clearpage

\begin{figure}[p]
\figurenum{24}
\epsscale{1.15}
\plotone{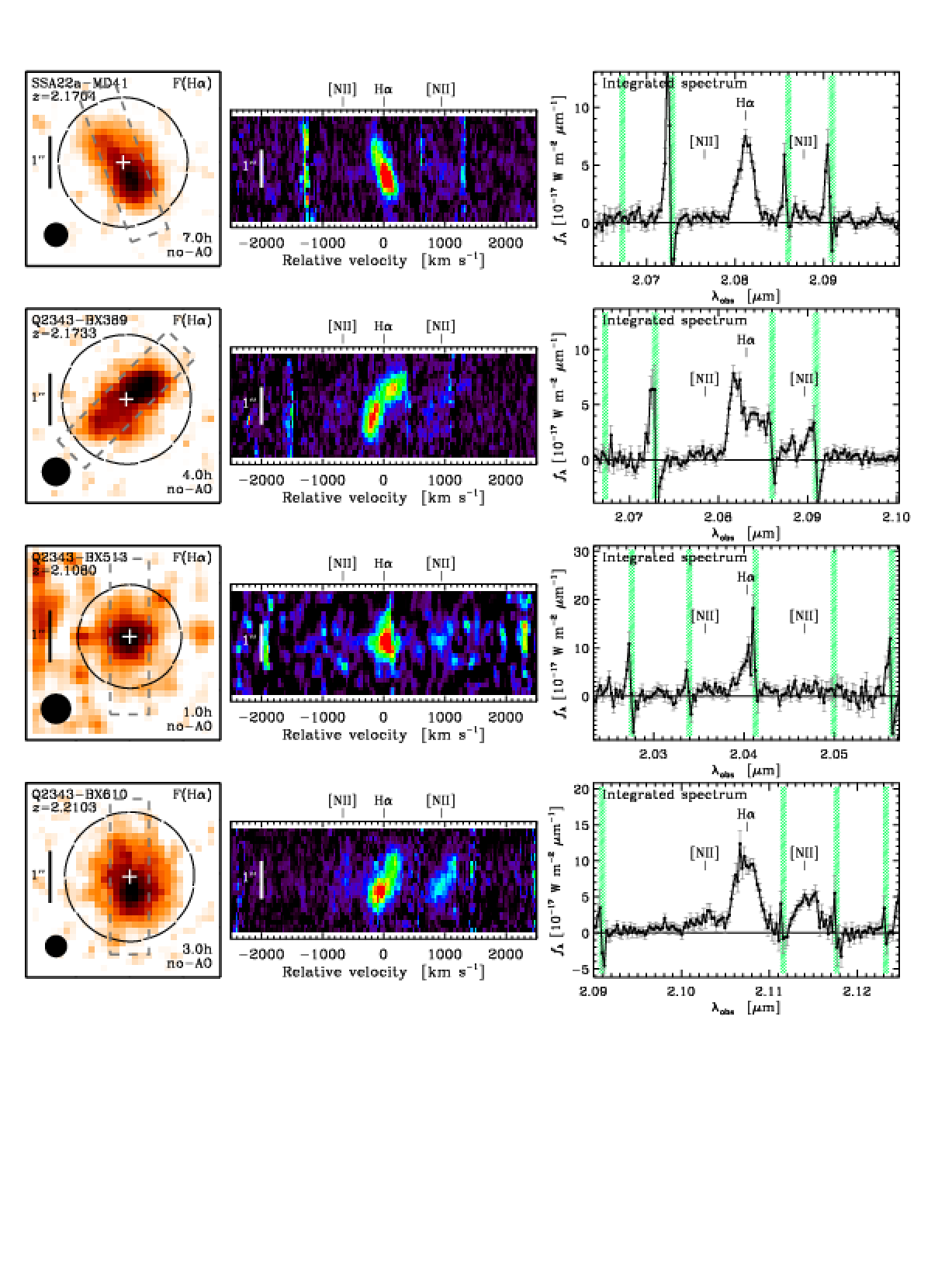}
\vspace{-0.0cm}
\caption{
\small
Same as Figure~\ref{fig-maps_bmbx1} for \bxbm\ galaxies
of the SINS H$\alpha$ sample taken from the NIRSPEC long-slit
sample of \citet{Erb06b}.
\label{fig-maps_bmbx3}
}
\end{figure}

\clearpage

\begin{figure}[p]
\figurenum{25}
\epsscale{1.15}
\plotone{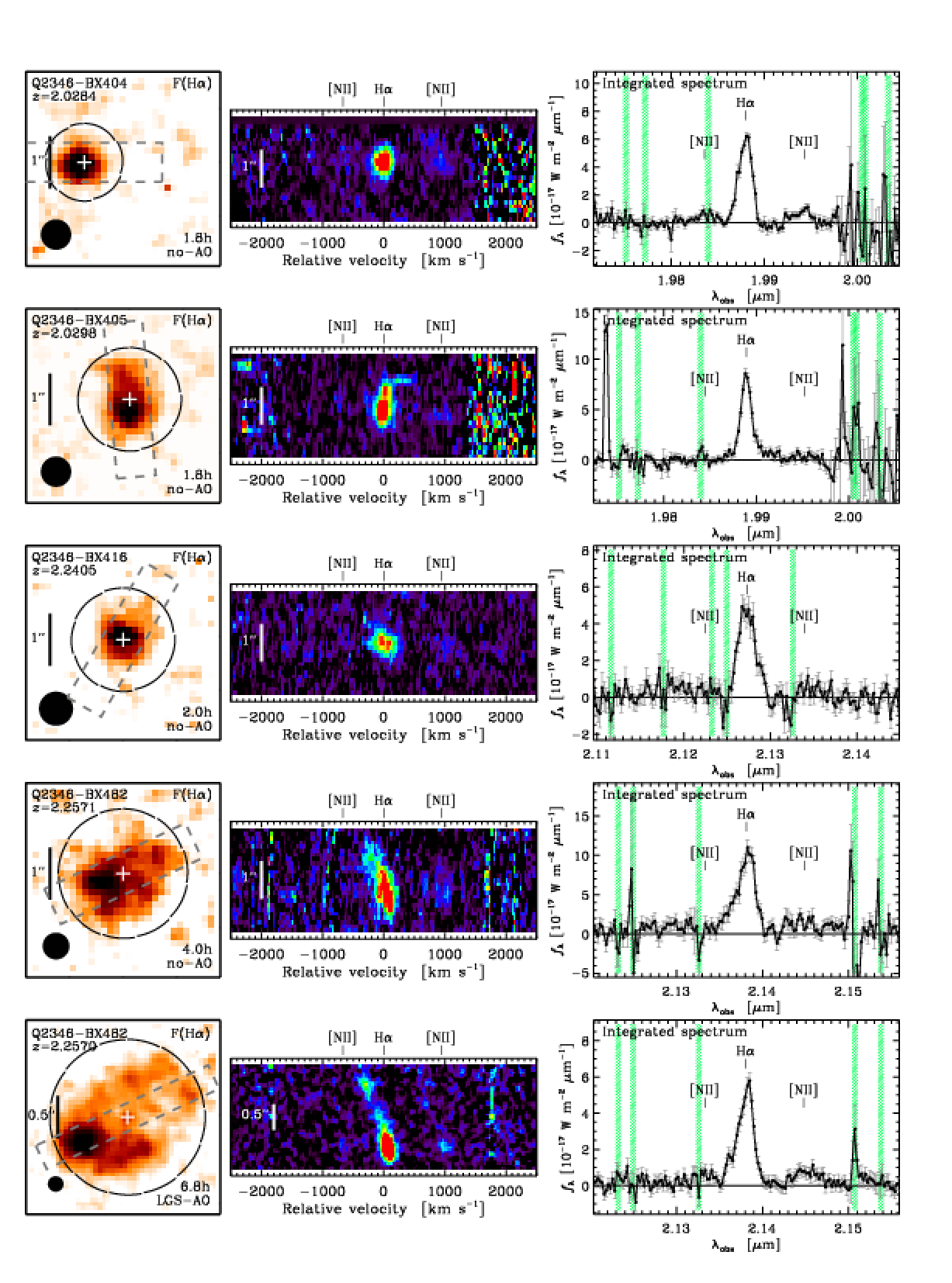}
\vspace{-0.0cm}
\caption{
\small
Same as Figure~\ref{fig-maps_bmbx1} for \bxbm\ galaxies
of the SINS H$\alpha$ sample taken from the NIRSPEC long-slit
sample of \citet{Erb06b}.
\label{fig-maps_bmbx4}
}
\end{figure}

\clearpage

\begin{figure}[p]
\figurenum{26}
\epsscale{1.15}
\plotone{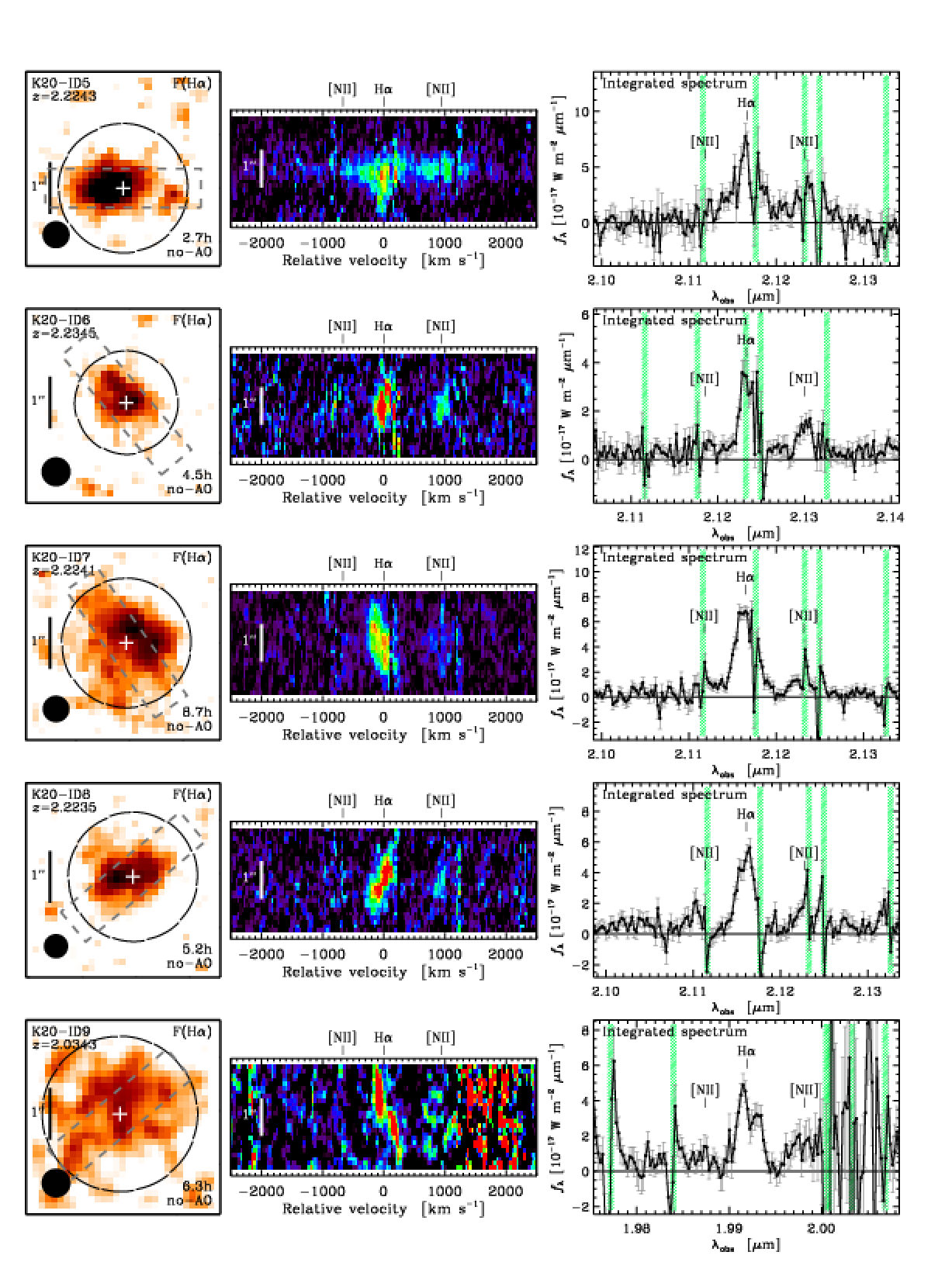}
\vspace{-0.0cm}
\caption{
\small
Same as Figure~\ref{fig-maps_bmbx1} for $K$-selected galaxies
of the SINS H$\alpha$ sample drawn from the K20 survey.
\label{fig-maps_k20}
}
\end{figure}

\clearpage

\begin{figure}[p]
\figurenum{27}
\epsscale{1.15}
\plotone{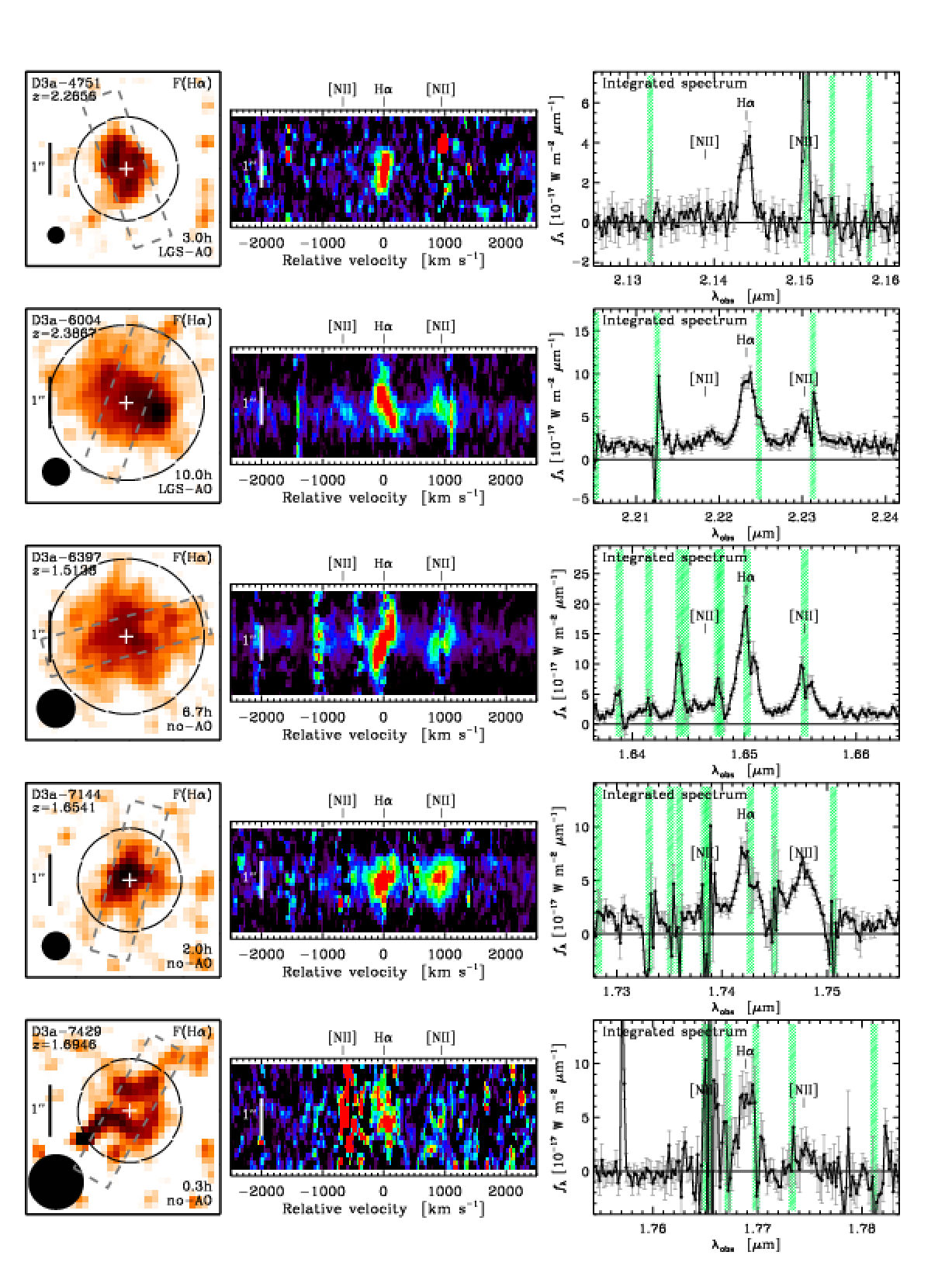}
\vspace{-0.0cm}
\caption{
\small
Same as Figure~\ref{fig-maps_bmbx1} for $BzK$-selected galaxies
of the SINS H$\alpha$ sample in the Deep3a field.
\label{fig-maps_deep3a1}
}
\end{figure}

\clearpage

\begin{figure}[p]
\figurenum{28}
\epsscale{1.15}
\plotone{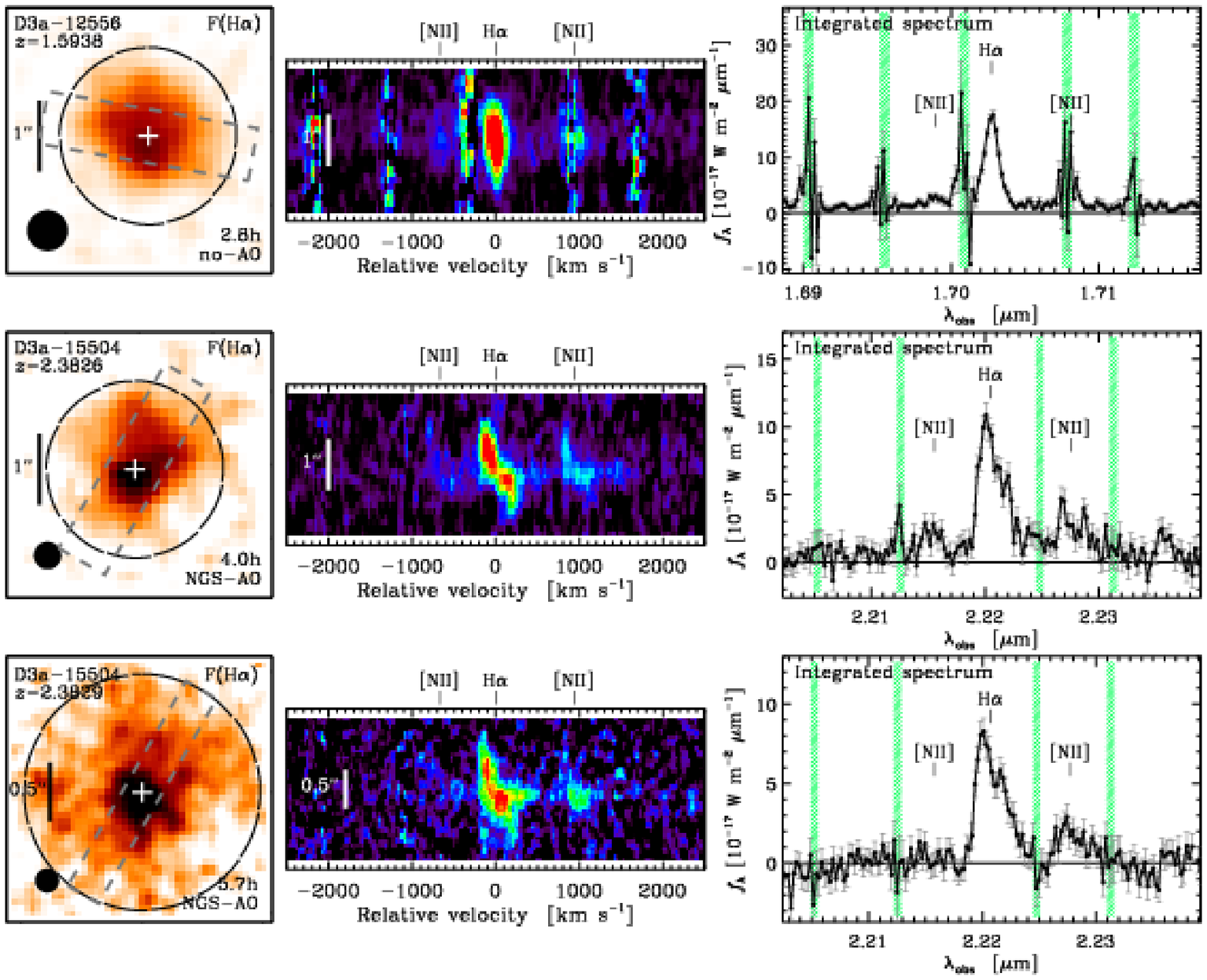}
\vspace{-0.0cm}
\caption{
\small
Same as Figure~\ref{fig-maps_bmbx1} for $BzK$-selected galaxies
of the SINS H$\alpha$ sample in the Deep3a field.
\label{fig-maps_deep3a2}
}
\end{figure}

\clearpage

\begin{figure}[p]
\figurenum{29}
\epsscale{1.15}
\plotone{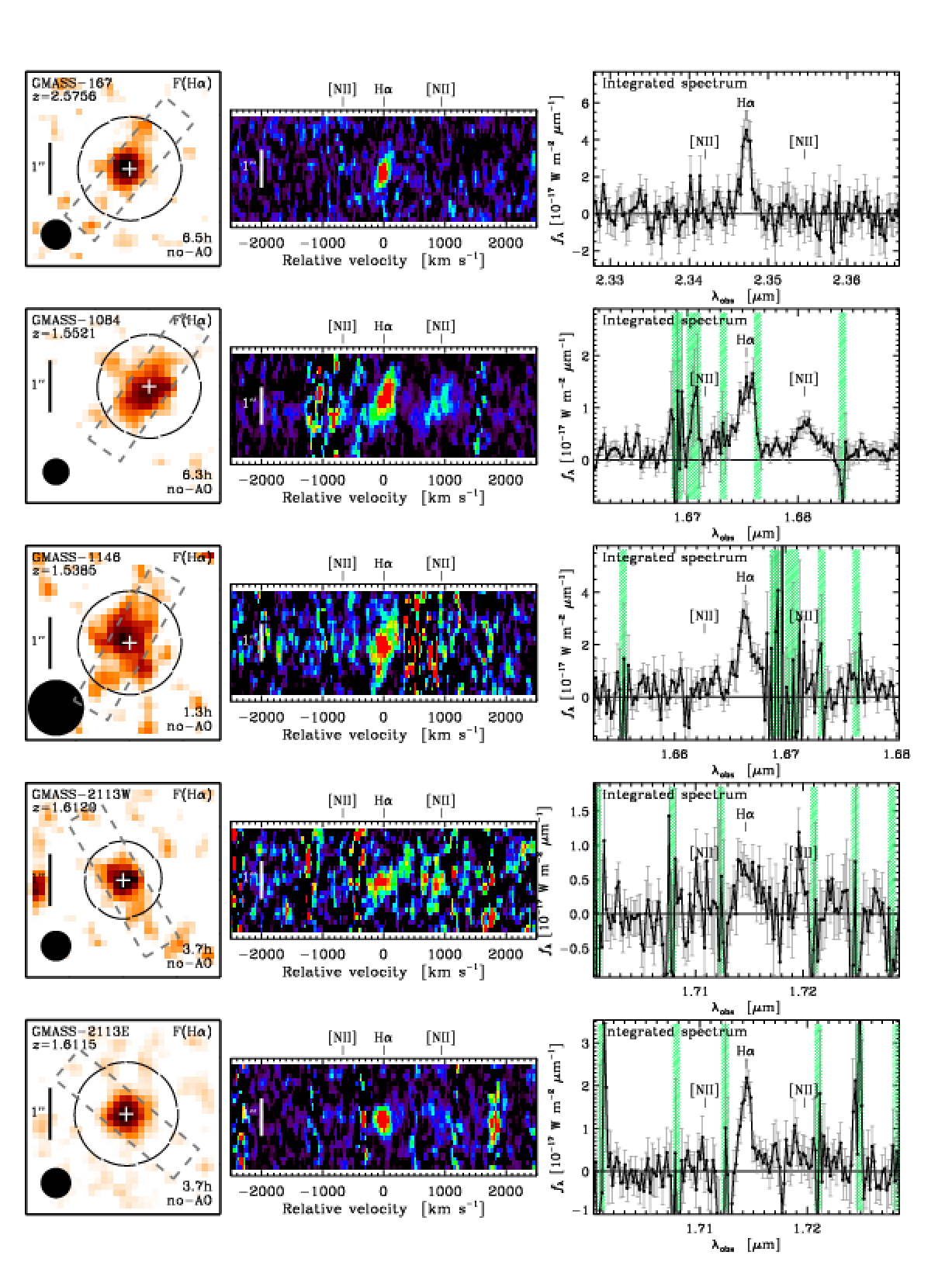}
\vspace{-0.0cm}
\caption{
\small
Same as Figure~\ref{fig-maps_bmbx1} for $\rm 4.5~\mu m$-selected
galaxies of the SINS H$\alpha$ sample drawn from the GMASS survey.
\label{fig-maps_gmass1}
}
\end{figure}

\clearpage

\begin{figure}[p]
\figurenum{30}
\epsscale{1.15}
\plotone{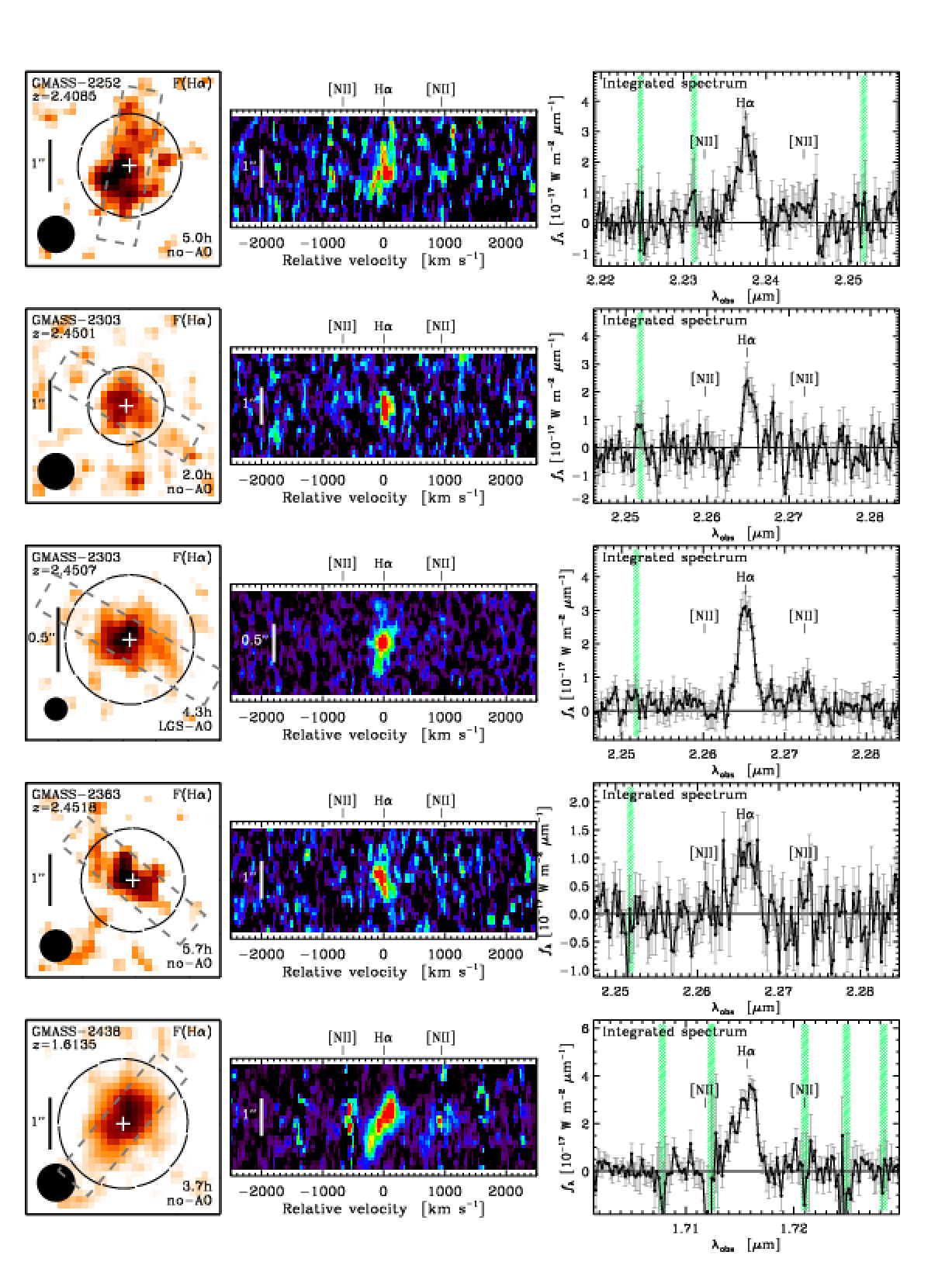}
\vspace{-0.0cm}
\caption{
\small
Same as Figure~\ref{fig-maps_bmbx1} for $\rm 4.5~\mu m$-selected
galaxies of the SINS H$\alpha$ sample drawn from the GMASS survey.
\label{fig-maps_gmass2}
}
\end{figure}

\clearpage

\begin{figure}[p]
\figurenum{31}
\epsscale{1.15}
\plotone{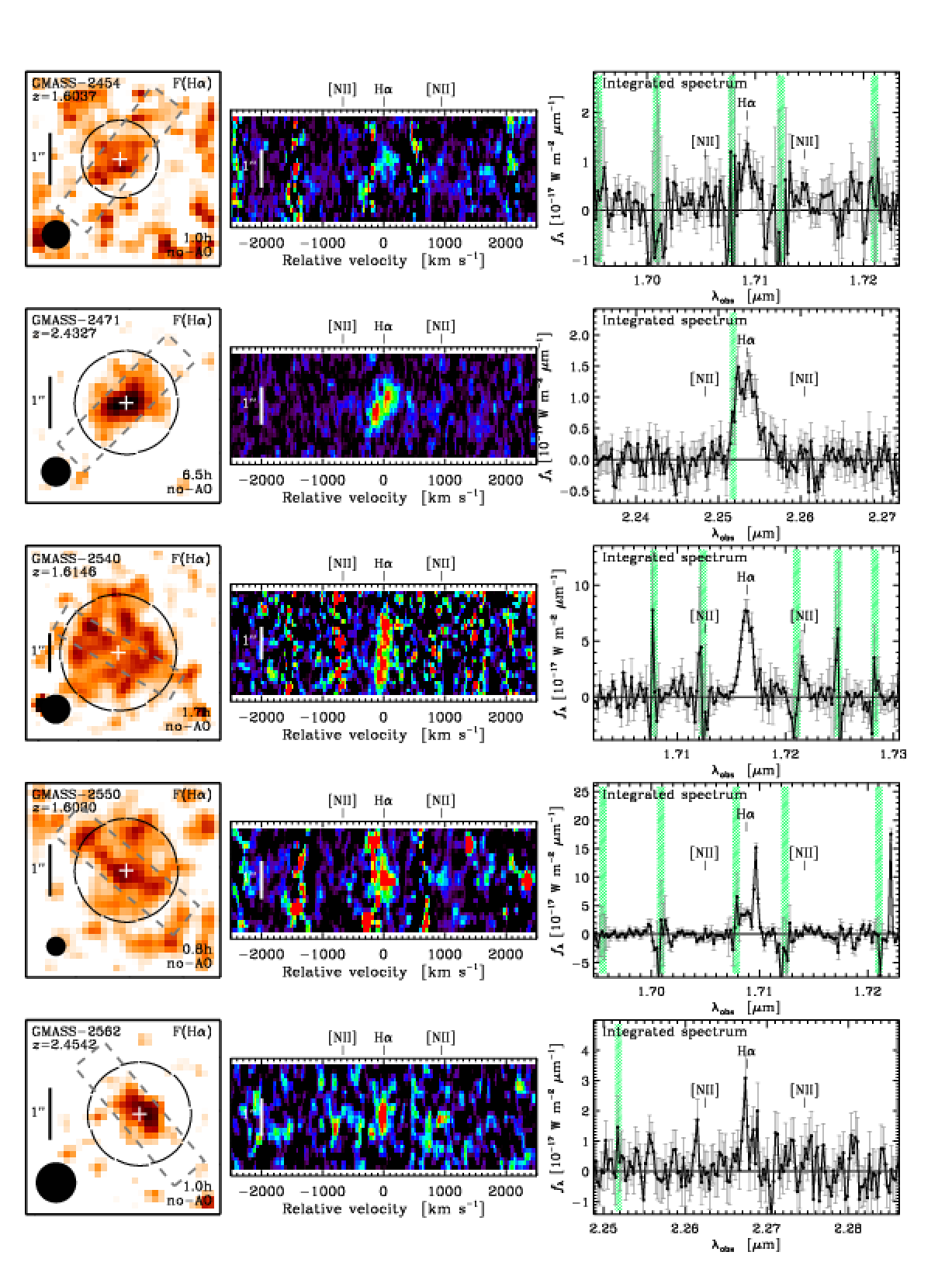}
\vspace{-0.0cm}
\caption{
\small
Same as Figure~\ref{fig-maps_bmbx1} for $\rm 4.5~\mu m$-selected
galaxies of the SINS H$\alpha$ sample drawn from the GMASS survey.
\label{fig-maps_gmass3}
}
\end{figure}

\clearpage

\begin{figure}[p]
\figurenum{32}
\epsscale{1.15}
\plotone{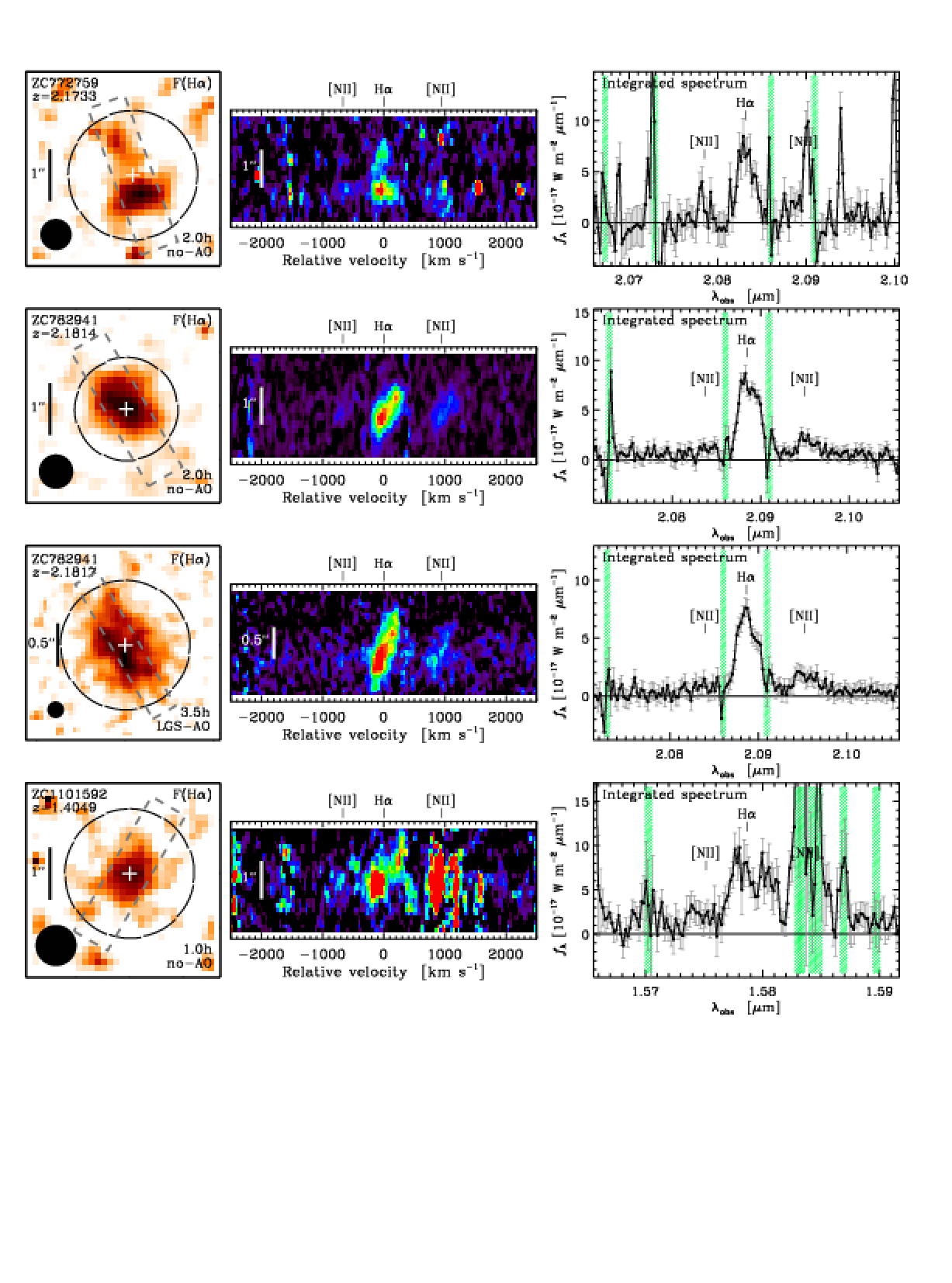}
\vspace{-0.0cm}
\caption{
\small
Same as Figure~\ref{fig-maps_bmbx1} for $BzK$-selected galaxies
of the SINS H$\alpha$ sample drawn from the $z$COSMOS survey.
\label{fig-maps_zcosmos}
}
\end{figure}

\clearpage

\begin{figure}[p]
\figurenum{33}
\epsscale{1.15}
\plotone{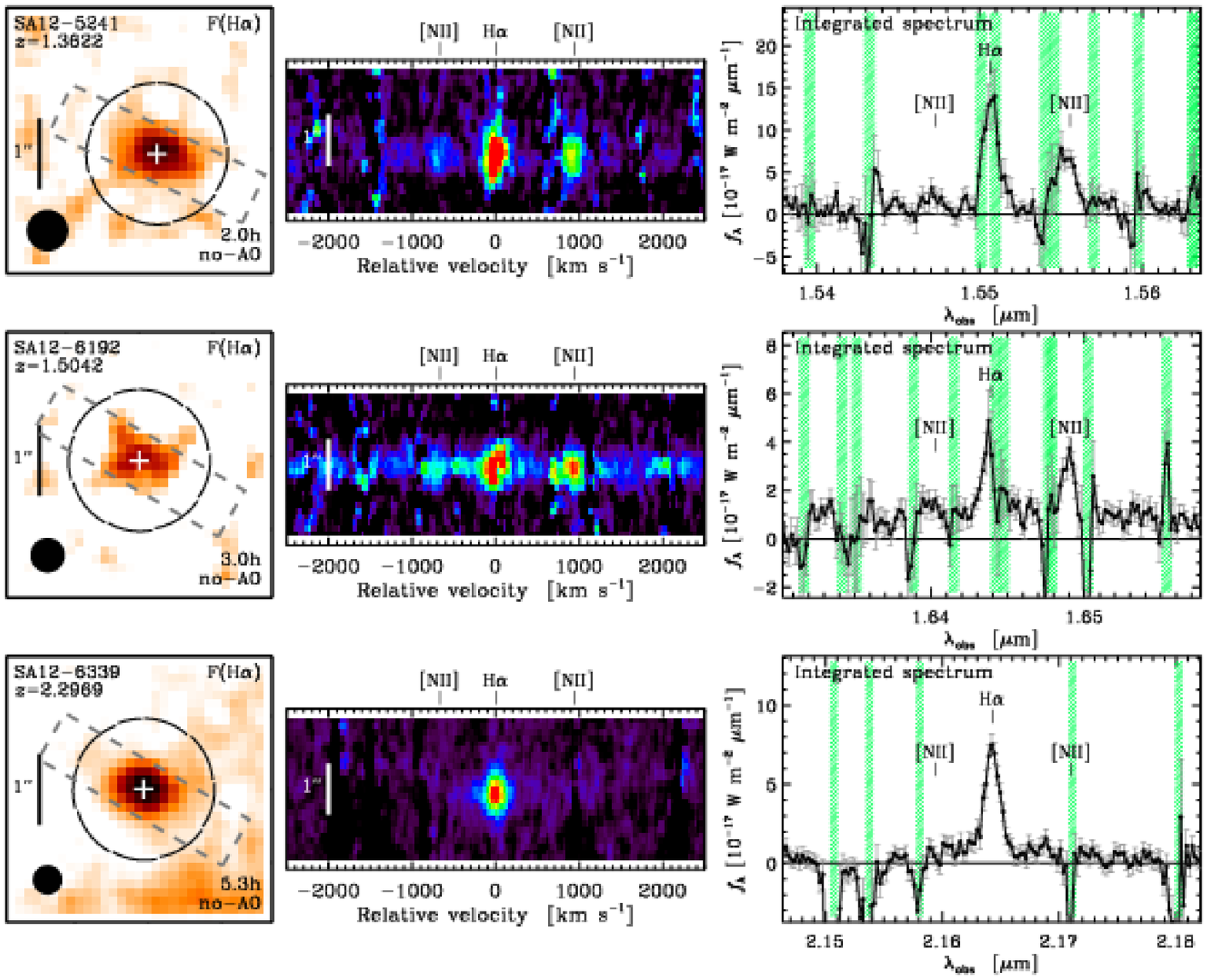}
\vspace{-0.0cm}
\caption{
\small
Same as Figure~\ref{fig-maps_bmbx1} for $K$-selected galaxies
of the SINS H$\alpha$ sample drawn from the GDDS survey.
\label{fig-maps_gdds1}
}
\end{figure}

\clearpage

\begin{figure}[p]
\figurenum{34}
\epsscale{1.15}
\plotone{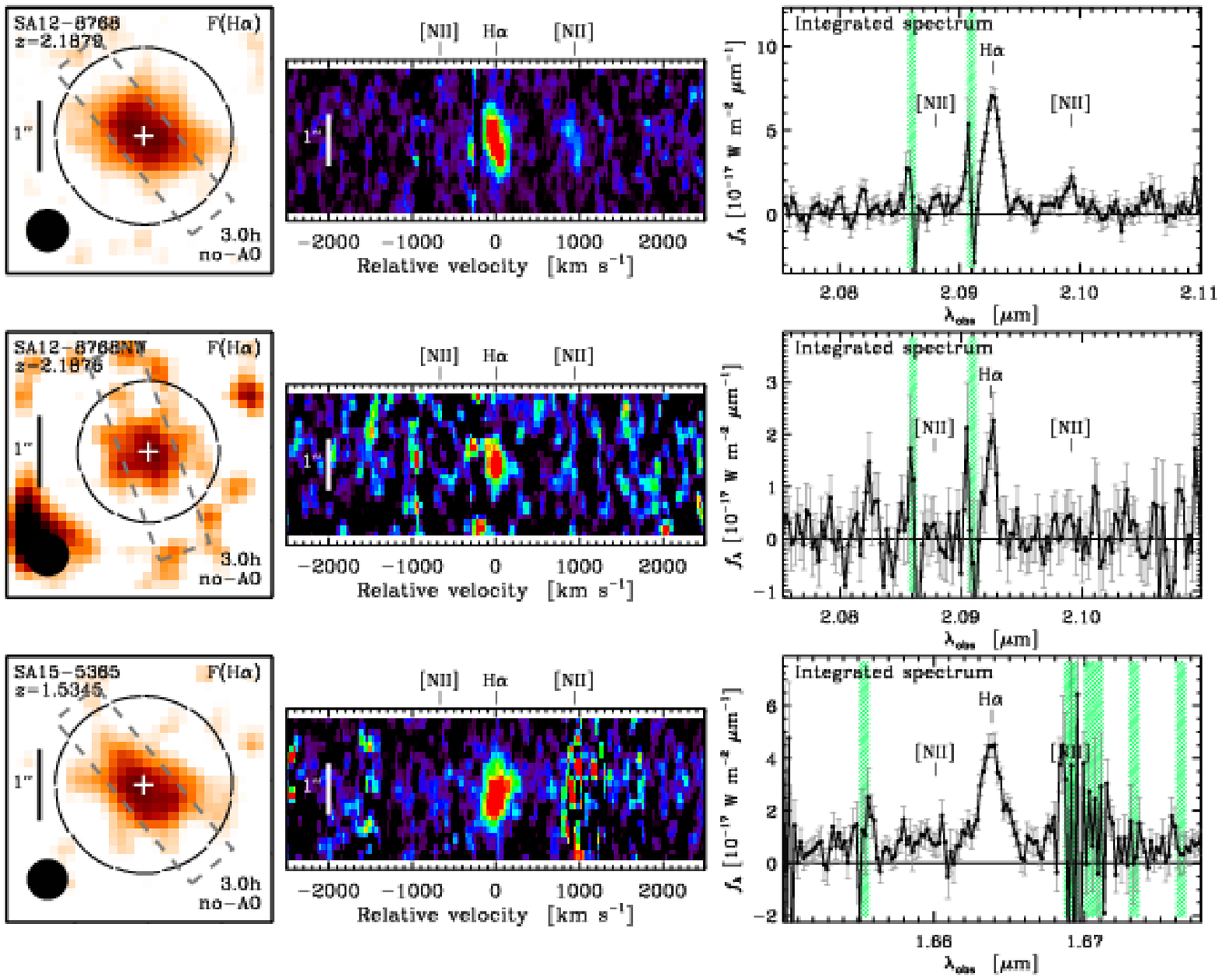}
\vspace{-0.0cm}
\caption{
\small
Same as Figure~\ref{fig-maps_bmbx1} for $K$-selected galaxies
of the SINS H$\alpha$ sample drawn from the GDDS survey.
\label{fig-maps_gdds2}
}
\end{figure}

\end{document}